%% file: LHCb-PAPER-2018-034.tex
\pdfoutput=1
\documentclass[12pt,a4paper]{article}
\usepackage{ifthen}
\newboolean{pdflatex}
\setboolean{pdflatex}{true}
\newboolean{articletitles}
\setboolean{articletitles}{true}
\newboolean{uprightparticles}
\setboolean{uprightparticles}{false}
\newboolean{inbibliography}
\setboolean{inbibliography}{false}

\def\paperauthors{LHCb collaboration}
\def\paperasciititle{Evidence for an eta_c(1S)pi^- resonance in B^0 -> eta_c(1S) K^+pi^- decays}
\def\papertitle{Evidence for an $\eta_c(1S) \pi^-$ resonance in $B^0 \to \eta_c(1S) K^+\pi^-$ decays}
\def\paperkeywords{{High Energy Physics}, {LHCb}, {QCD}, {exotics}, {B
  physics}, {quarkonium spectroscopy}, {branching fraction}}
\def\papercopyright{\the\year\ CERN for the benefit of the LHCb collaboration}
\def\paperlicence{CC-BY-4.0 licence}
\def\paperlicenceurl{https://creativecommons.org/licenses/by/4.0/}

\input{preamble}
\usepackage{longtable}
\usepackage{comment}
\usepackage{subfigure}
\usepackage[]{units}
\usepackage{booktabs}
\usepackage{rotating}
\usepackage{dcolumn}
\newcommand{\decaywo}{\decay{\Bd}{\etac \Kp \pim}}
\newcommand{\ppKpi}{\decay{\Bd}{\proton \antiproton \Kp \pim}}
\newcommand{\jpsiKpi}{\decay{\Bd}{\jpsi \Kp \pim}}
\newcommand{\mppKpi}{\ensuremath{m(\proton \antiproton \Kp \pim)}}
\newcommand{\mpp}{\ensuremath{m(\proton \antiproton)}}
\newcommand{\Kpi}{\ensuremath{\Kp \pim}}
\newcommand{\etacpi}{\ensuremath{\etac \pim}}
\newcommand{\MKpi}{\ensuremath{m^2(\Kp \pim)}}
\newcommand{\mKpi}{\ensuremath{m(\Kp \pim)}}
\newcommand{\etacK}{\ensuremath{\etac \Kp}}
\newcommand{\Metacpi}{\ensuremath{m^2(\etac \pim)}}

\newcommand{\metacpi}{\ensuremath{m(\etac \pim)}}
\newcommand{\metacK}{\ensuremath{m(\etac \Kp)}}

\usepackage[textwidth=2cm,colorinlistoftodos]{todonotes}

\begin{document}

\renewcommand{\thefootnote}{\fnsymbol{footnote}}
\setcounter{footnote}{1}

\input{title-LHCb-PAPER}

\renewcommand{\thefootnote}{\arabic{footnote}}
\setcounter{footnote}{0}
\pagestyle{plain}
\setcounter{page}{1}
\pagenumbering{arabic}

\input{introduction}

\input{detector}

\input{selection}

\input{branchingFraction}

\input{dalitzPlot}

\input{dalitzFit}

\input{systematics}

\input{results}

\input{acknowledgements}

\input{appendixB2etacKpi}

\addcontentsline{toc}{section}{References}
\setboolean{inbibliography}{true}
\bibliographystyle{LHCb}
\bibliography{standard,LHCb-PAPER,LHCb-CONF,LHCb-DP,LHCb-TDR,main}

\newpage
 
\newpage

\input{LHCb_Authorship_flat_24-Jul-2018.tex}

\end{document}

%% file: preamble.tex
\usepackage[top=1in, bottom=1.25in, left=1in, right=1in]{geometry}
\usepackage{microtype}
\usepackage{lineno}  % for line numbering during review
\usepackage{xspace} % To avoid problems with missing or double spaces after
\usepackage{caption} %these three command get the figure and table captions automatically small

\usepackage{graphicx}  % to include figures (can also use other packages)
\usepackage{color}
\usepackage{colortbl}
\graphicspath{{./figs/}} % Make Latex search fig subdir for figures

\usepackage{amsmath} % Adds a large collection of math symbols
\usepackage{amssymb}
\usepackage{amsfonts}
\usepackage{upgreek} % Adds in support for greek letters in roman typeset

\newcommand*\patchAmsMathEnvironmentForLineno[1]{%
\expandafter\let\csname old#1\expandafter\endcsname\csname #1\endcsname
\expandafter\let\csname oldend#1\expandafter\endcsname\csname
end#1\endcsname
 \renewenvironment{#1}%
   {\linenomath\csname old#1\endcsname}%
   {\csname oldend#1\endcsname\endlinenomath}%
}
\newcommand*\patchBothAmsMathEnvironmentsForLineno[1]{%
  \patchAmsMathEnvironmentForLineno{#1}%
  \patchAmsMathEnvironmentForLineno{#1*}%
}
\AtBeginDocument{%
\patchBothAmsMathEnvironmentsForLineno{equation}%
\patchBothAmsMathEnvironmentsForLineno{align}%
\patchBothAmsMathEnvironmentsForLineno{flalign}%
\patchBothAmsMathEnvironmentsForLineno{alignat}%
\patchBothAmsMathEnvironmentsForLineno{gather}%
\patchBothAmsMathEnvironmentsForLineno{multline}%
\patchBothAmsMathEnvironmentsForLineno{eqnarray}%
}

\usepackage{hyperxmp}

\usepackage[pdftex,
            pdfauthor={\paperauthors},
            pdftitle={\paperasciititle},
            pdfkeywords={\paperkeywords},
            pdfcopyright={Copyright (C) \papercopyright},
            pdflicenseurl={\paperlicenceurl}]{hyperref}

\usepackage[all]{hypcap} % Internal hyperlinks to floats.

\input{lhcb-symbols-def} % Add in the predefined LHCb symbols

\usepackage{cite} % Allows for ranges in citations
\usepackage{mciteplus}

%% file: lhcb-symbols-def.tex
\usepackage{xspace} 
\usepackage{upgreek}

\def\lhcb {\mbox{LHCb}\xspace}

\def\belle  {\mbox{Belle}\xspace}
\def\cleo   {\mbox{CLEO}\xspace}

\def\rich   {RICH\xspace}

\def\MagUp {\mbox{\em Mag\kern -0.05em Up}\xspace}

\ifthenelse{\boolean{uprightparticles}}%
{

 \def\Peta        {\ensuremath{\upeta}\xspace}

 \def\Ppi         {\ensuremath{\uppi}\xspace}

 \def\Ppsi        {\ensuremath{\uppsi}\xspace}

 \def\PDelta      {\ensuremath{\Delta}\xspace}                 
 \def\PXi      {\ensuremath{\Xi}\xspace}                 
 \def\PLambda      {\ensuremath{\Lambda}\xspace}                 
 \def\PSigma      {\ensuremath{\Sigma}\xspace}                 
 \def\POmega      {\ensuremath{\Omega}\xspace}                 
 \def\PUpsilon      {\ensuremath{\Upsilon}\xspace}                 
 
 \def\PB      {\ensuremath{\mathrm{B}}\xspace}                 
                  
 \def\PD      {\ensuremath{\mathrm{D}}\xspace}

 \def\PJ      {\ensuremath{\mathrm{J}}\xspace}                 
 \def\PK      {\ensuremath{\mathrm{K}}\xspace}

 \def\Pb      {\ensuremath{\mathrm{b}}\xspace}                 
 \def\Pc      {\ensuremath{\mathrm{c}}\xspace}

 \def\Pi      {\ensuremath{\mathrm{i}}\xspace}

 \def\Pp      {\ensuremath{\mathrm{p}}\xspace}

 \def\Ps      {\ensuremath{\mathrm{s}}\xspace}

}
{

 \def\Peta        {\ensuremath{\eta}\xspace}

 \def\Ppi         {\ensuremath{\pi}\xspace}

 \def\Ppsi        {\ensuremath{\psi}\xspace}                 
                  
 \mathchardef\PDelta="7101
 \mathchardef\PXi="7104
 \mathchardef\PLambda="7103
 \mathchardef\PSigma="7106
 \mathchardef\POmega="710A
 \mathchardef\PUpsilon="7107
                  
 \def\PB      {\ensuremath{B}\xspace}                 
                  
 \def\PD      {\ensuremath{D}\xspace}

 \def\PJ      {\ensuremath{J}\xspace}                 
 \def\PK      {\ensuremath{K}\xspace}

 \def\Pb      {\ensuremath{b}\xspace}                 
 \def\Pc      {\ensuremath{c}\xspace}

 \def\Pi      {\ensuremath{i}\xspace}

 \def\Pp      {\ensuremath{p}\xspace}

 \def\Ps      {\ensuremath{s}\xspace}

}

\makeatletter
\ifcase \@ptsize \relax% 10pt
  \newcommand{\miniscule}{\@setfontsize\miniscule{4}{5}}% \tiny: 5/6
\or% 11pt
  \newcommand{\miniscule}{\@setfontsize\miniscule{5}{6}}% \tiny: 6/7
\or% 12pt
  \newcommand{\miniscule}{\@setfontsize\miniscule{5}{6}}% \tiny: 6/7
\fi
\makeatother

\DeclareRobustCommand{\optbar}[1]{\shortstack{{\miniscule (\rule[.5ex]{1.25em}{.18mm})}
  \\ [-.7ex] $#1$}}

\def\squark    {{\ensuremath{\Ps}}\xspace}

\def\cquark    {{\ensuremath{\Pc}}\xspace}

\def\bquark    {{\ensuremath{\Pb}}\xspace}

\def\pion   {{\ensuremath{\Ppi}}\xspace}
\def\piz    {{\ensuremath{\pion^0}}\xspace}

\def\pip    {{\ensuremath{\pion^+}}\xspace}
\def\pim    {{\ensuremath{\pion^-}}\xspace}

\def\kaon    {{\ensuremath{\PK}}\xspace}
  \def\Kbar    {{\kern 0.2em\overline{\kern -0.2em \PK}{}}\xspace}

\def\KorKbar    {\kern 0.18em\optbar{\kern -0.18em K}{}\xspace}

\def\Kp      {{\ensuremath{\kaon^+}}\xspace}
\def\Km      {{\ensuremath{\kaon^-}}\xspace}

\def\Kstarz  {{\ensuremath{\kaon^{*0}}}\xspace}

  \def\Dbar    {{\kern 0.2em\overline{\kern -0.2em \PD}{}}\xspace}

\def\DorDbar    {\kern 0.18em\optbar{\kern -0.18em D}{}\xspace}

\def\Dzb     {{\ensuremath{\Dbar{}^0}}\xspace}

\def\B       {{\ensuremath{\PB}}\xspace}
\def\Bbar    {{\ensuremath{\kern 0.18em\overline{\kern -0.18em \PB}{}}}\xspace}

\def\BorBbar    {\kern 0.18em\optbar{\kern -0.18em B}{}\xspace}

\def\Bd      {{\ensuremath{\B^0}}\xspace}
\def\Bs      {{\ensuremath{\B^0_\squark}}\xspace}

\def\jpsi     {{\ensuremath{{\PJ\mskip -3mu/\mskip -2mu\Ppsi\mskip 2mu}}}\xspace}

\def\etac     {{\ensuremath{\Peta_\cquark}}\xspace}

  \def\Y#1S{\ensuremath{\PUpsilon{(#1S)}}\xspace}% no space before {...}!

\def\proton      {{\ensuremath{\Pp}}\xspace}
\def\antiproton  {{\ensuremath{\overline \proton}}\xspace}

\def\Lbar        {{\ensuremath{\kern 0.1em\overline{\kern -0.1em\PLambda}}}\xspace}
\def\LorLbar    {\kern 0.18em\optbar{\kern -0.18em \PLambda}{}\xspace}

\def\Lcbar   {{\ensuremath{\Lbar{}^-_\cquark}}\xspace}

\newcommand{\decay}[2]{\ensuremath{#1\!\to #2}\xspace}         % {\Pa}{\Pb \Pc}

\def\to                 {\ensuremath{\rightarrow}\xspace}

\def\AT#1     {\ensuremath{A_{\mathrm{T}}^{#1}}\xspace}           % 2

\def\C#1      {\ensuremath{\mathcal{C}_{#1}}\xspace}                       % 9
\def\Cp#1     {\ensuremath{\mathcal{C}_{#1}^{'}}\xspace}                    % 7
\def\Ceff#1   {\ensuremath{\mathcal{C}_{#1}^{\mathrm{(eff)}}}\xspace}        % 9  
\def\Cpeff#1  {\ensuremath{\mathcal{C}_{#1}^{'\mathrm{(eff)}}}\xspace}       % 7
\def\Ope#1    {\ensuremath{\mathcal{O}_{#1}}\xspace}                       % 2
\def\Opep#1   {\ensuremath{\mathcal{O}_{#1}^{'}}\xspace}                    % 7

\newcommand{\unit}[1]{\ensuremath{\mathrm{ \,#1}}\xspace}          % {kg}

\newcommand{\tev}{\ifthenelse{\boolean{inbibliography}}{\ensuremath{~T\kern -0.05em eV}}{\ensuremath{\mathrm{\,Te\kern -0.1em V}}}\xspace}
\newcommand{\gev}{\ensuremath{\mathrm{\,Ge\kern -0.1em V}}\xspace}
\newcommand{\mev}{\ensuremath{\mathrm{\,Me\kern -0.1em V}}\xspace}
\newcommand{\kev}{\ensuremath{\mathrm{\,ke\kern -0.1em V}}\xspace}
\newcommand{\ev}{\ensuremath{\mathrm{\,e\kern -0.1em V}}\xspace}
\newcommand{\gevc}{\ensuremath{{\mathrm{\,Ge\kern -0.1em V\!/}c}}\xspace}
\newcommand{\mevc}{\ensuremath{{\mathrm{\,Me\kern -0.1em V\!/}c}}\xspace}
\newcommand{\gevcc}{\ensuremath{{\mathrm{\,Ge\kern -0.1em V\!/}c^2}}\xspace}
\newcommand{\gevgevcccc}{\ensuremath{{\mathrm{\,Ge\kern -0.1em V^2\!/}c^4}}\xspace}
\newcommand{\mevcc}{\ensuremath{{\mathrm{\,Me\kern -0.1em V\!/}c^2}}\xspace}

\def\mum  {\ensuremath{{\,\upmu\mathrm{m}}}\xspace}

\def\fm   {\ensuremath{\mathrm{ \,fm}}\xspace}

\def\invfb   {\ensuremath{\mbox{\,fb}^{-1}}\xspace}

\newcommand{\chisq}{\ensuremath{\chi^2}\xspace}

\newcommand{\chisqip}{\ensuremath{\chi^2_{\text{IP}}}\xspace}

\newcommand{\chisqvtx}{\ensuremath{\chi^2_{\text{vtx}}}\xspace}

\def\gsim{{~\raise.15em\hbox{$>$}\kern-.85em
          \lower.35em\hbox{$\sim$}~}\xspace}
\def\lsim{{~\raise.15em\hbox{$<$}\kern-.85em
          \lower.35em\hbox{$\sim$}~}\xspace}

\newcommand{\Real}{\ensuremath{\mathcal{R}e}\xspace}

\def\sPlot{\mbox{\em sPlot}\xspace}

\def\sqs   {\ensuremath{\protect\sqrt{s}}\xspace}

\def\ptot       {\mbox{$p$}\xspace}
\def\pt         {\mbox{$p_{\mathrm{ T}}$}\xspace}

\def\evtgen     {\mbox{\textsc{EvtGen}}\xspace}

\def\geant      {\mbox{\textsc{Geant4}}\xspace}

\def\photos     {\mbox{\textsc{Photos}}\xspace}

\def\pythia     {\mbox{\textsc{Pythia}}\xspace}
\def\laura {\mbox{\textsc{Laura$^{++}$}}\xspace}
\def\jfit     {\mbox{\textsc{JFIT}}\xspace}

\def\roofit     {\mbox{\textsc{RooFit}}\xspace}

\def\tell1  {TELL1\xspace}
\def\ukl1   {UKL1\xspace}

\newcommand{\eg}{\mbox{\itshape e.g.}\xspace}
\newcommand{\ie}{\mbox{\itshape i.e.}\xspace}

%% file: title-LHCb-PAPER.tex
\begin{titlepage}
\pagenumbering{roman}

% Header ---------------------------------------------------
\vspace*{-1.5cm}
\centerline{\large EUROPEAN ORGANIZATION FOR NUCLEAR RESEARCH (CERN)}
\vspace*{1.5cm}
\noindent
\begin{tabular*}{\linewidth}{lc@{\extracolsep{\fill}}r@{\extracolsep{0pt}}}
\ifthenelse{\boolean{pdflatex}}% Logo format choice
{\vspace*{-1.5cm}\mbox{\!\!\!\includegraphics[width=.14\textwidth]{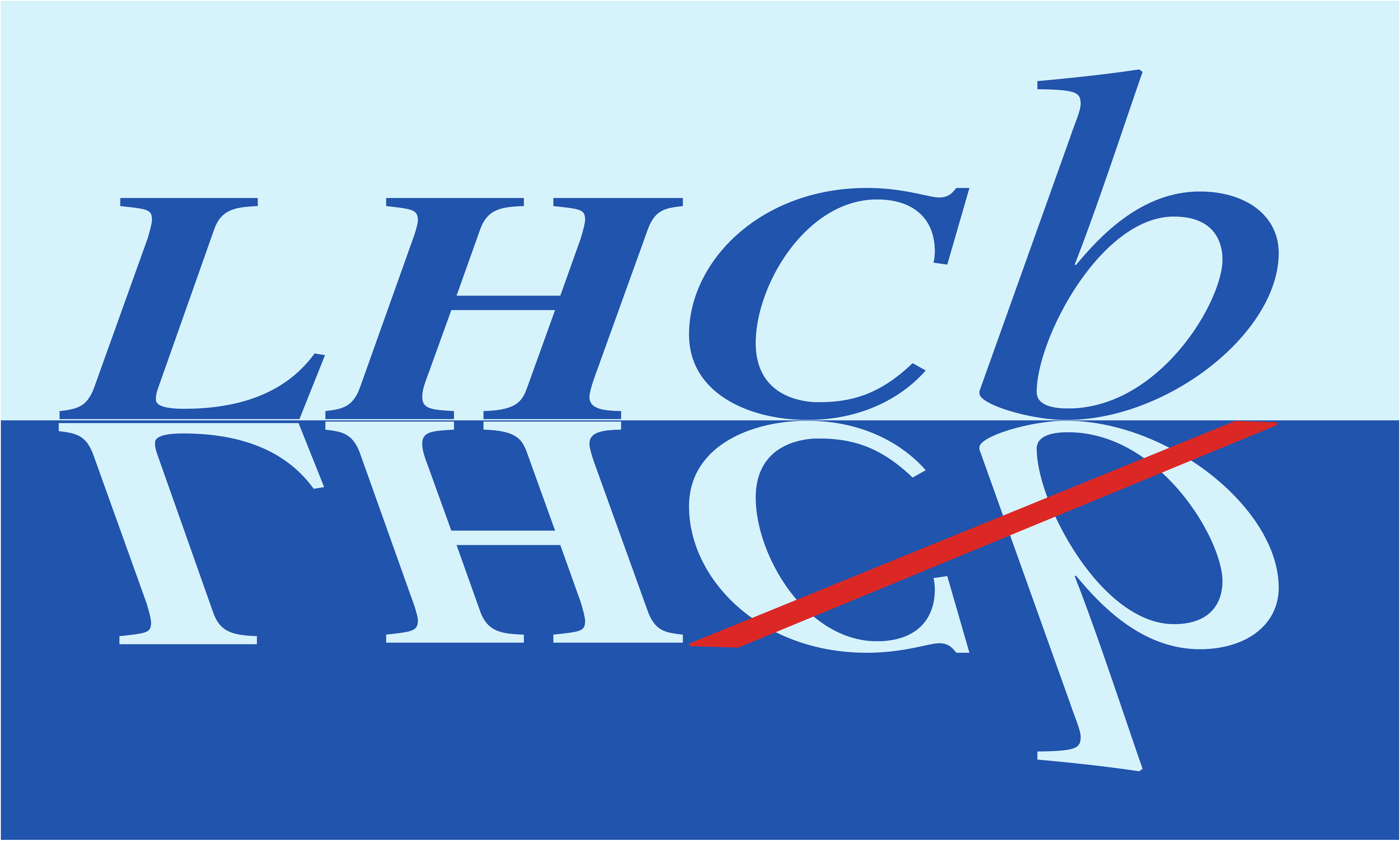}} & &}%
{\vspace*{-1.2cm}\mbox{\!\!\!\includegraphics[width=.12\textwidth]{lhcb-logo.eps}} & &}%
\\
 & & CERN-EP-2018-245 \\  % ID 
 & & LHCb-PAPER-2018-034 \\  % ID 
 & & December 18, 2018 \\  
 & & \\
\end{tabular*}

\vspace*{2.0cm}
%\vspace*{4.0cm}

% Title --------------------------------------------------
{\normalfont\bfseries\boldmath\huge
\begin{center}
  \papertitle 
\end{center}
}

\vspace*{2.0cm}

% Authors -------------------------------------------------
\begin{center}
\paperauthors\footnote{Authors are listed at the end of this paper.}
\end{center}

\vspace{\fill}

% Abstract -----------------------------------------------
\begin{abstract}
  \noindent
  A Dalitz plot analysis of \decay{\Bd}{\eta_c(1S) \Kp \pim} decays is performed
  using data samples of $pp$ collisions collected with the \lhcb
  detector at centre-of-mass energies of $\sqs=7,~8$ and $13\tev$,
  corresponding to a total integrated luminosity of $4.7 \invfb$. 
  A satisfactory description of the data is obtained when including a contribution representing an exotic $\eta_c(1S) \pi^-$ resonant state. 
  The significance of this exotic resonance is more than three standard deviations, while its
  mass and width are $4096 \pm 20~^{+18}_{-22} \mev$ and $152
  \pm 58~^{+60}_{-35} \mev$, respectively. The spin-parity
  assignments $J^P=0^+$ and $J^{P}=1^-$ are both consistent with the
  data.
  In addition, the first measurement of the \decay{\Bd}{\eta_c(1S) \Kp \pim} branching fraction is performed and gives

  \begin{center}
  $\displaystyle \mathcal{B}(\decay{\Bd}{\eta_c(1S) \Kp \pim}) = (5.73 \pm 0.24 \pm 0.13 \pm 0.66) \times 10^{-4}$,
  \end{center}
where the first uncertainty is statistical, the second systematic, and
the third is due to limited knowledge of external branching fractions.
\end{abstract}

\vspace{\fill}

\begin{center}
  Published in Eur.~Phys.~J.~{\bf C78} (2018) 1019
\end{center}

\vspace{\fill}

{\footnotesize 
\centerline{\copyright~\papercopyright. \href{\paperlicenceurl}{\paperlicence}.}}
\vspace*{2mm}

\end{titlepage}

%%%%%%%%%%%%%%%%%%%%%%%%%%%%%%%%
%%%%%  EOD OF TITLE PAGE  %%%%%%
%%%%%%%%%%%%%%%%%%%%%%%%%%%%%%%%

%  empty page follows the title page ----
\newpage
\setcounter{page}{2}
\mbox{~}

\cleardoublepage

%% file: introduction.tex
\section{Introduction}
\label{sec:Introduction}
Since the discovery of the $X(3872)$ state in 2003~\cite{PhysRevLett.91.262001}, several exotic hadron candidates have been observed, as reported in recent reviews~\cite{Chen:2016qju,Lebed:2016hpi,Esposito:2016noz,Guo:2017jvc,Ali:2017jda,RevModPhys.90.015003}.\footnote{The $X(3872)$ state has been recently renamed $\chi_{c1}(3872)$ in Ref.~\cite{PDG2018}}  The decay modes of these states indicate that they must contain a heavy quark-antiquark pair in their internal structure; however, they cannot easily be accommodated as an unassigned charmonium or bottomonium state due to either their mass, decay properties or electric charge, which are inconsistent with those of pure charmonium or bottomonium states. 
Different interpretations have been proposed about their nature~\cite{Chen:2016qju,Lebed:2016hpi,Esposito:2016noz}, including their quark composition and binding mechanisms. 
In order to improve the understanding of these hadrons, it is important to search for new exotic candidates, along with new production mechanisms and decay modes of already observed unconventional states.

The $Z_c(3900)^-$ state, discovered by the BESIII collaboration in the $\jpsi \pi^-$ final state~\cite{PhysRevLett.110.252001}, and confirmed by the \belle~\cite{PhysRevLett.110.252002} and \cleo~\cite{XIAO2013366} collaborations, can be interpreted as a hadrocharmonium state, where the compact heavy quark-antiquark pair interacts with the surrounding light quark mesonic excitation by a QCD analogue of the van der Waals force~\cite{PhysRevD.87.091501}. This interpretation of the $Z_c(3900)^-$ state predicts an as-yet-unobserved charged charmonium-like state with a mass of approximately $\unit[3800]{MeV}$ whose dominant decay mode is to the $\etacpi$ system.\footnote{Natural units with $\hbar = c = 1$ and the simplified notation $\etac$ to refer to the $\eta_c(1S)$ state are used throughout. In addition, the inclusion of charge-conjugate processes is always implied.} Alternatively, states like the $Z_c(3900)^-$ meson could be interpreted as analogues of quarkonium hybrids, where the excitation of the gluon field (the valence gluon) is replaced by an isospin-1 excitation of the gluon and light-quark fields~\cite{PhysRevLett.111.162003}. This interpretation, which is based on lattice QCD, predicts different multiplets of charmonium tetraquarks, comprising states with quantum numbers allowing the decay into the $\etacpi$ system. The $\etacpi$ system carries isospin $I=1$, $G$-parity $G=-1$, spin $J=L$ and parity $P=(-1)^{L}$, where $L$ is the orbital angular momentum between the \etac and the \pim mesons. Lattice QCD calculations~\cite{Liu:2012ze,Cheung:2016bym} predict the mass and quantum numbers of these states, comprising a $I^G(J^P)=1^-(0^+)$ state of mass $\unit[4025 \pm 49]{MeV}$, a $I^G(J^P)=1^-(1^-)$ state of mass $\unit[3770 \pm 42]{MeV}$, and a $I^G(J^P)=1^-(2^+)$ state of mass $\unit[4045 \pm 44]{MeV}$. The $Z_c(4430)^-$ resonance, discovered by the \belle collaboration~\cite{Mizuk:2009da} and confirmed by \lhcb~\cite{LHCb-PAPER-2014-014,LHCb-PAPER-2015-038}, could also fit into this scenario. Another prediction of a possible exotic candidate decaying to the $\etacpi$ system is provided by the diquark model~\cite{diQuarkModel}, where quarks and diquarks are the fundamental units to build a rich spectrum of hadrons, including the exotic states observed thus far. The diquark model predicts a $J^P=0^+$ candidate below the open-charm threshold that could decay into the $\etacpi$ final state. Therefore, the discovery of a charged charmonium-like meson in the $\etacpi$ system would provide important input towards understanding the nature of exotic hadrons.

\begin{figure}[tb]
   \centering
   \subfigure[]{
     \includegraphics[width=7.5cm, keepaspectratio]{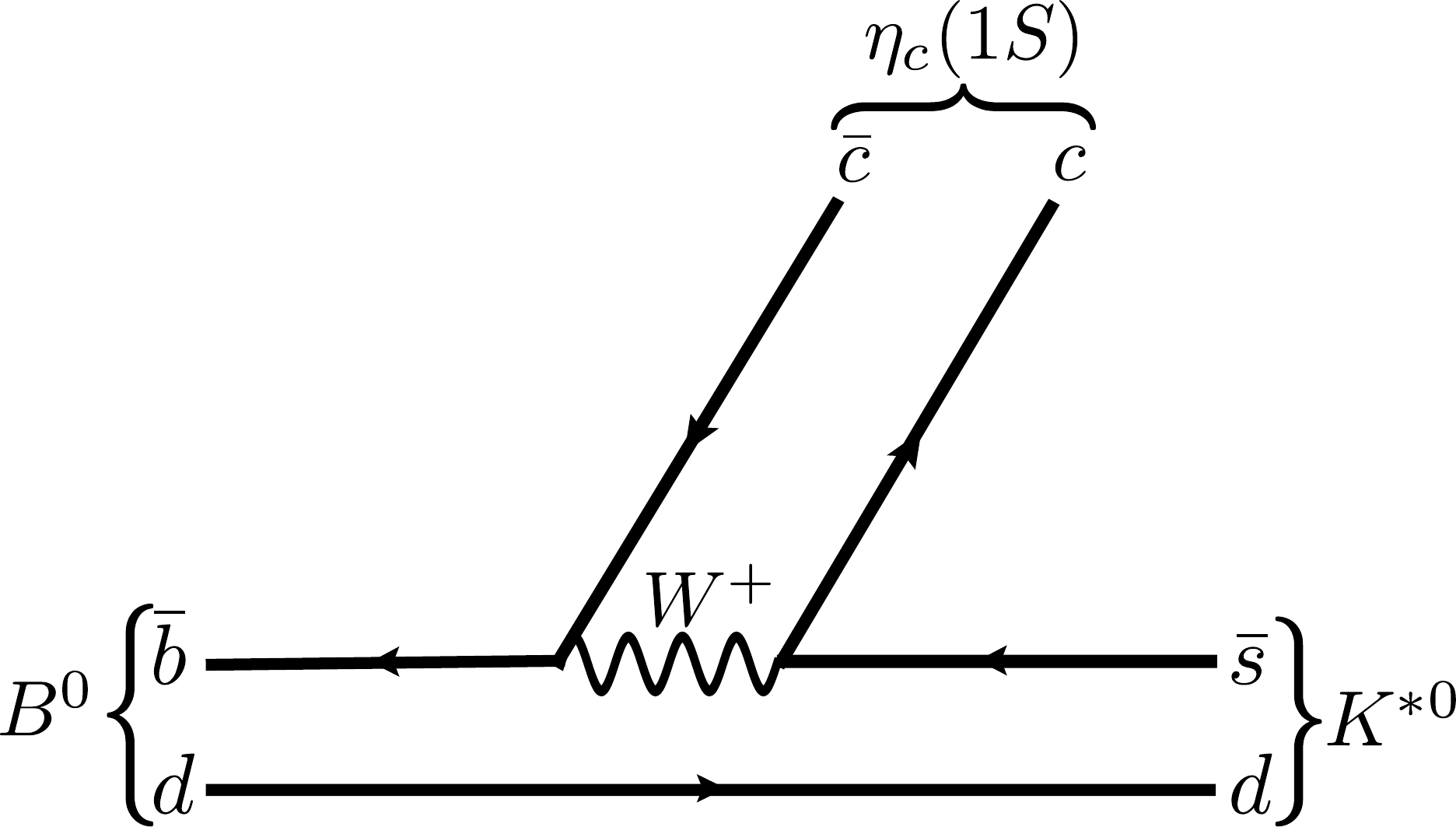}
     \label{B2etacKstar}}
   \subfigure[]{
     \includegraphics[width=7.5cm, keepaspectratio]{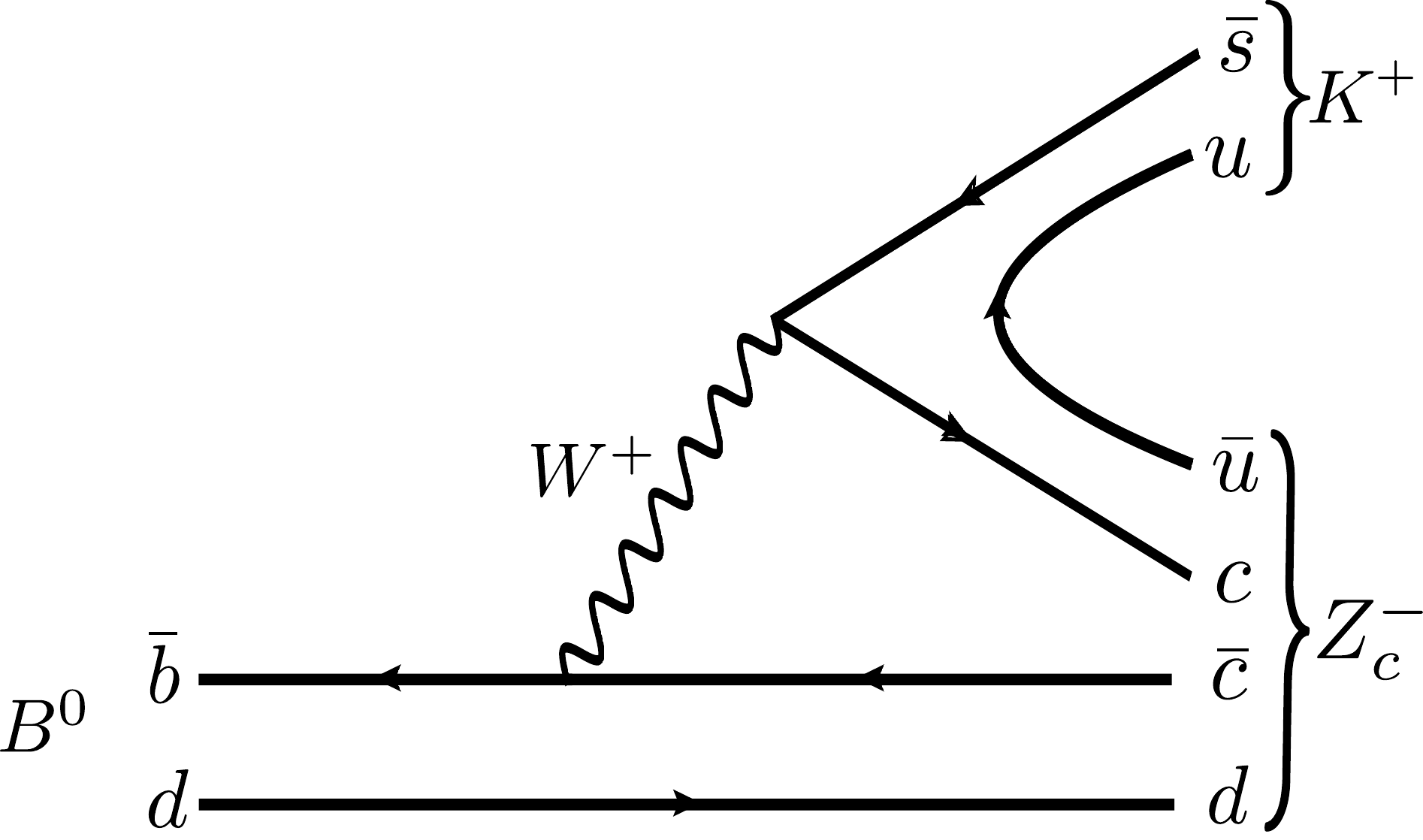}
     \label{B2ZK}}
   \label{diagrams}
   \caption{Feynman diagrams for \subref{B2etacKstar} $B^0 \to \eta_cK^{*0}$ and \subref{B2ZK} $B^0 \to Z_c^-K^+$ decay sequences.}
\end{figure}

In this article, the $\decaywo$ decay
is studied for the first time, with the \etac meson reconstructed using the \proton\antiproton decay mode. The decay is expected to proceed through \decay{\Kstarz}{\Kp \pim} intermediate states, where \Kstarz refers to any neutral kaon resonance,
following the diagram shown in Fig.~\ref{B2etacKstar}.
If the decay also proceeds through exotic resonances in the \etacpi system, denoted by $Z_c^-$ states in the following, a diagram like that shown in Fig.~\ref{B2ZK} would contribute.
The $\decaywo$ decay involves only pseudoscalar mesons, hence it is fully described by two independent kinematic quantities. Therefore, the Dalitz plot (DP) analysis technique~\cite{dalitzPlot} can be used to completely characterise the decay.

The data sample used corresponds to an integrated luminosity of $4.7 \invfb$ of \proton\proton collision data collected with the \lhcb detector at centre-of-mass energies of $\sqs=7,~8$ and $13\tev$ in 2011, 2012 and 2016, respectively. Data collected in 2011 and 2012 are referred to as Run~1 data, while data collected in 2016 are referred to as Run~2 data.

This paper is organised as follows. A brief description of the \lhcb detector as well as the reconstruction and simulation software is given in Sec.~\ref{sec:Detector}. The selection of \ppKpi candidates is described in Sec.~\ref{sec:Selection}, and the first measurement of the $\decaywo$ branching fraction is presented in Sec.~\ref{sec:BranchingFraction}. An overview of the DP analysis formalism is given in Sec.~\ref{sec:DalitzPlot}. Details of the implementation of the DP fit are presented in Sec.~\ref{sec:DalitzFit}. The evaluation of systematic uncertainties is given in Sec.~\ref{sec:Systematics}. The results are summarised in Sec.~\ref{sec:Results}.

%% file: detector.tex
\section{Detector and simulation}
\label{sec:Detector}
The \lhcb detector~\cite{Alves:2008zz,LHCb-DP-2014-002} is a single-arm forward spectrometer covering the \mbox{pseudorapidity} range $2<\eta <5$, designed for the study of particles containing \bquark or \cquark quarks. The detector includes a high-precision tracking system consisting of a silicon-strip vertex detector surrounding the $pp$ interaction region, a large-area silicon-strip detector located upstream of a dipole magnet with a bending power of about $4{\mathrm{\,Tm}}$, and three stations of silicon-strip detectors and straw drift tubes placed downstream of the magnet. The tracking system provides a measurement of the momentum, \ptot, of charged particles with a relative uncertainty that varies from 0.5\% at low momentum to 1.0\% at 200\gev. The minimum distance of a track to a primary vertex (PV), the impact parameter, is measured with a resolution of $(15+29/\pt)\mum$, where \pt is the component of the momentum transverse to the beam, in\,\gev. Different types of charged hadrons are distinguished using information from two ring-imaging Cherenkov (\rich) detectors. Photons, electrons and hadrons are identified by a calorimeter system consisting of scintillating-pad and preshower detectors, an electromagnetic calorimeter and a hadronic calorimeter. Muons are identified by a system composed of alternating layers of iron and multiwire proportional chambers. 

The online event selection is performed by a trigger~\cite{LHCb-DP-2012-004}, which consists of a hardware stage, based on information from the calorimeter and muon systems, followed by a software stage, which applies a full event reconstruction. At the hardware trigger stage, events are required to have a hadron with high transverse energy in the calorimeters. The software trigger requires a two-, three- or four-tracks secondary vertex with a significant displacement from any PV. At least one charged particle must have a large transverse momentum and be inconsistent with originating from a PV. A multivariate algorithm~\cite{BBDT,LHCb-PROC-2015-018} is used to identify secondary vertices that are consistent with $b$-hadron decays. 

Simulated events, generated uniformly in the phase space of the $\ppKpi$ or $\decaywo$ decay modes, are used to develop the selection, to validate the fit models and to evaluate the efficiencies entering the branching fraction measurement and the DP analysis. In the simulation, $pp$ collisions are generated using \pythia~\cite{Sjostrand:2007gs, *Sjostrand:2006za} with a specific \lhcb configuration~\cite{LHCb-PROC-2010-056}. Decays of hadronic particles are described by \evtgen~\cite{Lange:2001uf}, in which final-state radiation is generated using \photos~\cite{Golonka:2005pn}. The interaction of the generated particles with the detector, and its response, are implemented using the \geant toolkit~\cite{Allison:2006ve, *Agostinelli:2002hh} as described in Ref.~\cite{LHCb-PROC-2011-006}.

%% file: selection.tex
\section{Selection}
\label{sec:Selection}
An initial offline selection comprising loose criteria is applied to reconstructed particles, where the associated trigger decision was due to the \Bd candidate. The final-state tracks are required to have $p > \unit[1500]{MeV}$, $\pt > \unit[300]{MeV}$, and to be inconsistent with originating from any PV in the event. Loose particle identification (PID) criteria are applied, requiring the particles to be consistent with either the proton, kaon or pion hypothesis. 
All tracks are required to be within the acceptance of the \rich detectors ($2.0 < \eta < 4.9$). 
Moreover, protons and antiprotons are required to have momenta larger than $\unit[8]{GeV}$ ($\unit[11]{GeV}$) to avoid kinematic regions where proton-kaon separation is limited for Run~1 (Run~2) data.

The $\Bd$ candidates are required to have a small $\chisqip$ with respect to a PV, where $\chisqip$ is defined as the difference in the vertex-fit $\chisq$ of a given PV reconstructed with and without the candidate under consideration. 
The PV providing the smallest $\chisqip$ value is associated to the $\Bd$ candidate. 
The $\Bd$ candidate is required to be consistent with originating from this PV by applying a criterion on the direction angle (DIRA) between the $\Bd$ candidate momentum vector and the distance vector between the PV to the $\Bd$ decay vertex. 
When building the $\Bd$ candidates, the resolution 
on kinematic quantities such as the $\mpp$ distribution, and the Dalitz variables that will be defined in Sec.~\ref{sec:DalitzPlot},
is improved by performing a kinematic fit~\cite{Hulsbergen:2005pu} in which the $\Bd$ candidate is constrained to originate from its associated PV, and its reconstructed invariant mass is constrained to the known $\Bd$ mass~\cite{PDG2018}. 

A boosted decision tree~(BDT)~\cite{Breiman,AdaBoost} algorithm is used to further suppress the combinatorial background that arises when unrelated particles are combined to form a \Bd candidate. The training of the BDT is performed using simulated $\ppKpi$ decays as the signal sample and candidates from the high-mass data sideband as the background sample, defined as the $5450 < \mppKpi < 5550\mev$ range. The input variables to the BDT classifiers are the same for Run~1 and~2 samples and comprise typical discriminating variables of $b$-hadron decays: the vertex-fit $\chisqvtx$, $\chisqip$, DIRA and flight distance significance of the reconstructed $\Bd$ candidates; the maximum distance of closest approach between final-state particles; and the maximum and minimum $p$ and \pt of the proton and antiproton.

The requirements placed on the output of the BDT algorithm and PID variables are simultaneously optimised to maximise the figure of merit defined as $S/\sqrt{S+B}$. Here $S$ is the observed $\ppKpi$ yield before any BDT selection multiplied by the efficiency of the BDT requirement evaluated using simulated decays, while $B$ is the the combinatorial background yield. The training of the BDT and the optimisation of the selection are performed separately for Run~1 and~2 data to accommodate for differences in the two data-taking periods.

%% file: branchingFraction.tex
\section{Branching fraction measurement}
\label{sec:BranchingFraction}
The measurement of the \decaywo branching fraction is performed relative to that of the \jpsiKpi normalisation channel, where the \jpsi meson is also reconstructed in the \proton\antiproton decay mode.
A two-stage fit procedure to the combined Run~1 and~2 data sample is used. In the first stage, an extended unbinned maximum-likelihood (UML) fit is performed to the \mppKpi\ distribution in order to separate the \ppKpi and background contributions. The \roofit package~\cite{RooFit} is used to perform the fit, and the \sPlot 
technique~\cite{Pivk:2004ty} is applied to assign weights for each candidate to subtract the background contributions.
In the second stage, a weighted UML fit to the \proton\antiproton invariant-mass spectrum is performed to disentangle the \etac, \jpsi, and nonresonant (NR) contributions. 
The efficiency-corrected yield ratio is
\begin{equation} \label{ratio}
R = \frac{N_{\etac}}{N_{\jpsi}} \times \frac{\epsilon_{\jpsi}}{\epsilon_{\etac}},
\end{equation} 
where $N_{\etac}$ and $N_{\jpsi}$ are the observed \etac and \jpsi yields, while $\epsilon_{\etac}$ and $\epsilon_{\jpsi}$ are the total efficiencies, which are obtained from a combination of simulated and calibration samples. The \decaywo branching fraction is determined as 
\begin{equation} \label{formulaBF}
\mathcal{B}(\decaywo) = R \times \mathcal{B}(\jpsiKpi) \times \frac{\mathcal{B}(\decay{\jpsi}{\proton\antiproton})}{\mathcal{B}(\decay{\etac}{\proton\antiproton})},
\end{equation} 
where \mbox{$\mathcal{B}(\jpsiKpi) = (1.15 \pm 0.05) \times 10^{-3}$}, \mbox{$\mathcal{B}(\decay{\jpsi}{\proton\antiproton}) = (2.121 \pm 0.029) \times 10^{-3}$} and \mbox{$\mathcal{B}(\decay{\etac}{\proton\antiproton}) = (1.52 \pm 0.16) \times 10^{-3}$} are the external branching fractions taken from Ref.~\cite{PDG2018}.

\subsection{Signal and normalisation yields}\label{bfYields}
The first-stage UML fit to the \mppKpi\ distribution is performed in the 5180--5430\mev range. The \ppKpi signal decays, \decay{\Bs}{\proton\antiproton\Kp\pim} decays and various categories of background are present in this range. In addition to the combinatorial background, partially reconstructed backgrounds are present originating from $b$-hadron decays with additional particles that are not part of the reconstructed decay chain, such as a \piz meson or a photon. Another source of background is $b$-hadron decays where one of the final-state particles has been incorrectly identified, which includes the decays \decay{\Bd}{\proton\antiproton\pip\pim} and \decay{\Bs}{\proton\antiproton\Kp\Km}. The \decay{\Dzb}{\Kp\pim} and \decay{\Lcbar}{\antiproton\Kp\pim} decays are removed by excluding the mass range 1845--1885\mev in the $m(\Kp\pim)$ distribution and the range 2236--2336\mev in the $m(\antiproton\Kp\pim)$ distribution, respectively. The latter veto also removes partially reconstructed $b$-hadron decays.

\begin{figure}[t]
  \centering
  \includegraphics[width=14cm, keepaspectratio]{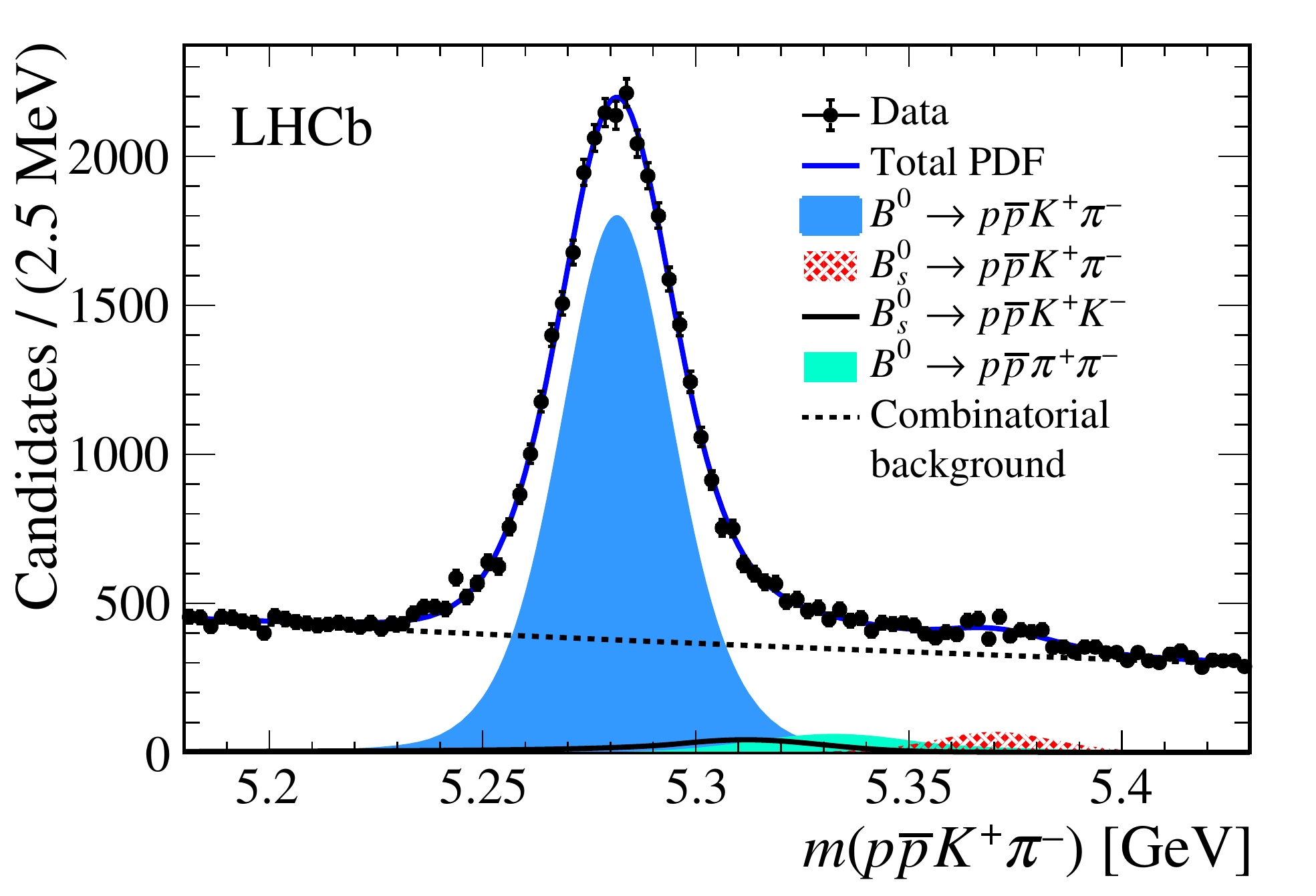}
  \caption{Distribution of the \proton\antiproton\Kp\pim invariant mass. The solid blue curve is the projection of the total fit result. The components are shown in the legend.}
  \label{B2ppKpiFit}
\end{figure}

Both the \ppKpi and \decay{\Bs}{\proton\antiproton\Kp\pim} components are modelled by Hypatia functions~\cite{Santos:2013gra}. The Hypatia distribution is a generalisation of the Crystall Ball function~\cite{Skwarnicki:1986xj}, where the Gaussian core of the latter is replaced by a hyperbolic core to take into account the distortion on the measured mass due to different sources of uncertanty. The Hypatia functions share a common resolution parameter, while the tail parameters are fixed to the values obtained from the corresponding simulated sample. The distributions of the misidentified \mbox{\decay{\Bd}{\proton\antiproton\pip\pim}} and \mbox{\decay{\Bs}{\proton\antiproton\Kp\Km}} backgrounds are described by Crystal Ball functions, with parameters fixed to the values obtained from simulation. The combinatorial background is modelled using an exponential function. The masses of the \Bd and \Bs mesons, the resolution parameter of the Hypatia functions, the slope of the exponential function, and all the yields, are free to vary in the fit to the data. Using the information from the fit to the \mppKpi\ distribution, shown in Fig.~\ref{B2ppKpiFit}, \ppKpi signal weights are computed and the background components are subtracted using the \sPlot technique~\cite{Pivk:2004ty}. About $3.0 \times 10^4$ \Bd decays are observed. Correlations between the \proton\antiproton and \proton\antiproton\Kp\pim invariant-mass variables for both signal and background are found to be negligible.

\begin{figure}[t]
  \centering
  \includegraphics[width=7.5cm, keepaspectratio]{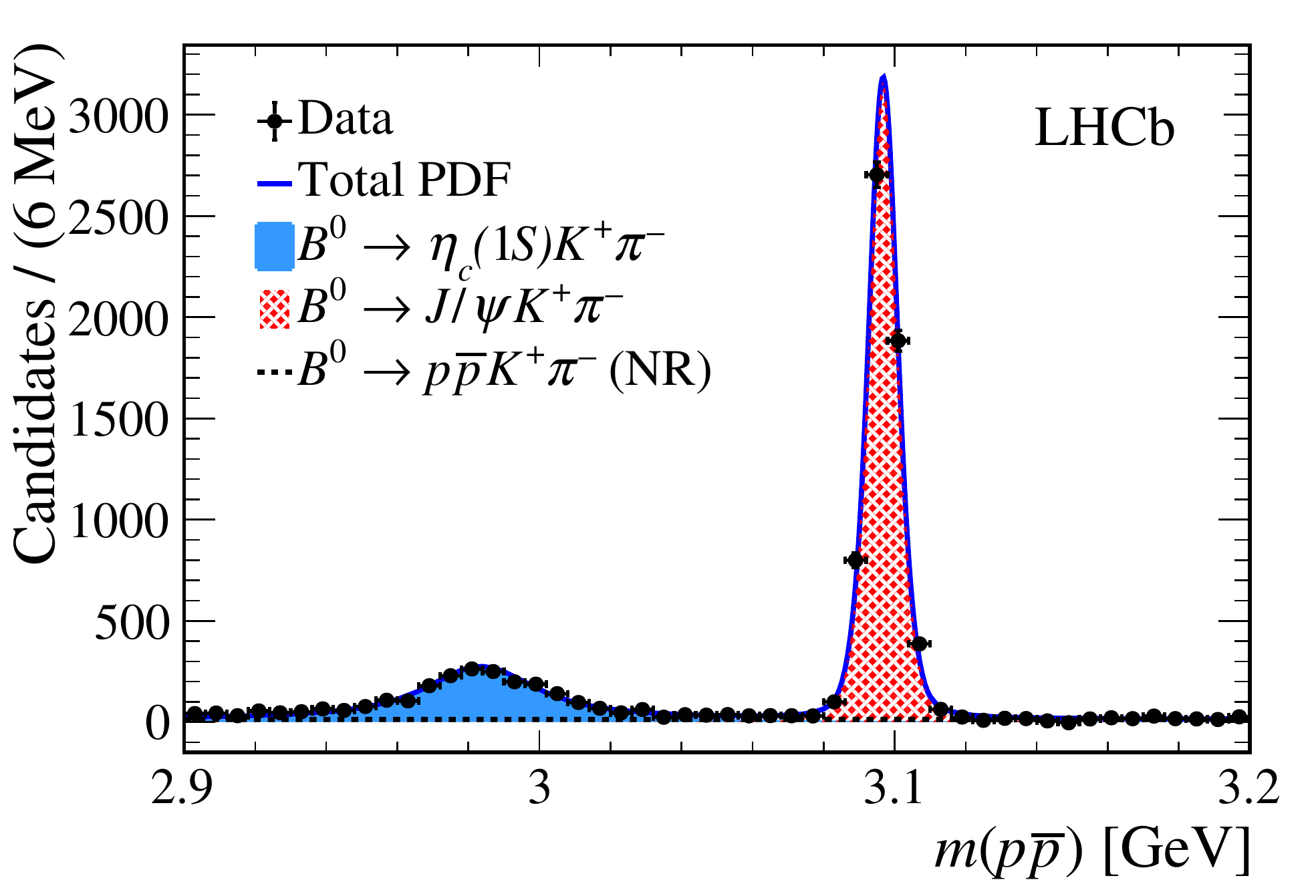}\quad
  \includegraphics[width=7.5cm, keepaspectratio]{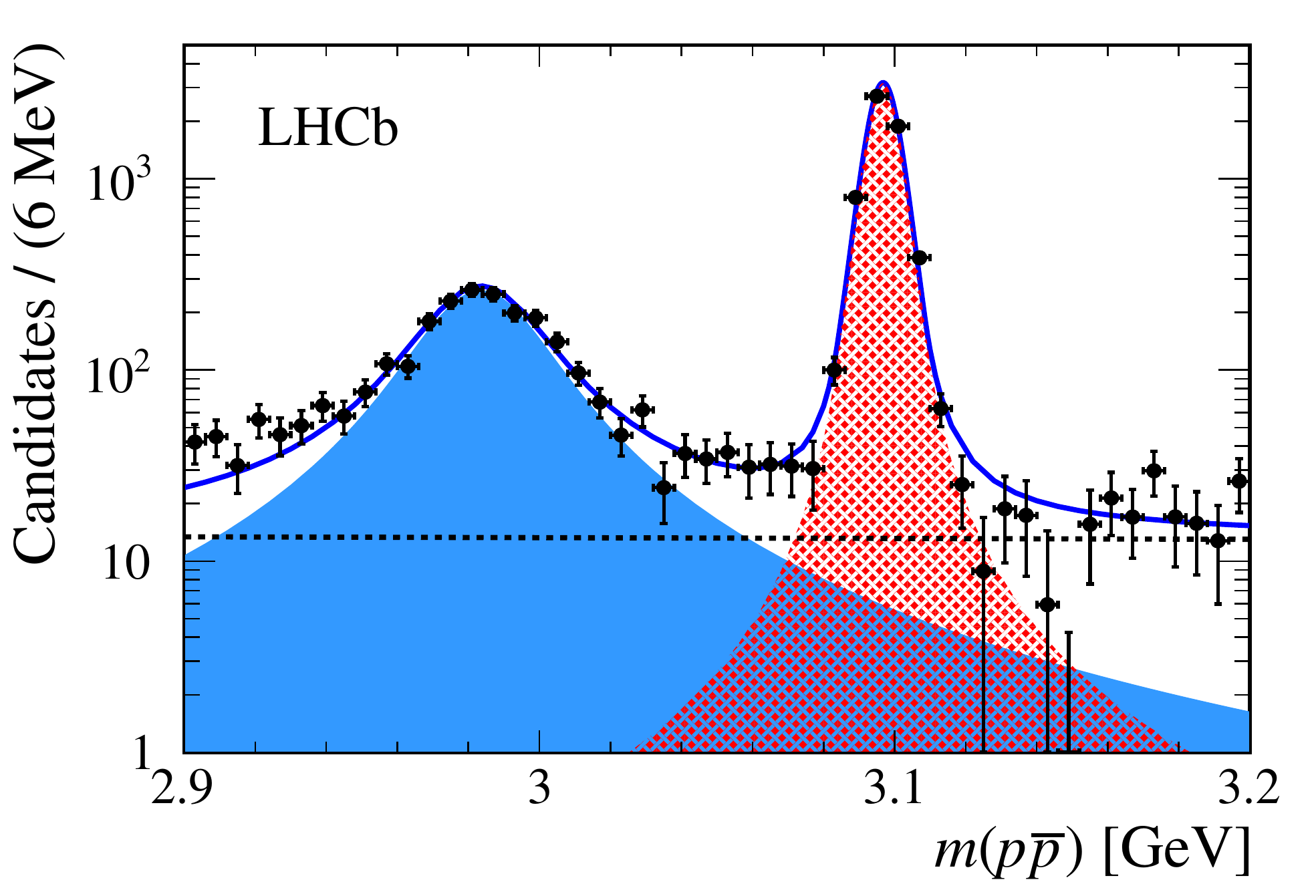}
  \caption{Distribution of the \proton\antiproton invariant mass in (left) linear and (right) logarithmic vertical-axis scale for weighted \ppKpi candidates obtained by using the \sPlot technique. The solid blue curve is the projection of the total fit result. The full azure, tight-cross-hatched red and dashed-black line areas show the \etac, \jpsi and NR \proton\antiproton contributions, respectively.}
  \label{ppsfit}
\end{figure}

The second-stage UML fit is then performed to the weighted \proton\antiproton invariant-mass distribution in the mass range 2700--3300\mev, which includes \etac, \jpsi, and NR \mbox{\ppKpi} contributions. The \proton\antiproton invariant-mass distribution of \etac candidates is described by the convolution of a nonrelativistic Breit--Wigner function and a Gaussian function describing resolution effects. Using simulated samples, the \proton\antiproton invariant-mass resolution is found to be $\approx \unit[5]{MeV}$. Given the width $\Gamma_{\etac} = \unit[32.0 \pm 0.8]{MeV}$~\cite{PDG2018}, the impact of the detector resolution on the \etac lineshape is small. The \jpsi resonance, having a small natural width, is parametrised using an Hypatia function, with tail parameters fixed to the values obtained from the corresponding simulated sample. The same resolution parameter is used for the \etac and \jpsi contributions, which is free to vary in the fit to the data. The \etac and \jpsi masses are also floating, while the \etac natural width is Gaussian constrained to the known value~\cite{PDG2018}. The NR \ppKpi contribution is parametrised with an exponential function with the slope free to vary in the fit. All yields are left unconstrained in the fit.
A possible term describing the interference between the \etac resonance and the NR \proton\antiproton S-wave is investigated and found to be negligible.
The result of the fit to the weighted \proton\antiproton invariant-mass distribution is shown in Fig.~\ref{ppsfit}. The yields of the \decaywo and \jpsiKpi fit components, entering Eq.\eqref{ratio}, are $2105 \pm 75$ and $5899 \pm 86$, respectively.

\subsection{Ratio of efficiencies}
The ratio of efficiencies of Eq.~\eqref{ratio} is obtained from \decaywo and \mbox{\jpsiKpi} simulated samples, both selected using the same criteria used in data.  Since these decays have the same final-state particles and similar kinematic distributions, the ratio of efficiencies is expected to be close to unity. 
The efficiencies are computed as the product of the geometrical acceptance of the \lhcb detector, the reconstruction efficiency and the efficiency of the offline selection criteria, including the trigger and PID requirements. The efficiency of the PID requirements is obtained using calibration samples of pions, kaons and protons, as a function of the particle momentum, pseudorapidity and the multiplicity of the event, \eg the number of charged particles in the event~\cite{LHCb-DP-2018-001}. The final ratio of efficiencies is given by

\begin{equation}
\frac{\epsilon_{\jpsi}}{\epsilon_{\etac}} = 1.000 \pm 0.013,
\end{equation}
which is compatible with unity as expected.

\subsection{Systematic uncertainties}
Table~\ref{systSummaryBF} summarises the systematic uncertainties on the measurement of the ratio $R$ of Eq.~\eqref{ratio}. Since the kinematic distributions of the signal and normalisation channel are similar, the uncertainties corresponding to the reconstruction and selection efficiencies largely cancel in the ratio of branching fractions. A new value of the ratio $R$ is computed for each source of systematic uncertainty, and its difference with the nominal value is taken as the associated systematic uncertainty. The overall systematic uncertainty is assigned by combining all contributions in quadrature.

The systematic uncertainty arising from fixing the shape parameters of the Hypatia functions used to parametrise the \Bd and \jpsi components is evaluated by repeating the fits and varying all shape parameters simultaneously. These shape parameters are varied according to normal distributions, taking into account the correlations between the parameters and with variances related to the size of the simulated samples.

To assign a systematic uncertainty arising from the model used to describe the detector resolution, the fits are repeated for each step replacing the Hypatia functions by Crystal Ball functions, whose parameters are obtained from simulation.

The systematic uncertainty associated to the parametrisation of the NR \mbox{\ppKpi} contribution is determined by replacing the exponential function with a linear function.

The systematic uncertainty associated to the determination of the efficiency involves contributions arising from the weighting procedure of the calibration samples used to determine the PID efficiencies. The granularity of the binning in the weighting procedure is halved and doubled.

The free shape parameters in the first stage UML fit lead to uncertainties that are not taken into account by the \sPlot technique. In order to estimate this effect, these parameters are varied within their uncertainties and the signal weights are re-evaluated. The variations on the ratio $R$ resulting from the second stage UML fit are found to be negligible.

\begin{table}[t]
\caption{Relative systematic uncertainties on the ratio $R$ of Eq.~\eqref{ratio}. The total systematic uncertainty is obtained from the quadratic sum of the individual sources.}
\begin{center}
\begin{tabular}{l c}
\midrule
Source & Systematic uncertainty (\%)\\
\midrule
Fixed shape parameters & 0.8\\
Resolution model & 0.3\\
NR $p\bar{p}$ model & 1.7\\
Efficiency ratio & 1.1\\
\midrule
Total & 2.2\\
\midrule
\end{tabular}
\end{center}
\label{systSummaryBF}
\end{table}

\subsection{Results}
The ratio $R$ is determined to be 
\begin{equation*}\label{ResultRatio}
R = 0.357 \pm 0.015 \pm 0.008, 
\end{equation*}
where the first uncertainty is statistical and the second systematic. The statistical uncertainty includes contributions from the per-candidate weights obtained using the \sPlot technique. The value of $R$ is used to compute the \decaywo branching fraction using Eq.~\eqref{formulaBF} which gives 
\begin{equation*} \label{resultBF}
\mathcal{B}(\decaywo) = (5.73 \pm 0.24 \pm 0.13 \pm 0.66) \times 10^{-4},
\end{equation*}
where the first uncertainty is statistical, the second systematic, and the third is due to the limited knowledge of the external branching fractions.

%% file: dalitzPlot.tex
\section{Dalitz plot formalism}
\label{sec:DalitzPlot}
The phase space for a three-body decay involving only pseudoscalar
particles can be represented in a DP, where two of the three possible
two-body invariant-mass-squared combinations, here $\MKpi$ and $\Metacpi$,
are used to define the DP axes. However, given the sizeable natural
width of the \etac meson, the invariant mass $m(\proton\antiproton)$
is used instead of the known value of the \etac mass~\cite{PDG2018} to
compute the kinematic quantities such as $m^2(\etac\Kp)$,
$m^2(\etac\pim)$ and the helicity angles.

The isobar model~\cite{PhysRev.135.B551,PhysRev.166.1731,PhysRevD.11.3165} is
used to write the decay amplitude as a coherent sum of
amplitudes from resonant and NR intermediate processes as
\begin{equation}\label{isobarAmplitude}
\mathcal{A}[\MKpi,\Metacpi] = \sum_{j=1}^{N}c_j\mathcal{F}_j[\MKpi,\Metacpi] ,
\end{equation}
where $c_j$ are complex coefficients giving the relative contribution
of each intermediate process. The $\mathcal{F}_j[\MKpi,\Metacpi]$
complex functions describe the resonance dynamics and are normalised such that the
integral of their squared magnitude over the DP is unity
\begin{equation}\label{normal}
\int_{\text{DP}}  |\mathcal{F}_j[\MKpi,\Metacpi]|^2
\text{d}\MKpi~\text{d}\Metacpi = 1.
\end{equation}
Each $\mathcal{F}_j[\MKpi,\Metacpi]$ contribution is composed of the product of several factors. 
For a $\Kpi$ resonance, for instance,
\begin{equation}\label{dynAmplitude}
\mathcal{F}[\MKpi,\Metacpi] = \mathcal{N} \times X(|\vec{p}| r_{\text{BW}}) \times X(|\vec{q}|r_{\text{BW}}) \times Z(\vec{p},\vec{q}) \times T[\mKpi],
\end{equation}
where $\displaystyle \mathcal{N}$ is a normalisation constant and
$\vec{p}$ and $\vec{q}$ are the momentum of the accompanying particle (the \etac meson in this case) and
the momentum of one of the resonance decay products, respectively,
both evaluated in the $\Kpi$ rest frame. The $X(z)$ terms are the
Blatt--Weisskopf barrier factors~\cite{Blatt:1952ije} reported in
Appendix~\ref{Blatt-Weisskopf}. The barrier radius, $r_{\text{BW}}$,
is taken to be $\unit[4]{GeV^{-1}}$ (corresponding to $\sim 0.8 \fm$) for all resonances. 
The $Z(\vec{p},\vec{q})$ term describes the angular probability distribution in the Zemach tensor formalism~\cite{PhysRev.133.B1201,PhysRev.140.B97}, given by the equations reported in Appendix~\ref{Zemach}. 
The function $T[\mKpi]$ of Eq.~\eqref{dynAmplitude} is the mass
lineshape. Most of the resonant contributions are
described by the relativistic Breit--Wigner (RBW) function
\begin{equation}
T(m) = \frac{1}{m_0^2 - m^2-im_0\Gamma(m)},
\end{equation}
where the mass-dependent width is given by
\begin{equation}
\Gamma(m) = \Gamma_0 \left ( \frac{|\vec{q}|}{q_0} \right )^{(2L+1)} \left (\frac{m_0}{m} \right )X^2(|\vec{q}| r_{\text{BW}}) 
\end{equation}
and $q_0$ is the value of $|\vec{q}|$ for $m=m_0$, $m_0$ being the pole mass of the resonance.

The amplitude parametrisations using RBW functions lead to unitarity
violation within the isobar model if there are overlapping resonances
or if there is a significant interference with a NR
component, both in the same partial wave~\cite{Meadows:2007jm}. This is the case for the
$\Kpi$ S-wave at low $\Kpi$ mass, where the $K^*_0(1430)^0$ resonance
interferes strongly with a slowly varying NR S-wave
component. Therefore, the $\Kpi$ S-wave at low mass is modelled using a
modified LASS lineshape~\cite{ASTON1988493}, given by
\begin{equation}
T(m) = \frac{m}{|\vec{q}|\cot \delta_B - i |\vec{q}|} + e^{2i\delta_B}\frac{m_0\Gamma_0\frac{m_0}{q_0}}{m_0^2 - m^2 -im_0\Gamma_0\frac{|\vec{q}|}{m}\frac{m_0}{q_0}},
\label{LASSpdf}
\end{equation}
with
\begin{equation}
\cot \delta_B = \frac{1}{a |\vec{q}|} + \frac{1}{2}r |\vec{q}|,
\end{equation}
and where $m_0$ and $\Gamma_0$ are the pole mass and width of the $K^*_0(1430)^0$ state, 
and $a$ and $r$ are the scattering length and the
effective range, respectively. The parameters $a$ and $r$ depend on the production
mechanism and hence on the decay under study. The slowly varying
part (the first term in Eq.~\eqref{LASSpdf}) is not well modelled at
high masses and it is set to zero for $\mKpi$ values above
$\unit[1.7]{GeV}$.

The probability density function for signal events across the DP, neglecting
reconstruction effects, can be written as
\begin{equation}\label{density}
\mathcal{P}_{\text{sig}}[\MKpi,\Metacpi] = \frac{|\mathcal{A}|^2}{\int_{\text{DP}} |\mathcal{A}|^2\text{d}\MKpi~ \text{d}\Metacpi}, 
\end{equation}
where the dependence of $\mathcal{A}$ on the DP position has been
suppressed for brevity. The natural width of the \etac meson is set to zero when computing the DP normalisation shown in the denominator of Eq.~\eqref{density}. The effect of this simplification is determined when assessing the systematic uncertainties as described in Sec.~\ref{sec:Systematics}.

The complex coefficients, given by $c_j$ in Eq.~\eqref{isobarAmplitude},
depend on the choice of normalisation, phase convention and amplitude
formalism. Fit fractions and interference fit fractions are
convention-independent quantities that can be directly compared between different analyses. The fit fraction is defined as the integral of the amplitude
for a single component squared divided by that of the coherent matrix
element squared for the complete DP,
\begin{equation}
\text{FF}_i = \frac{\int_{\text{DP}}  |c_i\mathcal{F}_i[\MKpi,\Metacpi]|^2 \text{d}
  \MKpi~\text{d}\Metacpi}{\int_{\text{DP}} 
  |\mathcal{A}[\MKpi,\Metacpi]|^2 \text{d}\MKpi~\text{d}\Metacpi}.
\end{equation}
In general, the fit fractions do not sum to unity due to the possible
presence of net constructive or destructive interference over the whole DP area. This effect can be described by
interference fit fractions defined for $i<j$ by
\begin{equation}
\text{FF}_{ij} = \frac{\int_{\text{DP}} 2 \Real \left [
    c_ic_j^*\mathcal{F}_i \mathcal{F}_j^*
  \right ]\text{d}\MKpi~\text{d}\Metacpi}{\int_{\text{DP}} 
  |\mathcal{A}|^2\text{d}\MKpi~\text{d}\Metacpi},
\end{equation}
where the dependence of $\mathcal{F}_i^{(*)}$ and $\mathcal{A}$ on the
DP position is omitted.

%% file: dalitzFit.tex
\section{Dalitz plot fit}
\label{sec:DalitzFit}
The \laura package~\cite{Back:2017zqt} is used to perform the unbinned DP fit, with the
Run~1 and~2 subsamples fitted simultaneously using the \jfit
framework~\cite{Ben-Haim:2014afa}. The free parameters in the
amplitude fit are in common between the two subsamples, while the signal and background yields and the maps describing the efficiency variations across the phase space, are different. Within the DP fit, the signal corresponds to $\decaywo$ decays, while the background comprises both combinatorial background and NR \mbox{$\ppKpi$} contributions. The likelihood function is given by
\begin{equation}
\mathcal{L} = \prod_i^{N_c} \left [ \sum_k N_k \mathcal{P}_k [m_i^{2}(K^+\pi^-),m_i^{2}(\eta_c \pi^-)] \right ],
\end{equation}
where the index $i$ runs over the $N_c$ candidates, $k$ runs over the
signal and background components, and $N_k$ is the
yield of each component. The procedure to determine the signal and
background yields is described in Sec.~\ref{sec:Yields}. The
probability density function for the signal,
$\mathcal{P}_{\text{sig}}$, is given by Eq.~\eqref{density}  where the
$|\mathcal{A}[\MKpi,\Metacpi]|^2$ term is multiplied by the
efficiency function described in Sec.~\ref{sec:Efficiency}. In order
to avoid problems related to the imperfect parametrisation of the
efficiencies at the DP borders, a veto of $\pm \unit[70]{MeV}$ is
applied around the DP, \ie to the phase space boundaries of the
$\mKpi$, $\metacpi$ and $\metacK$ distributions. This veto is used
when determining the signal and background yields, and the probability
density functions for the background, obtained as described in
Sec.~\ref{sec:Background}. The $\Kpi$ mass resolution is $\approx
\unit[5]{MeV}$, which is much smaller than the $K^*(892)^0$ meson
width $\Gamma_{K^*(892)^0} \approx \unit[50]{MeV}$, the
narrowest contribution to the DP; therefore, the resolution has negligible
effects and is not considered further. The amplitude fits are repeated
many times with randomised initial values to ensure the absolute
minimum is found.

\subsection{Signal and background yields}
\label{sec:Yields}
There is a non-negligible fraction of NR $\ppKpi$ decays in the region
of the \etac meson. In order to
separate the contributions of $\decaywo$ and NR
$\ppKpi$ decays, a two-dimensional (2D) UML fit to the $\mppKpi$ and
$\mpp$ distributions is performed in the domain $5220 < \mppKpi <
5340\mev$ and $2908 < \mpp < 3058\mev$. These ranges are chosen to
avoid the misidentified decays reported in Sec.~\ref{bfYields}, and
they also define the DP fit domain. 
The Run~1 and~2 2D mass fits
are performed separately. 
The $\mppKpi$ distributions of $\decaywo$ signal and NR $\ppKpi$ decays are described by Hypatia
functions. The $\mppKpi$ distribution of the combinatorial background
is parametrised using an exponential function. The $\mpp$
distribution of $\decaywo$ signal decays is described by the same model described in Sec.~\ref{bfYields}.
A possible component where genuine \etac mesons are
combined with random kaons and pions from the PV is investigated but
found to be negligible. The \Bd meson mass, the $\mppKpi$ resolution,
the value of $m_{\eta_c}$, the slopes of the exponential functions, and the yields, are
free to vary in the 2D mass fits. The $\mpp$ resolution and the
\etac meson natural width are Gaussian constrained to the value obtained in
the fit to the weighted $\mpp$ distribution of Sec.~\ref{bfYields},
and to the known value~\cite{PDG2018}, respectively.

\begin{table}[!t]
\caption{Yields of the components in the 2D mass fit to the
  joint [$\mppKpi$, $\mpp$] distribution for the Run~1 and~2 subsamples.}
\begin{center}
\begin{tabular}{l D{,}{\pm}{-1} D{,}{\pm}{-1}}
\midrule
Component & \multicolumn{1}{c}{\text{Run~1}} &
                                               \multicolumn{1}{c}{\text{Run~2}}\\
\midrule
$\decaywo$ & 805\hspace{1mm},\hspace{1mm}48 &
                                              1065\hspace{1mm},\hspace{1mm}56\\
\noindent
$\ppKpi$ (NR) & 234\hspace{1mm},\hspace{1mm}48 &
                                                 273\hspace{1mm},\hspace{1mm}56\\
Combinatorial background & 409\hspace{1mm},\hspace{1mm}36 & 498\hspace{1mm},\hspace{1mm}41\\
\midrule
\end{tabular}
\end{center}
\label{2Dyields}
\end{table}

\begin{figure}[t]
  \centering
    \includegraphics[width=0.45\linewidth]{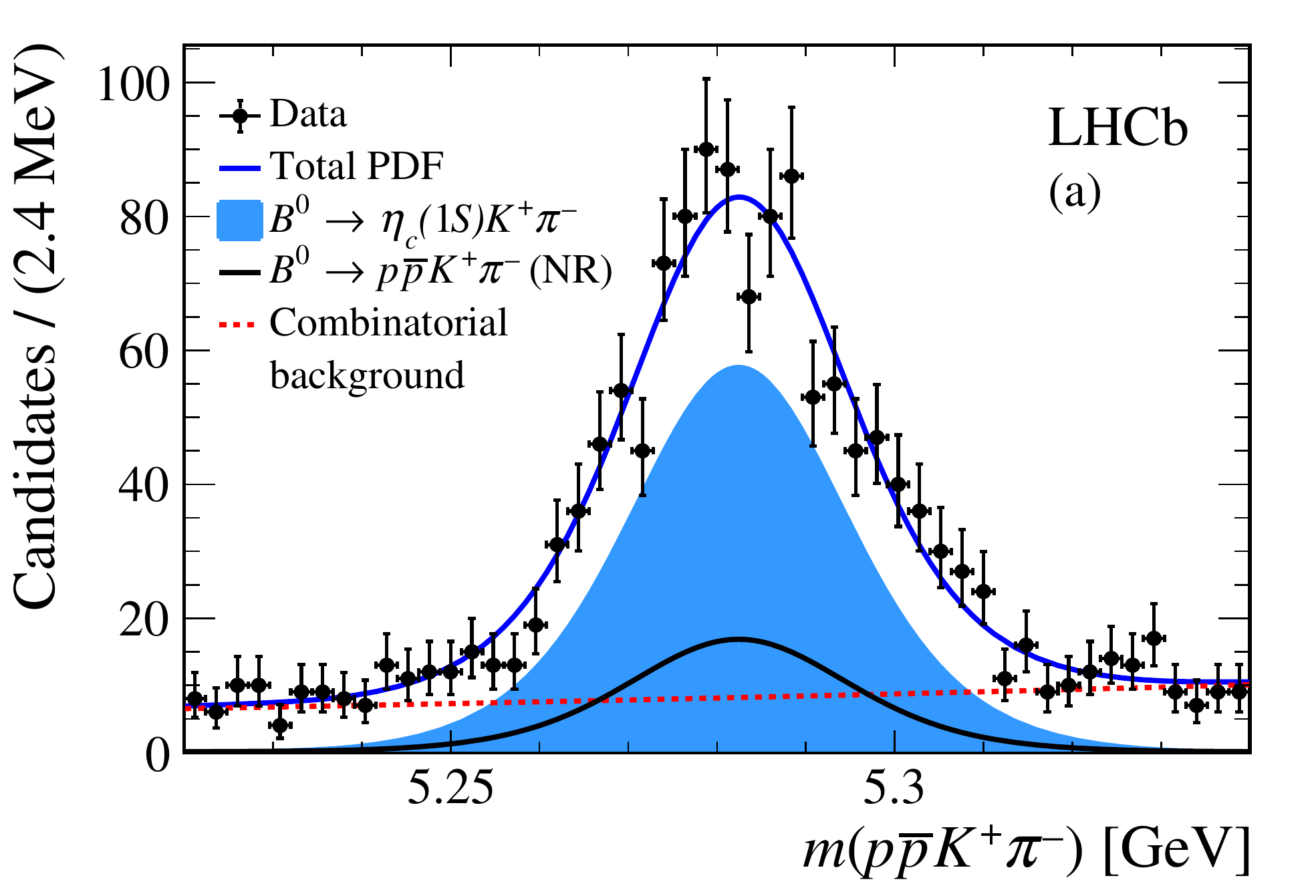}
    \includegraphics[width=0.45\linewidth]{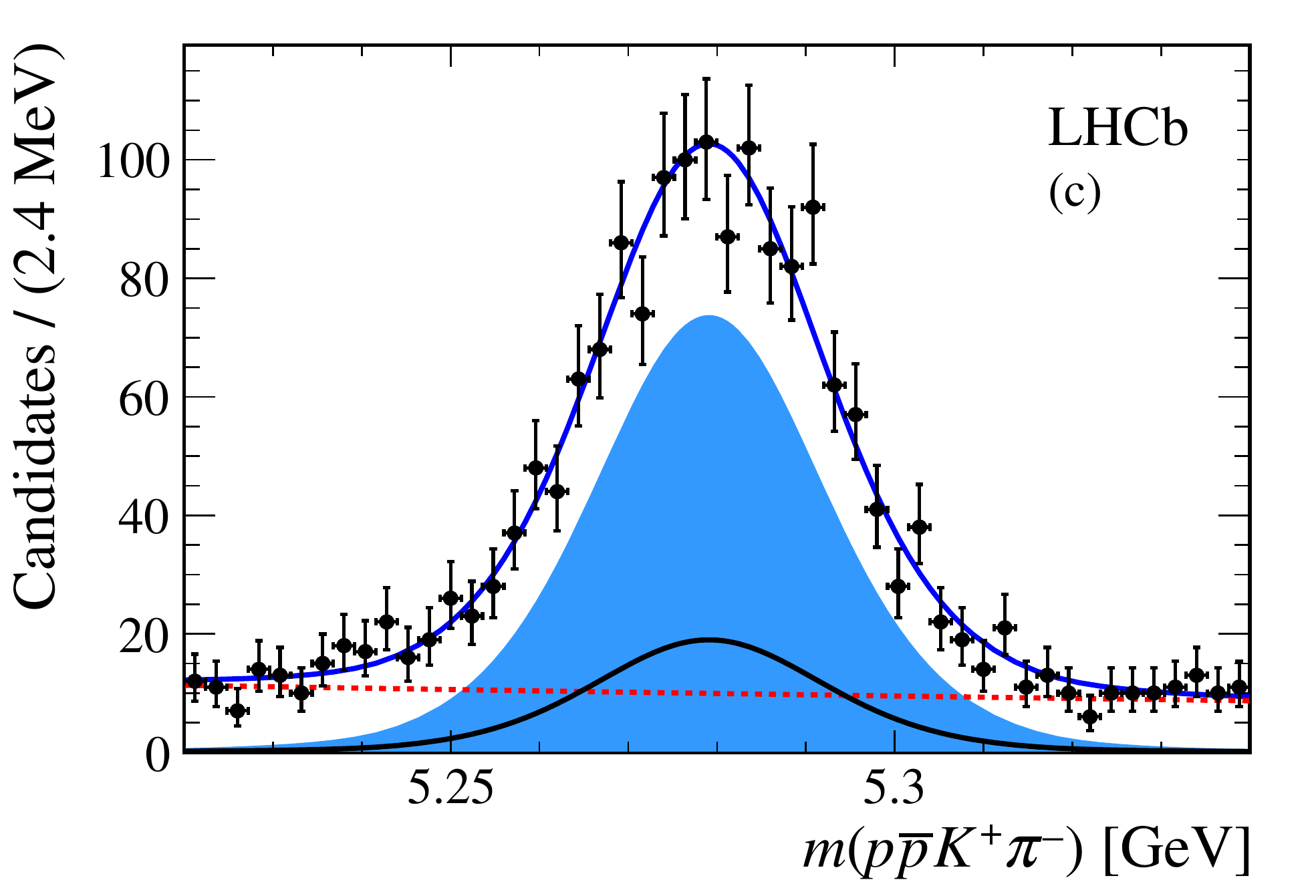}
    \includegraphics[width=0.45\linewidth]{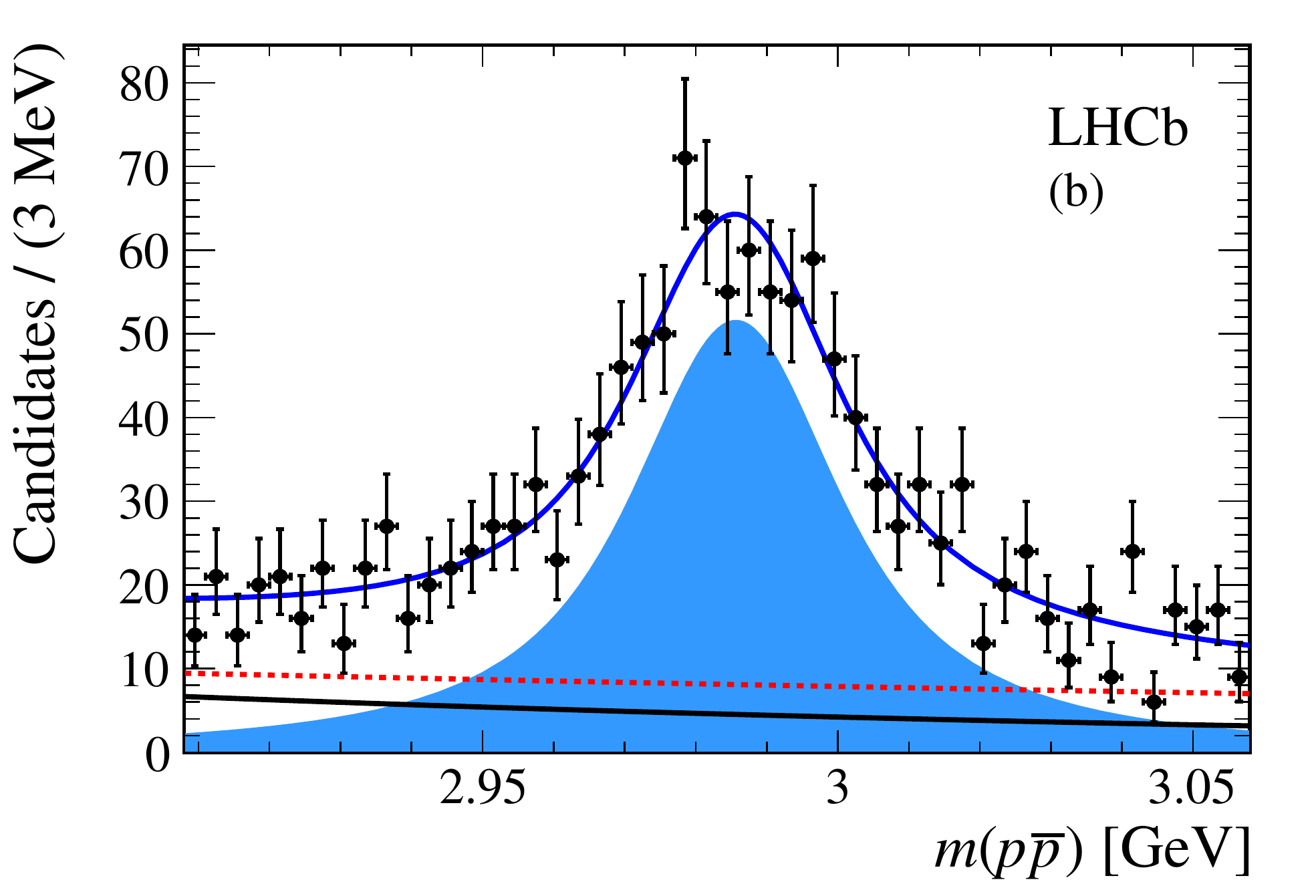}
    \includegraphics[width=0.45\linewidth]{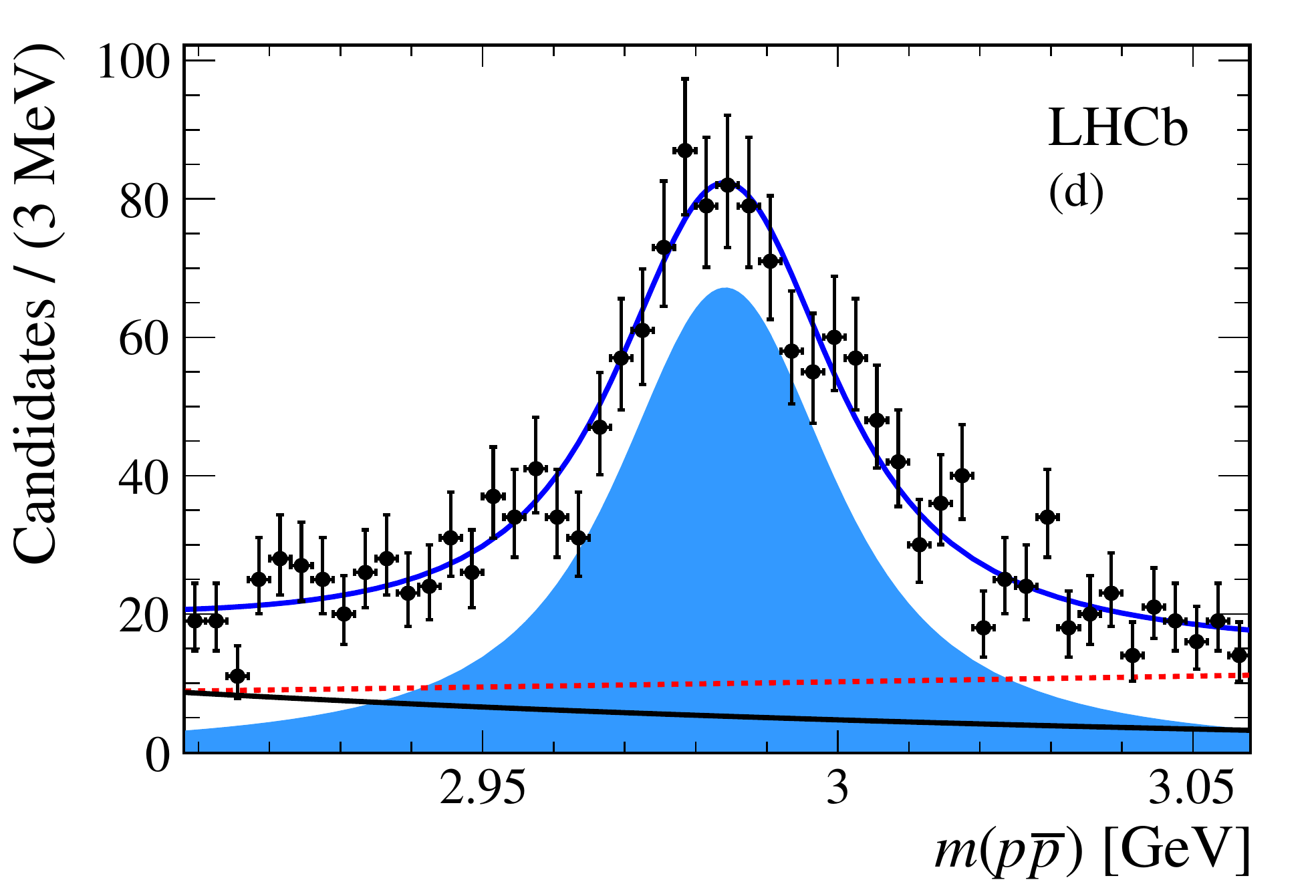}
  \caption{Results of the 2D mass fit to the joint [$\mppKpi$, $\mpp$]
    distribution for the (a)~Run~1 $\mppKpi$
    projection, (b) Run~1 $\mpp$ projection, (c) Run~2 $\mppKpi$
    projection,  and (d) Run~2
    $\mpp$ projection. The legend is shown in the top left plot.
    }
  \label{2Dfit}
\end{figure}

The yields of all fit components are reported in
Table~\ref{2Dyields}. Figure~\ref{2Dfit} shows the result of the 2D
mass fits for the Run~1 and~2 subsamples that yield a total of
approximately 2000 $\decaywo$ decays. The total yield of the
$\decaywo$ component is lower than that reported in
Sec.~\ref{bfYields} since the fit ranges are reduced. The goodness of
fit is validated using pseudoexperiments to determine the 2D pull, \ie
the difference between the fit model and data divided by the uncertainty.

\subsection{Parametrisation of the backgrounds}
\label{sec:Background}
The probability density functions for the combinatorial and
NR background categories are obtained from the DP
distribution of each background source, represented with a uniformly
binned 2D histogram. In order to avoid artefacts related
to the curved boundaries of the DP, the
histograms are built in terms of the Square Dalitz plot (SDP) parametrised by the variables $m'$ and $\theta'$
which are defined in the range 0 to 1 and are given by

\begin{align}\label{mPrime}
m' &\equiv \frac{1}{\pi} \arccos \left ( 2 \frac{\mKpi - m^{\min}_{\Kpi}}{m^{\max}_{\Kpi} - m^{\min}_{\Kpi}} - 1\right ),\\ 
\theta' &\equiv \frac{1}{\pi} \theta(\Kpi),
\end{align}
where $\displaystyle  m^{\max}_{\Kpi} = m_{B^0} - m_{\eta_c}$,
$\displaystyle  m^{\min}_{\Kpi} =
m_{K^+} + m_{\pi^-}$ are the kinematic boundaries of $\mKpi$ allowed
in the $\decaywo$ decay, and $\displaystyle \theta(\Kpi)$ is the
helicity angle of the $\Kpi$ system (the angle between the $K^+$ and
the \etac mesons in the $\Kpi$ rest frame).

\begin{figure}[t]
\begin{center}
    \includegraphics[width=0.45\linewidth]{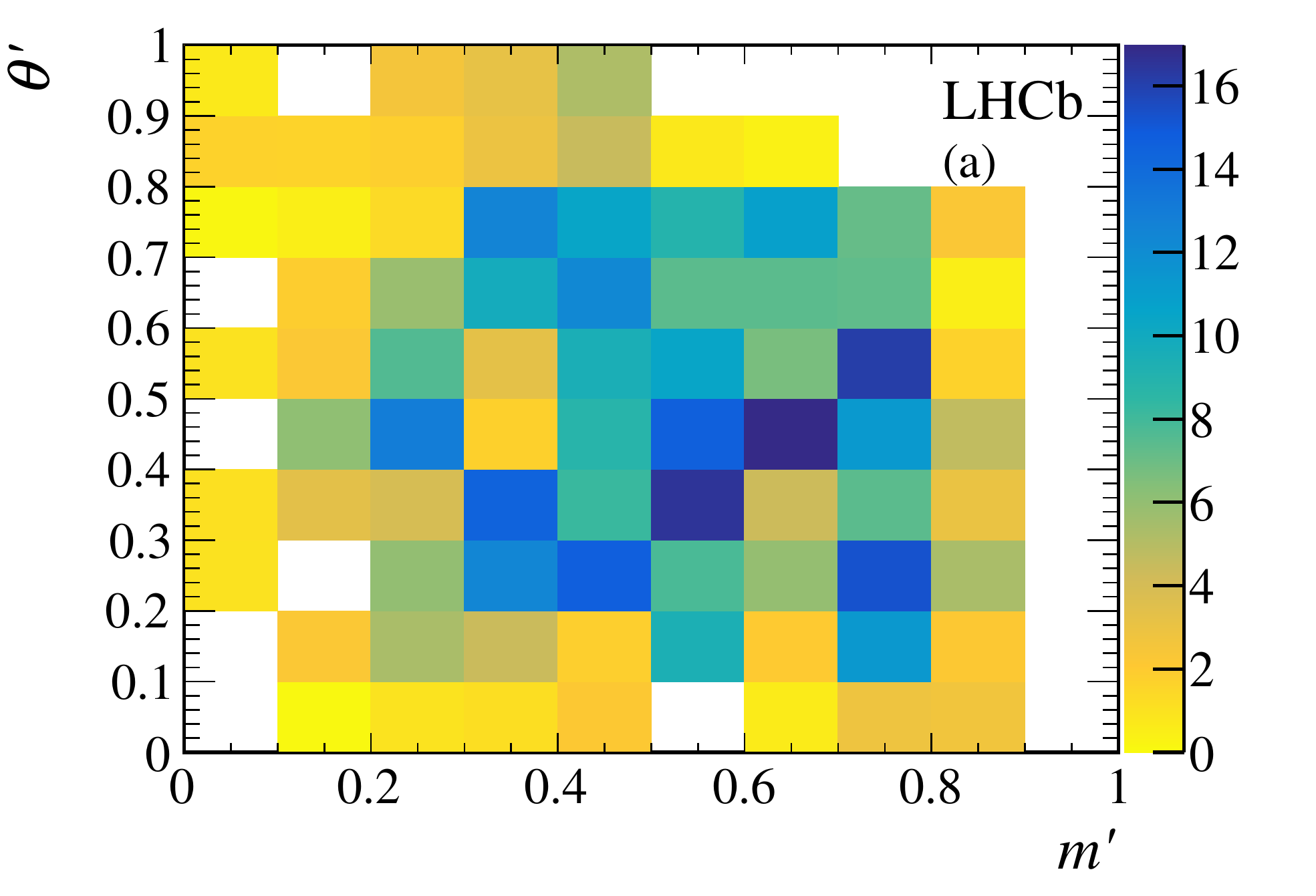}\quad
    \includegraphics[width=0.45\linewidth]{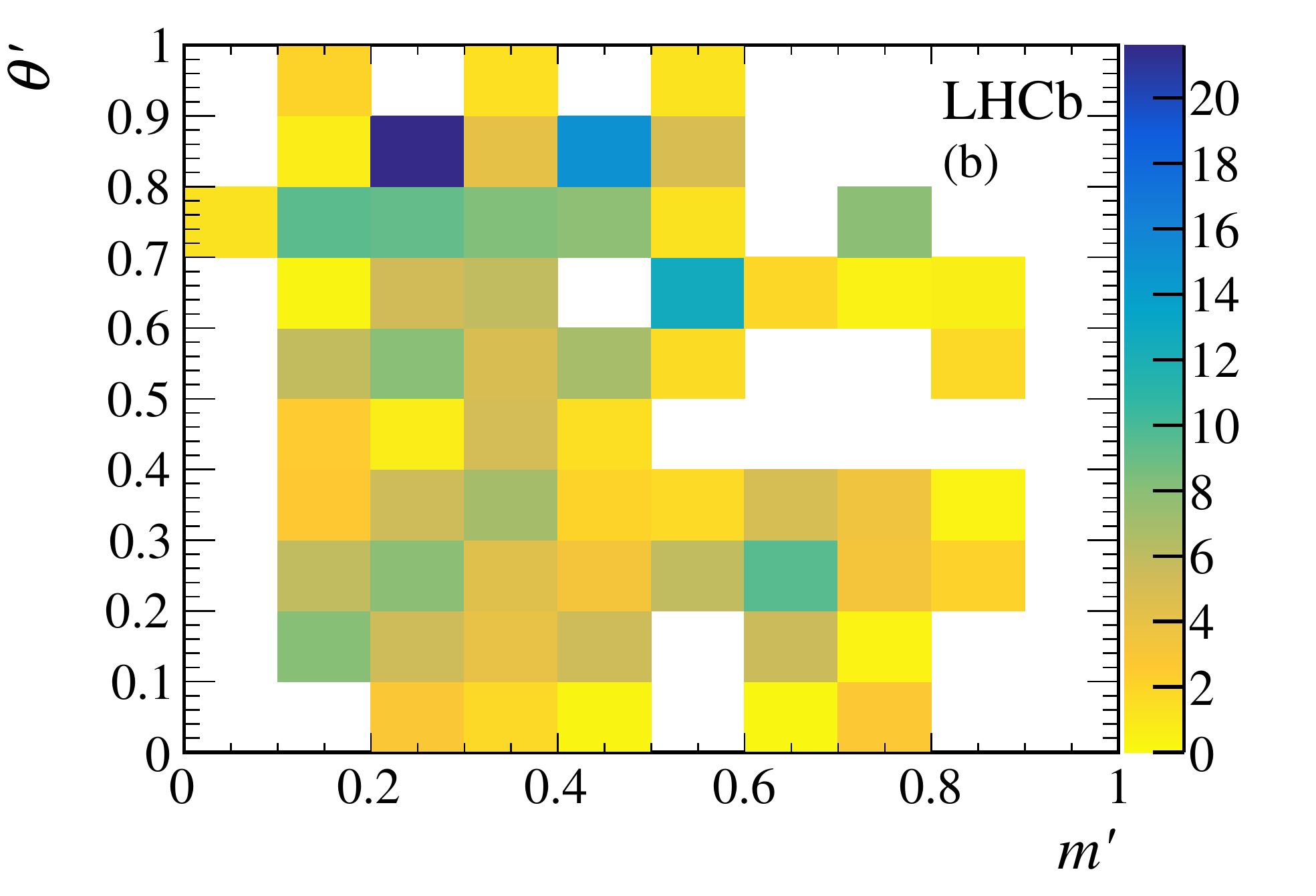}
\end{center}\vskip-1.5em
  \caption{SDP distributions used in the DP fit to the Run~2 subsample
    for (a) combinatorial background and (b) NR $\ppKpi$ background.}
  \label{bkgHistos}
\end{figure}

The combinatorial and NR background histograms are filled
using the weights obtained by applying the \sPlot technique to the joint
[$\mppKpi$,~$\mpp$] distribution, merging the Run~1 and~2 data
samples. Each histogram is scaled for the corresponding yield in the two subsamples. The combinatorial and NR
background histograms for the Run~2 subsample are shown in
Fig.~\ref{bkgHistos}. Statistical fluctuations in the histograms due
to the limited size of the samples are smoothed by applying a 2D cubic
spline interpolation. 

\begin{figure}[!h]
  \centering
  \includegraphics[width=15cm, keepaspectratio]{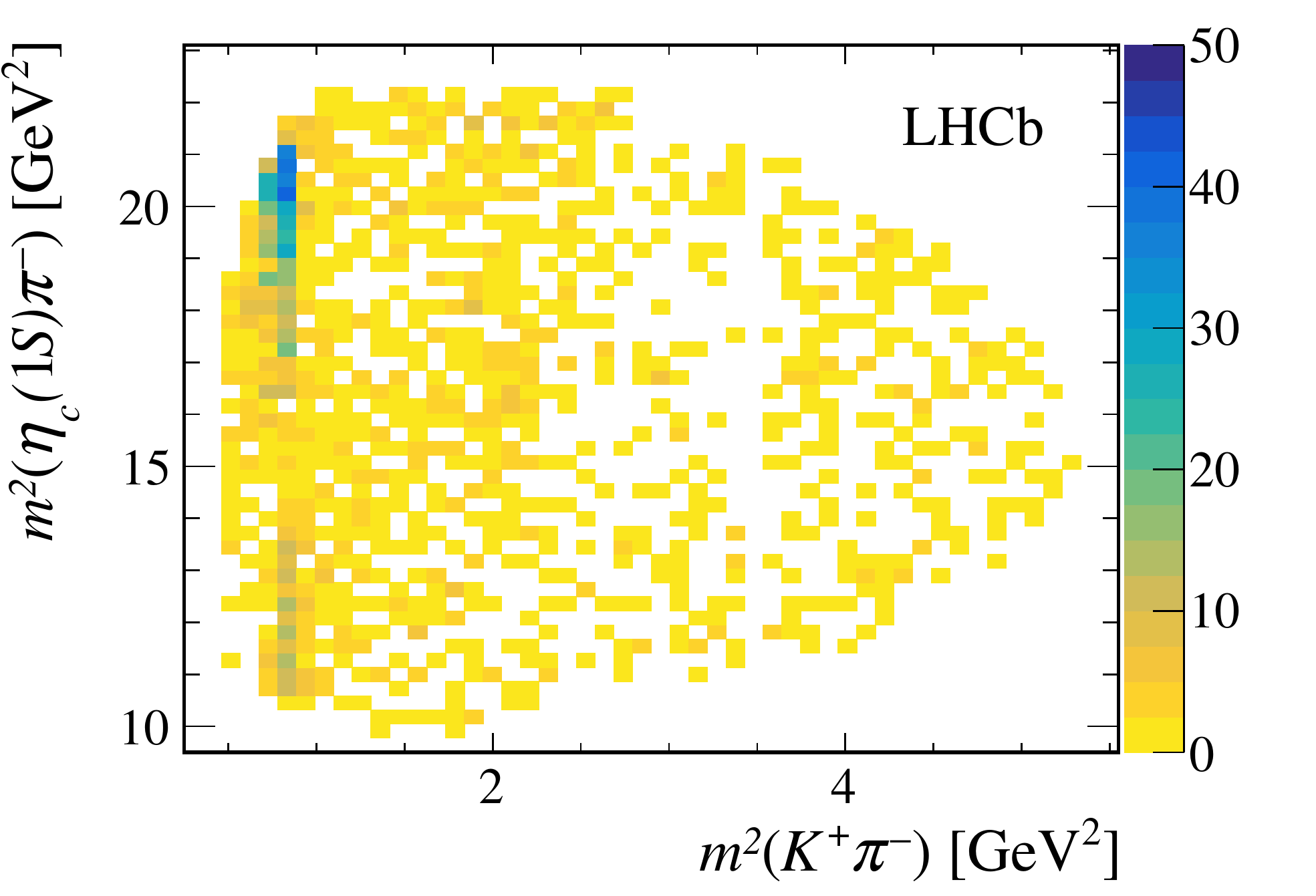}\quad
  \includegraphics[width=15cm, keepaspectratio]{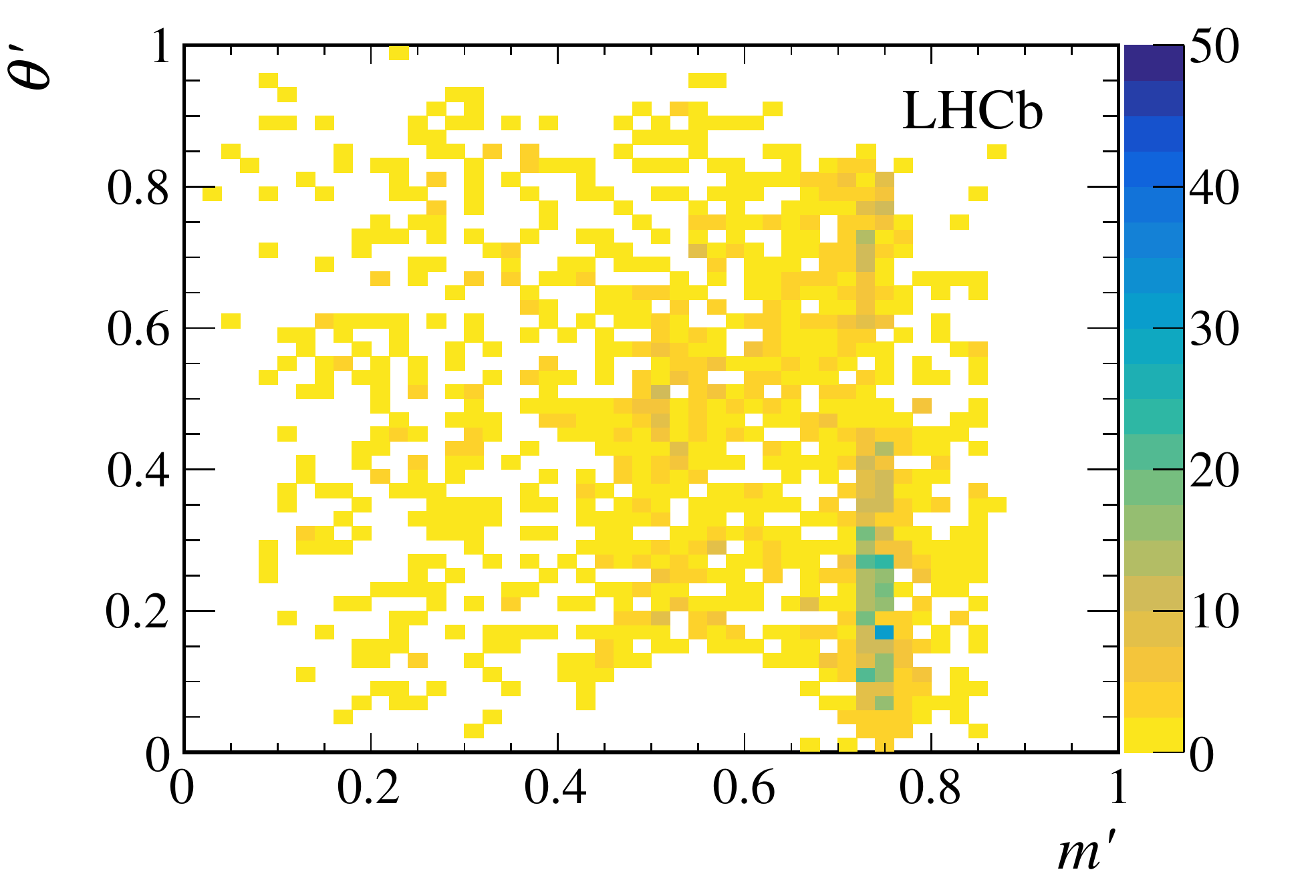}
  \caption{Background-subtracted (top) DP and (bottom) SDP
    distributions corresponding to the total data sample used in the
    analysis. The structure corresponding to the $K^*(892)^0$
    resonance is evident. The veto of \decay{\Bd}{\etac \Kp \pim}
    decays in the \Dzb region is visible in the DP.}
  \label{DPtotal}
\end{figure}

The 2D mass fit described in Sec.~\ref{sec:Yields} is repeated to the
combined Run~1 and~2 data sample, and the \sPlot technique is
applied to determine the background-subtracted DP and SDP
distributions shown in Fig.~\ref{DPtotal}.

\subsection{Signal efficiency}
\label{sec:Efficiency}
\begin{figure}[t]
\begin{center}
    \includegraphics[width=0.45\linewidth]{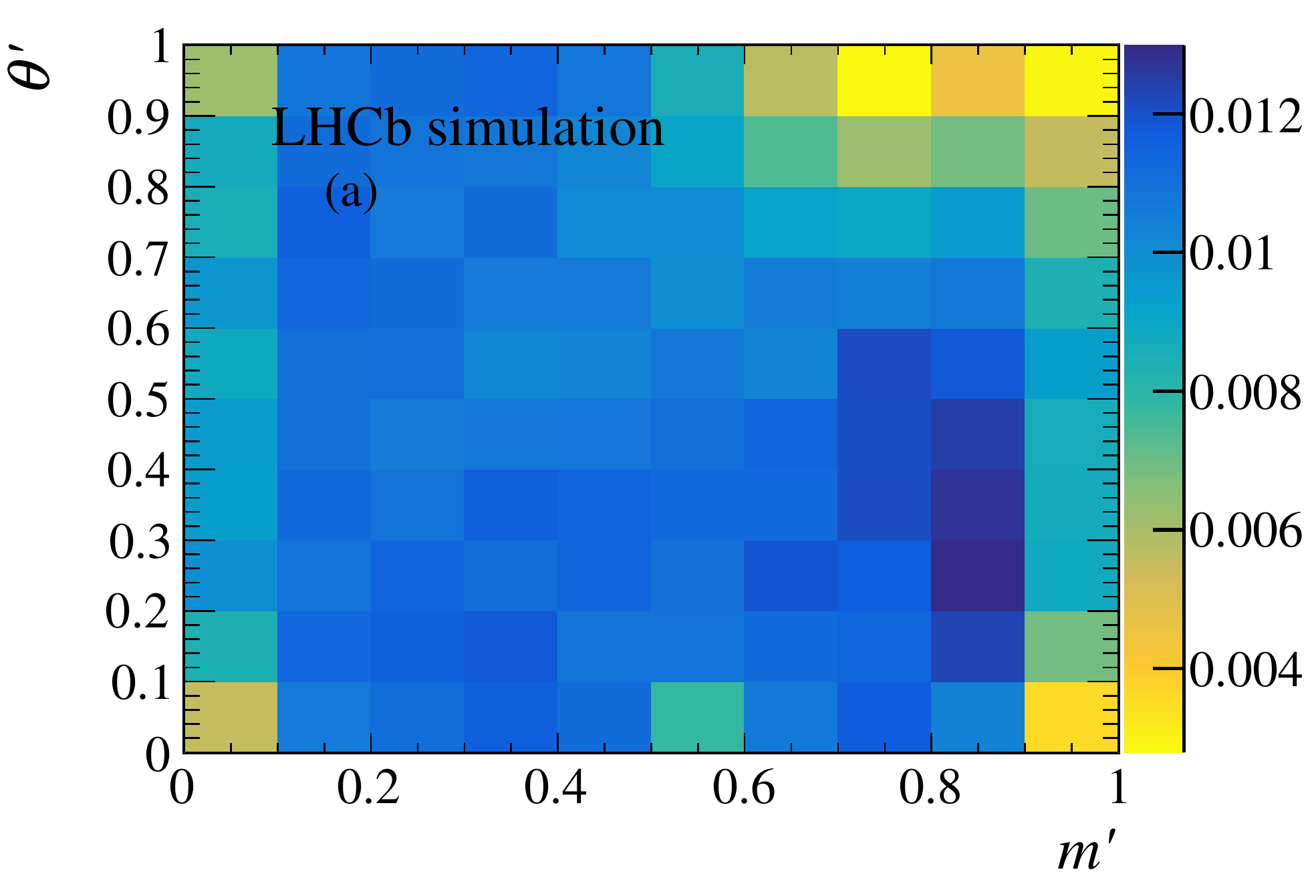}\quad
    \includegraphics[width=0.45\linewidth]{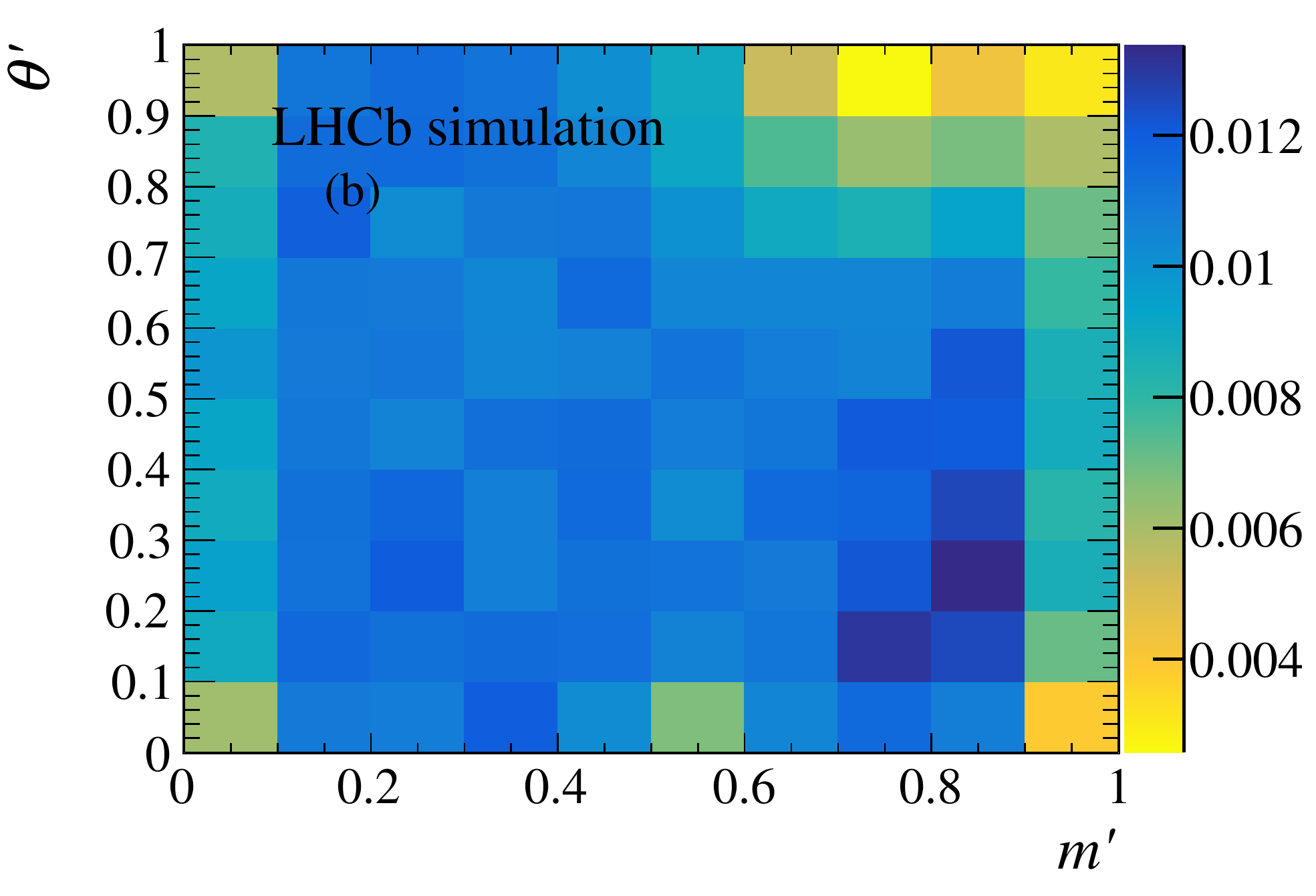}
\end{center}\vskip-1.5em
  \caption{$\decaywo$ signal efficiency across the SDP for the (a) Run~1
    and (b) Run~2 samples.}
  \label{efficiencyMaps}
\end{figure}

Efficiency variation across the SDP is caused by the
detector acceptance and by the trigger and offline selection
requirements. The efficiency variation is evaluated with simulated
samples generated uniformly across the SDP. Corrections
are applied for known differences between data and simulation
in PID efficiencies. The effect of the vetoes in the phase space is
separately accounted for by the \laura package, setting to zero the
signal efficiency within the vetoed regions. Therefore, the vetoes
corresponding to the \Dzb meson and the phase-space border are not
applied when constructing the numerator of the efficiency
histogram. The efficiency is studied separately for the Run~1 and~2 subsamples, and the resulting efficiency maps are shown in
Fig.~\ref{efficiencyMaps}. Lower efficiency in
regions with a low-momentum track is due to geometrical effects. Statistical
fluctuations in the histograms due to the limited size of the simulated
samples are smoothed by applying a 2D cubic spline interpolation. 

\subsection{\boldmath Amplitude model with only $\Kpi$ contributions}
\label{sec:DefaultModel}
In the absence of contributions from exotic resonances, only $\Kpi$ resonances are expected as intermediate states. The established $\Kstarz \to \Kp\pim$ mesons reported in Ref.~\cite{PDG2018} with $m(\Kstarz) \lesssim m(\Bd)-m(\etac)$, {\em i.e.}\
with masses within or slightly above the phase space boundary in
\decaywo decays, are used as a guide when building the model. Only
those amplitudes providing significant improvements in the description
of the data are included. This model is referred to as the baseline
model and comprises the resonances shown in Table~\ref{KstarStates}.

\begin{table}[t]
\caption{Resonances included in the baseline model, where
  parameters and uncertainties are taken from Ref.~\cite{PDG2016}. The
LASS lineshape also parametrise the $\Kpi$ S-wave in \decay{\Bd}{\etac \Kp \pim} NR decays.}
\begin{center}
\begin{tabular}{l D{,}{\pm}{-1} D{,}{\pm}{-1} c c}
\toprule
Resonance & \multicolumn{1}{c}{\text{Mass} $[\unit[]{MeV}]$} & \multicolumn{1}{c}{\text{Width} $[\unit[]{MeV}]$} & \multicolumn{1}{c}{$J^P$} & \multicolumn{1}{c}{\text{Model}}\\
\midrule
$K^*(892)^0$ & 895.55\hspace{1mm},\hspace{1mm}0.20 & 47.3\hspace{1mm},\hspace{1mm}0.5 & $1^-$ & RBW\\
$K^*(1410)^0$ & 1414\hspace{1mm},\hspace{1mm}15 & 232\hspace{1mm},\hspace{1mm}21 & $1^-$ & RBW\\ 
$K^*_0(1430)^0$ & 1425\hspace{1mm},\hspace{1mm}50 & 270\hspace{1mm},\hspace{1mm}80 & $0^+$ & LASS\\
$K^*_2(1430)^0$ & 1432.4\hspace{1mm},\hspace{1mm}1.3 & 109\hspace{1mm},\hspace{1mm}5 & $2^+$ & RBW\\
$K^*(1680)^0$ & 1717\hspace{1mm},\hspace{1mm}27 & 322\hspace{1mm},\hspace{1mm}110 & $1^-$ & RBW\\
$K^*_0(1950)^0$ & 1945\hspace{1mm},\hspace{1mm}22 & 201\hspace{1mm},\hspace{1mm}90 & $0^+$ & RBW\\
\bottomrule
\end{tabular}
\end{center}
\label{KstarStates}
\end{table}

The S-wave at low $\Kpi$ mass is modelled with the LASS probability density function. The real and imaginary parts of the complex coefficients $c_j$
introduced in Eq.~\eqref{isobarAmplitude} are free parameters of the
fit, except for the $K^*(892)^0$ component, which is taken
as the reference amplitude. Other free parameters in the fit are the scattering length ($a$) and the
effective range ($r$) parameters of the LASS function, defined in Eq.~\eqref{LASSpdf}. The
mass and width of the $K^*_0(1430)^0$ meson are Gaussian constrained to the known values~\cite{PDG2018}. 

\begin{figure}[!h]
\begin{center}
    \includegraphics[width=0.45\linewidth]{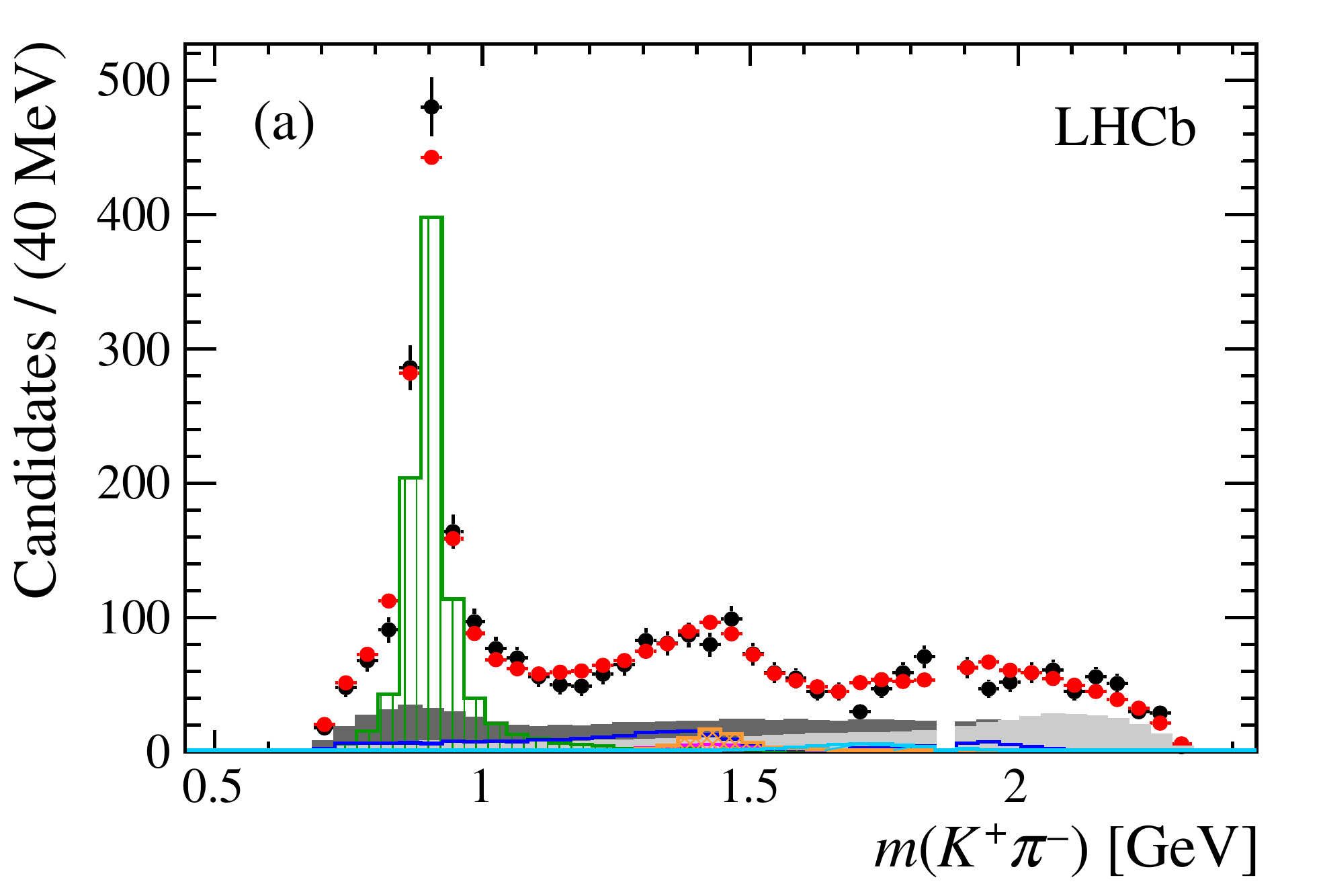}\quad
    \includegraphics[width=0.45\linewidth]{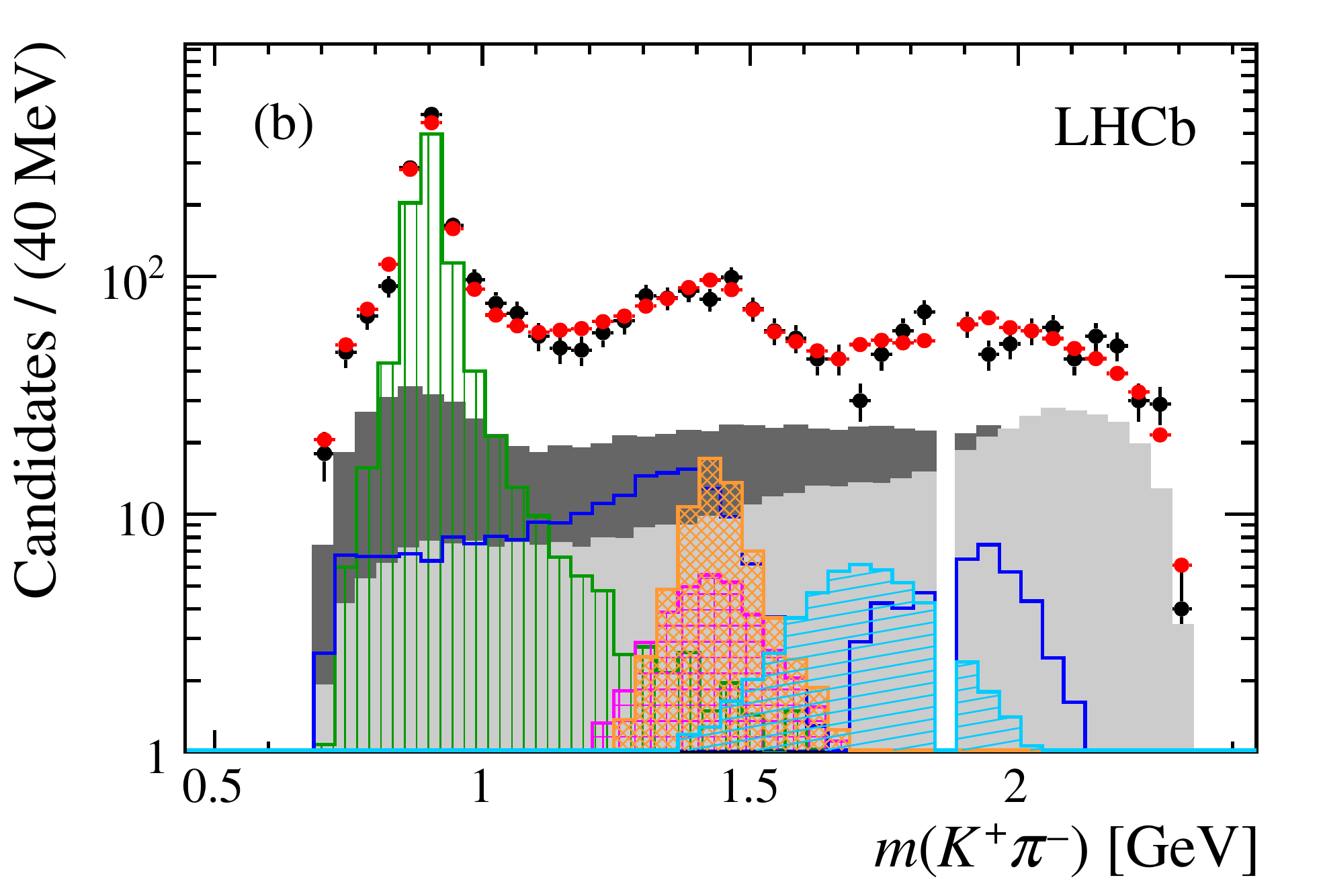}\quad
    \includegraphics[width=0.45\linewidth]{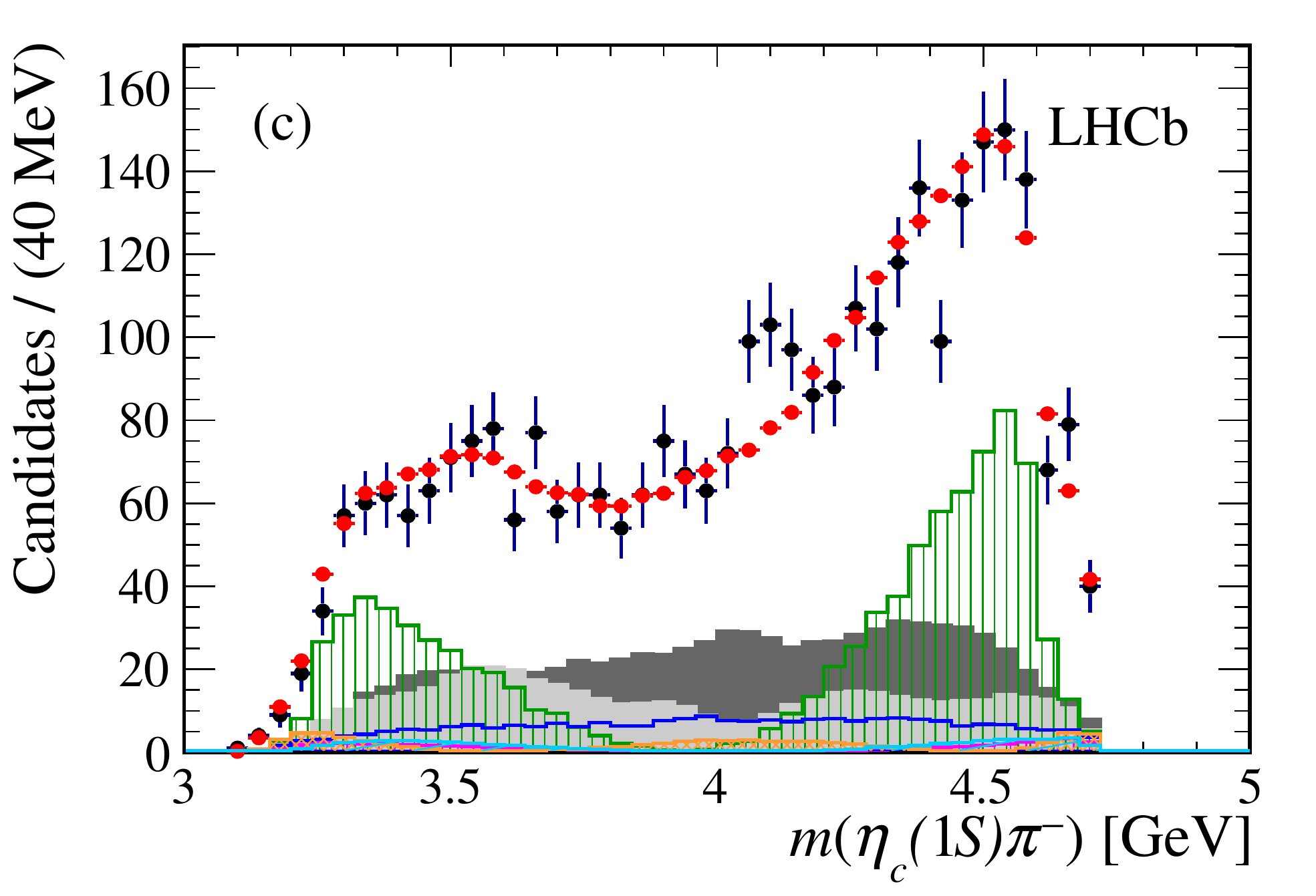}\quad
    \includegraphics[width=0.45\linewidth]{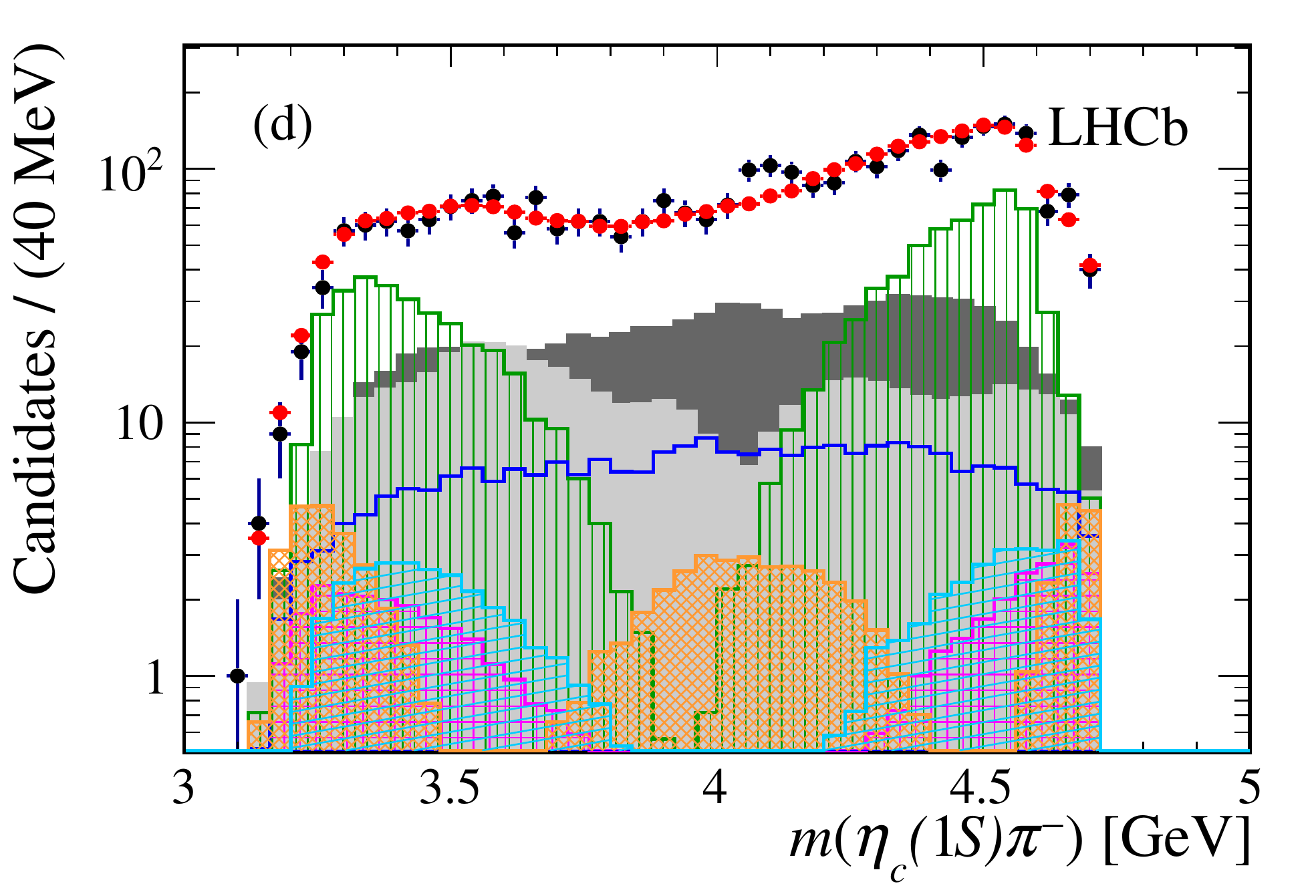}\quad
    \includegraphics[width=0.45\linewidth]{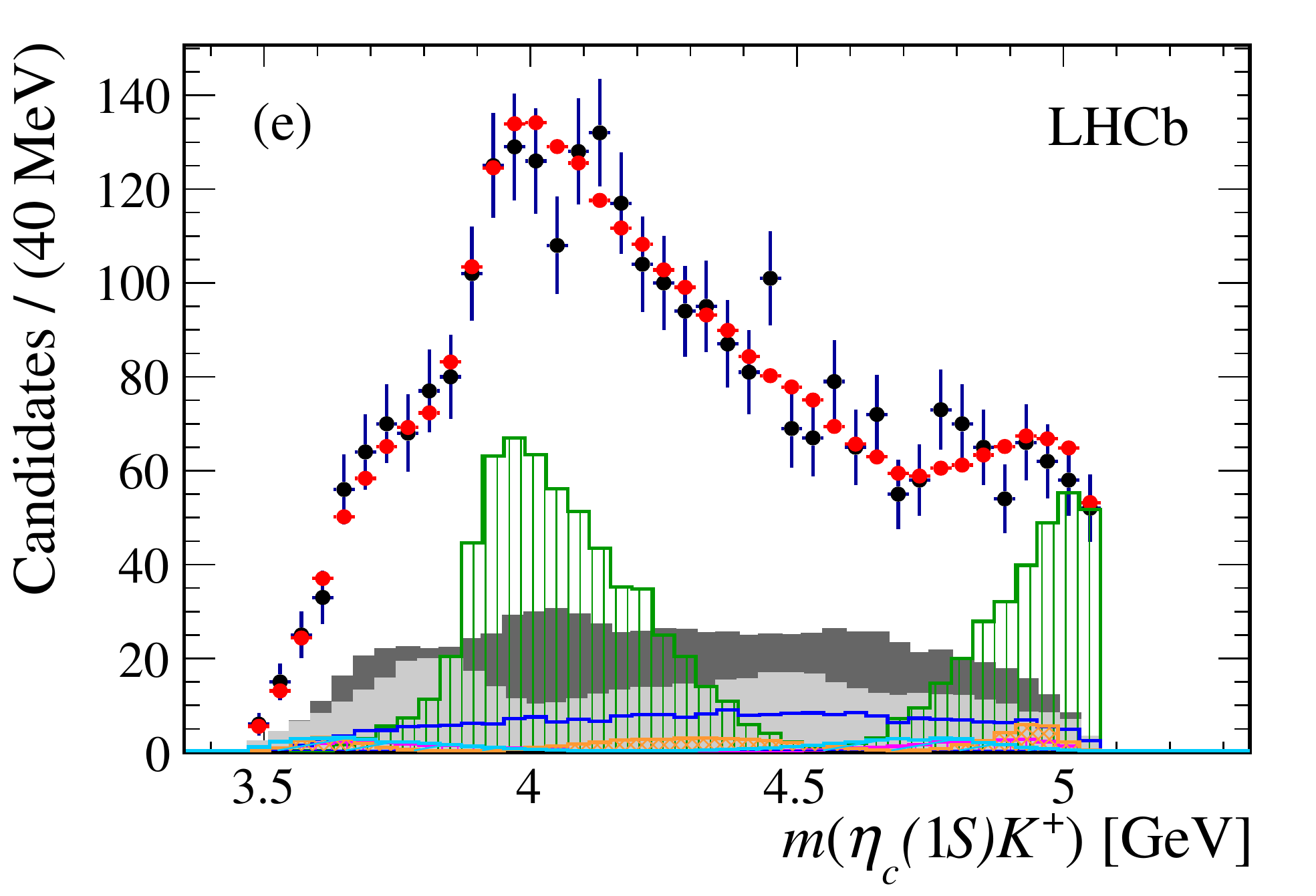}\quad
    \includegraphics[width=0.45\linewidth]{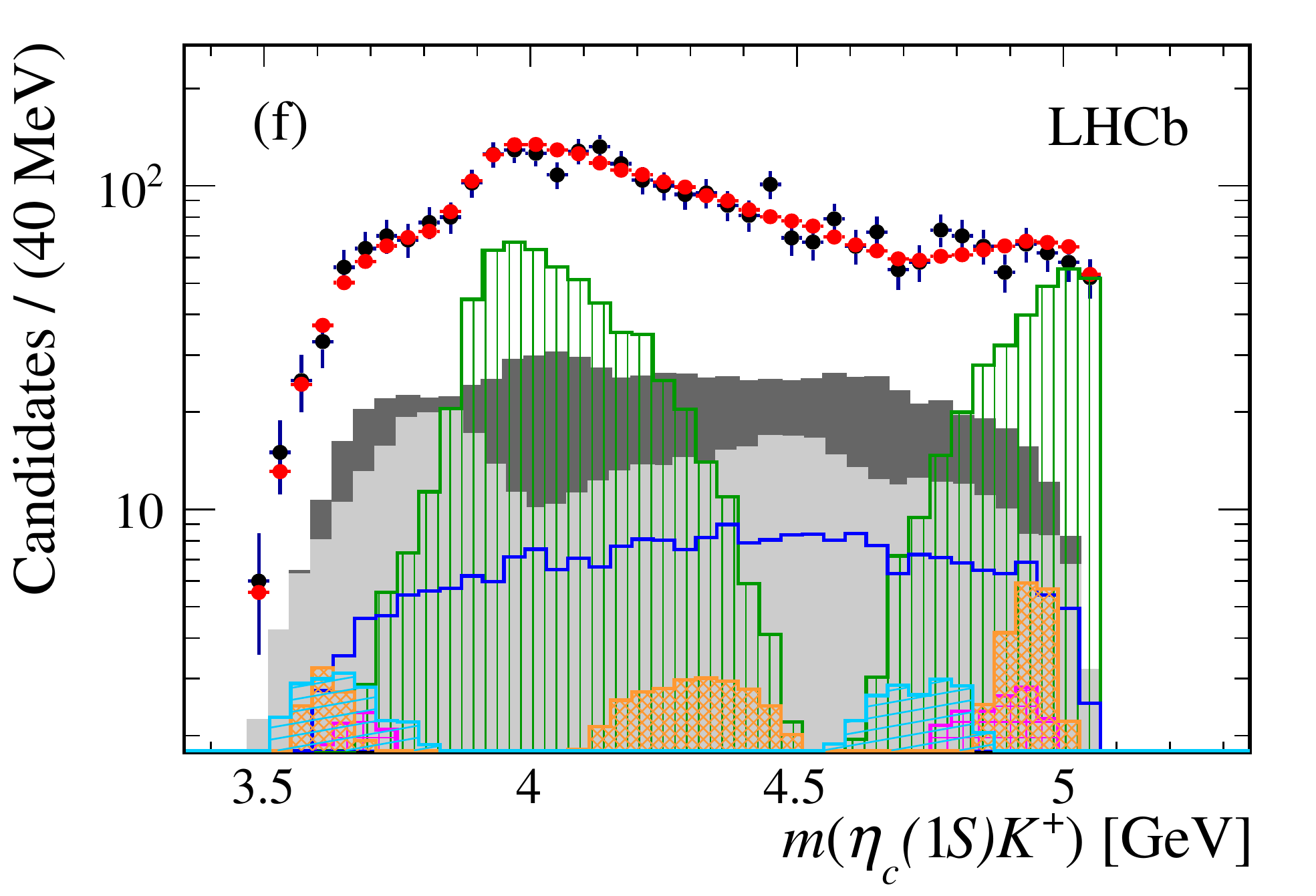}\quad
    \includegraphics[width=0.8\linewidth]{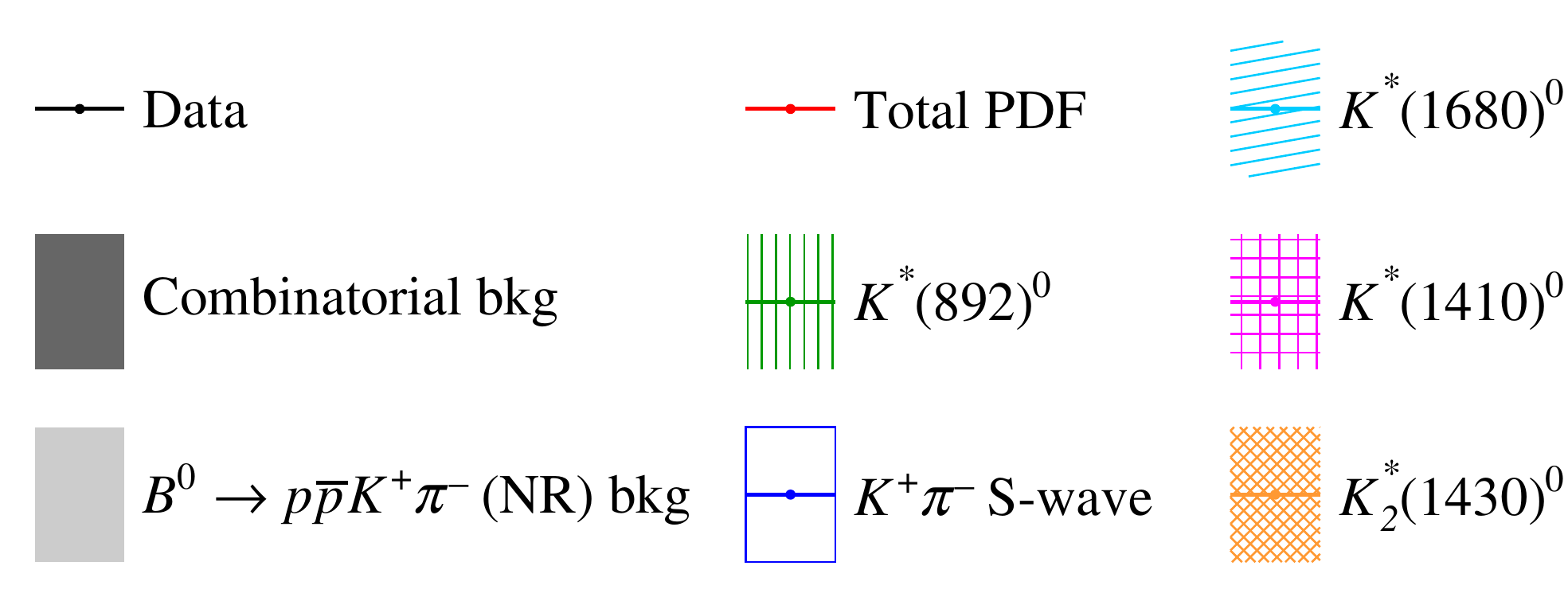}
\end{center}
  \caption{Projections of the data and amplitude fit using the
    baseline model onto~(a) $\mKpi$,~(c) $\metacpi$ and~(e) $\metacK$,
    with the same projections shown in~(b),~(d) and~(f) with a
    logarithmic vertical-axis scale. The veto of
    \decay{\Bd}{\proton\antiproton\Dzb} decays is visible in
    plot~(b). The $\Kpi$ S-wave component comprises the LASS and
    $K^*_0(1950)^0$ meson contributions. The components are
    described in the legend at the bottom.}
  \label{Kstar}
\end{figure}

While it is possible to describe the $\mKpi$ and $\metacK$ distributions well with $\Kpi$ contributions alone, the fit
projection onto the $\metacpi$ distribution does not provide a good description of data, as shown in Fig.~\ref{Kstar}. In particular, a discrepancy around $\metacpi \approx \unit[4.1]{GeV}$ is evident. 

A \chisq variable is computed as a quantitative determination of the
fit quality, using an adaptive 2D binning schema to obtain 144 equally
populated bins. The baseline model yields a
$\chisq/\text{ndof}=195/129=1.5$ value, where ndof is the number of
degrees of freedom. Including additional $\Kpi$ resonant states does not lead to significant improvements in the description of the data. These include established states such as the $K^*_3(1780)^0$ and $K^*_4(2045)^0$ mesons, the high mass $K^*_5(2380)^0$ resonance which falls outside the phase space limits, and the $K^*_2(1980)^0$ state which has not been seen in the $\Kpi$ final state thus far. The unestablished P-, D- and F-wave $\Kpi$ states predicted by the Godfrey--Isgur model~\cite{PhysRevD.32.189} to decay into the $\Kpi$ final state were also tested.

\subsection{\boldmath Amplitude model with $\Kpi$ and $\etacpi$
  contributions}\label{nominalModel}
A better description of the data is obtained by adding an exotic $Z_c^- \to
\etacpi$ component to the $\Kpi$ contributions of
Table~\ref{KstarStates}. The resulting signal model consists of eight
amplitudes: seven resonances and one NR term. The $\Kpi$ amplitudes are modelled in the same way as in the baseline model. Alternative models for the $\Kpi$ S-wave are used to
assign systematic uncertainties as discussed in
Sec.~\ref{sec:Systematics}. In addition to the free parameters used in the baseline model, the isobar coefficients, mass and width of the $Z^-_c$ resonance
are left floating. 

A likelihood-ratio test is used to discriminate between any pair of amplitude models based on the log-likelihood difference $\Delta ( -2 \ln \mathcal{L})$~\cite{significance}. Three quantum number hypotheses are probed for the $Z_c^-$ resonance, repeating the amplitude fit for the $J^P=0^+~,1^-~\text{and}~2^+$ assignments. The variations of the $\Delta ( -2 \ln \mathcal{L})$ value with respect to the baseline model are $\Delta ( -2 \ln \mathcal{L})=22.8,~41.4$, and 7.0, respectively. The model providing the best description of the data, referred to below as the nominal fit model, is obtained with the addition of a $Z^-_c$ candidate with $J^P=1^-$. The $J^P=2^+$ assignment is not considered further given the small variation in $\ln \mathcal{L}$ with respect to the additional four free parameters.

The LASS parameters obtained in the nominal fit model are
\mbox{$m_{K^{*}_0(1430)^0} = \unit[1427 \pm 21]{MeV}$},
\mbox{$\Gamma_{K^{*}_0(1430)^0} = \unit[256 \pm 33]{MeV}$}, \mbox{$a =
  \unit[3.1 \pm 1.0]{GeV^{-1}}$} and \mbox{$r = \unit[7.0 \pm
  2.4]{GeV^{-1}}$}. The parameters of the $Z_c^-$ candidate obtained in the nominal fit
model are \mbox{$m_{Z_c^-} =\unit[4096 \pm 20]{MeV}$} and
\mbox{$\Gamma_{Z_c^-} = \unit[152 \pm 58]{MeV}$}.
The values of the complex coefficients and fit fractions returned by the nominal fit model are shown in
Table~\ref{isobarResults}. The statistical uncertainties on all parameters of interest are calculated using large samples of simulated pseudoexperiments generated from the fit results in order to take into account the correlations between parameters and to guarantee the correct coverage of the uncertainties. 

\begin{table}[tb]
\caption{Complex coefficients and fit fractions determined from the DP
  fit using the nominal model. Uncertainties are statistical only.}
\begin{center}
\begin{tabular}{l D{,}{\pm}{-1} D{,}{\pm}{-1} D{,}{\pm}{-1}}
\toprule
Amplitude & \multicolumn{1}{c}{\text{Real part}} & \multicolumn{1}{c}{\text{Imaginary part}} & \multicolumn{1}{c}{\text{Fit fraction (\%)}}\\
\midrule
\vspace{1mm}
$B^0 \to \eta_c K^*(892)^0$ & \multicolumn{1}{c}{\hspace{1mm} 1 (fixed)} &
                                                       \multicolumn{1}{c}{\hspace{1mm} 0 (fixed)} & 51.4\hspace{1mm},\hspace{1mm}1.9\\
\vspace{1mm}
$B^0 \to \eta_c K^*(1410)^0$ & 0.17\hspace{1mm},\hspace{1mm}0.07 & 0.11\hspace{1mm},\hspace{1mm}0.08 & 2.1\hspace{1mm},\hspace{1mm}1.1\\
\vspace{1mm}
$B^0 \to \eta_c K^+ \pi^-$ (NR)  & -0.45\hspace{1mm},\hspace{1mm}0.08 & 0.01\hspace{1mm},\hspace{1mm}0.09 & 10.3\hspace{1mm},\hspace{1mm}1.4\\
\vspace{1mm}
$B^0 \to \eta_c  K^*_0(1430)^0$ & -0.62\hspace{1mm},\hspace{1mm}0.09 & -0.33\hspace{1mm},\hspace{1mm}0.25 & 25.3\hspace{1mm},\hspace{1mm}3.5\\
\vspace{1mm}
$B^0 \to \eta_c  K^*_2(1430)^0$ & 0.16\hspace{1mm},\hspace{1mm}0.06 & -0.23\hspace{1mm},\hspace{1mm}0.05 & 4.1\hspace{1mm},\hspace{1mm}1.5\\
\vspace{1mm}
$B^0 \to \eta_c  K^*(1680)^0$ & -0.11\hspace{1mm},\hspace{1mm}0.08 & -0.18\hspace{1mm},\hspace{1mm}0.06 & 2.2\hspace{1mm},\hspace{1mm}2.0\\
\vspace{1mm}
$B^0 \to \eta_c K^*_0(1950)^0$ & 0.27\hspace{1mm},\hspace{1mm}0.04 & 0.04\hspace{1mm},\hspace{1mm}0.14 & 3.8\hspace{1mm},\hspace{1mm}1.8\\
\vspace{1mm}
$B^0 \to Z_c(4100)^{-} K^+$ & -0.25\hspace{1mm},\hspace{1mm}0.04 & -0.01\hspace{1mm},\hspace{1mm}0.08 & 3.3\hspace{1mm},\hspace{1mm}1.1\\
\bottomrule
\end{tabular}
\end{center}
\label{isobarResults}
\end{table}

\begin{figure}[!h]
  \begin{center}
    \includegraphics[width=0.45\linewidth]{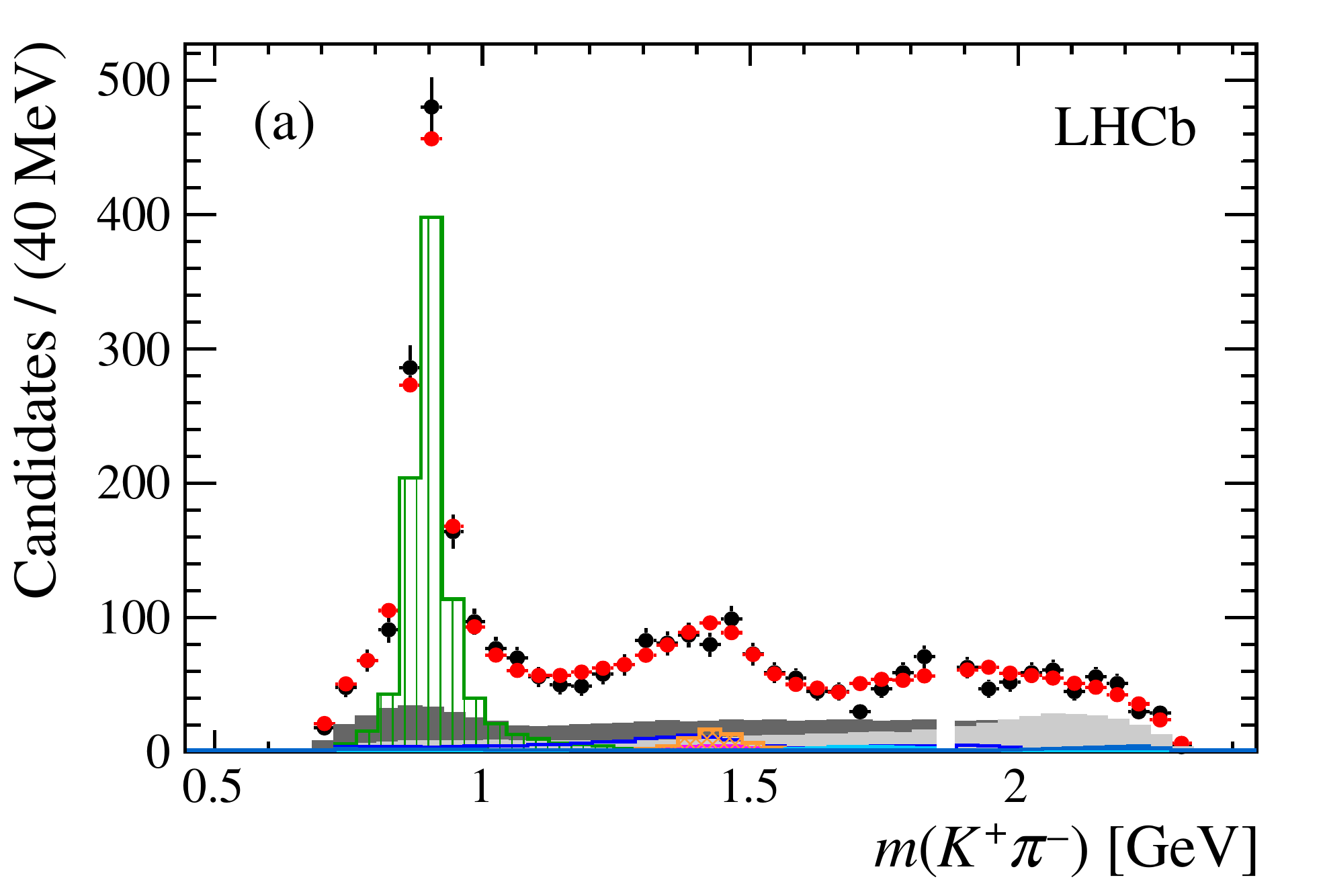}\quad
    \includegraphics[width=0.45\linewidth]{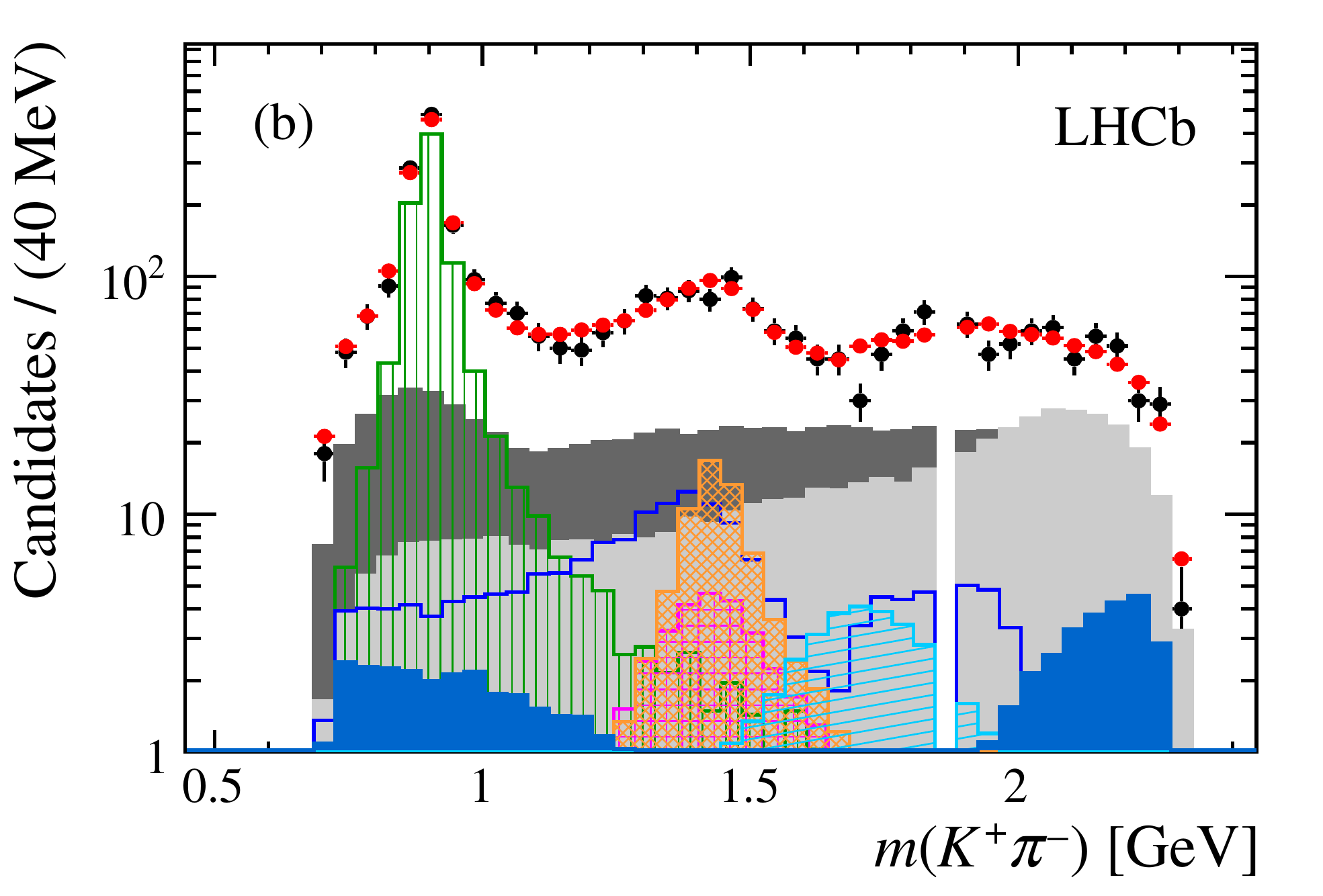}\quad
    \includegraphics[width=0.45\linewidth]{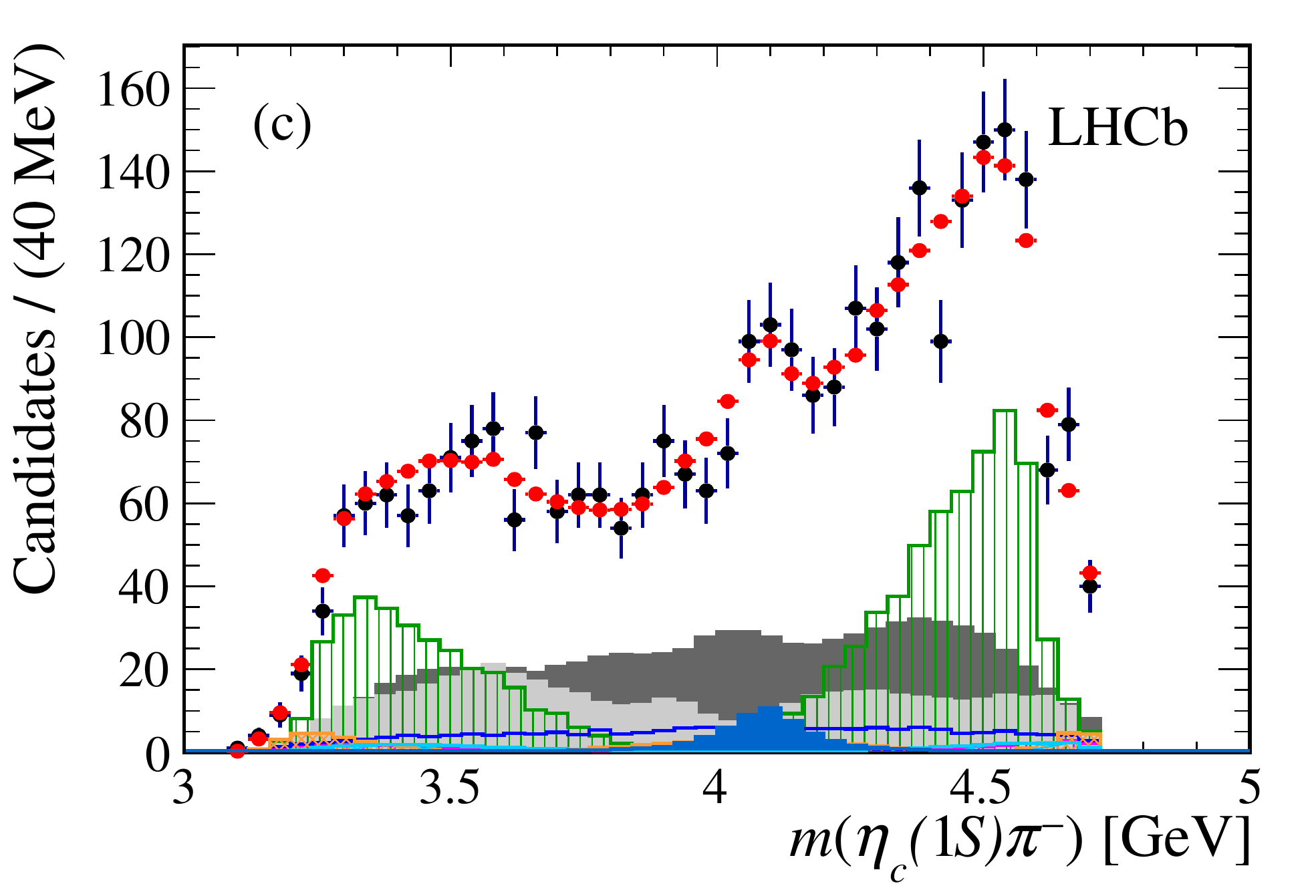}\quad
    \includegraphics[width=0.45\linewidth]{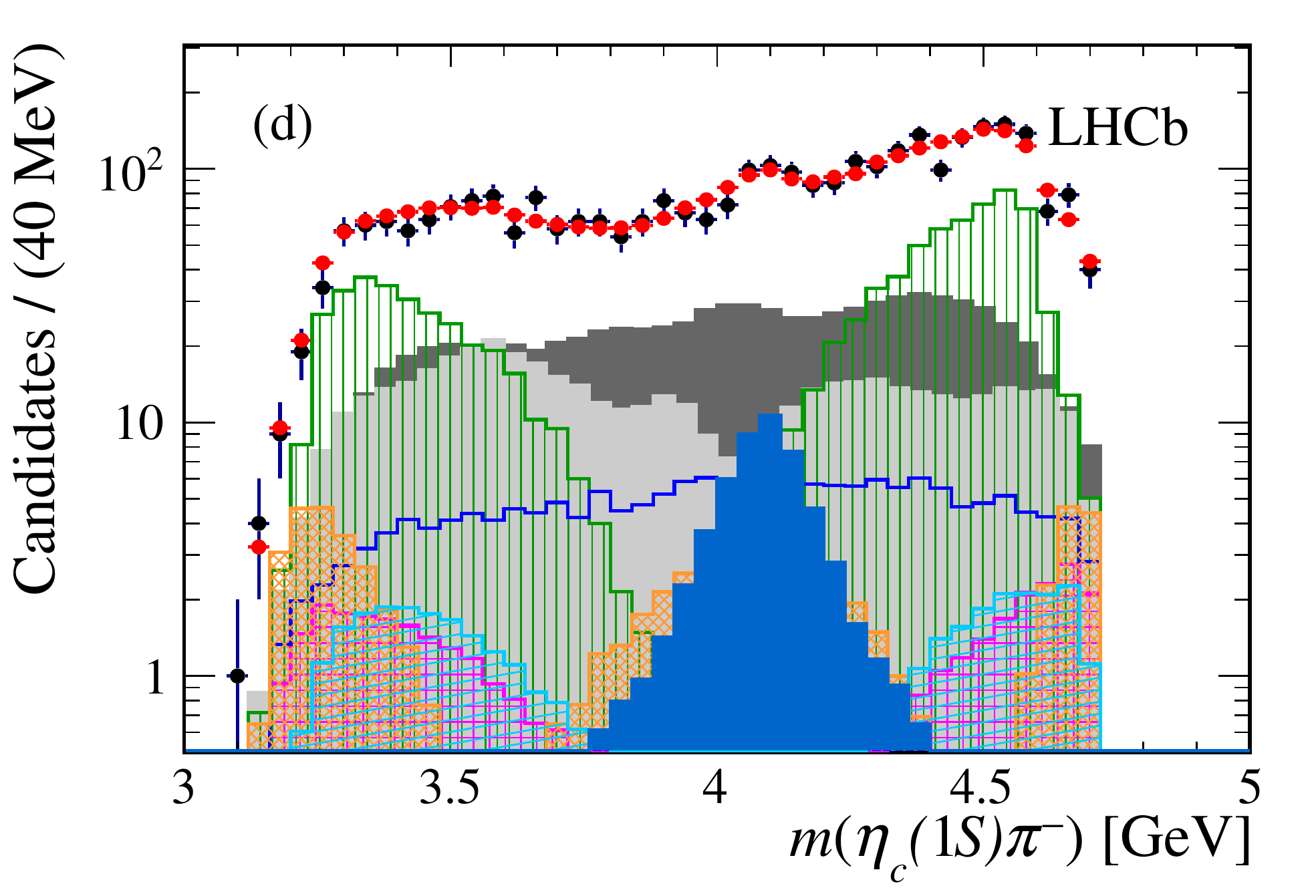}\quad
    \includegraphics[width=0.45\linewidth]{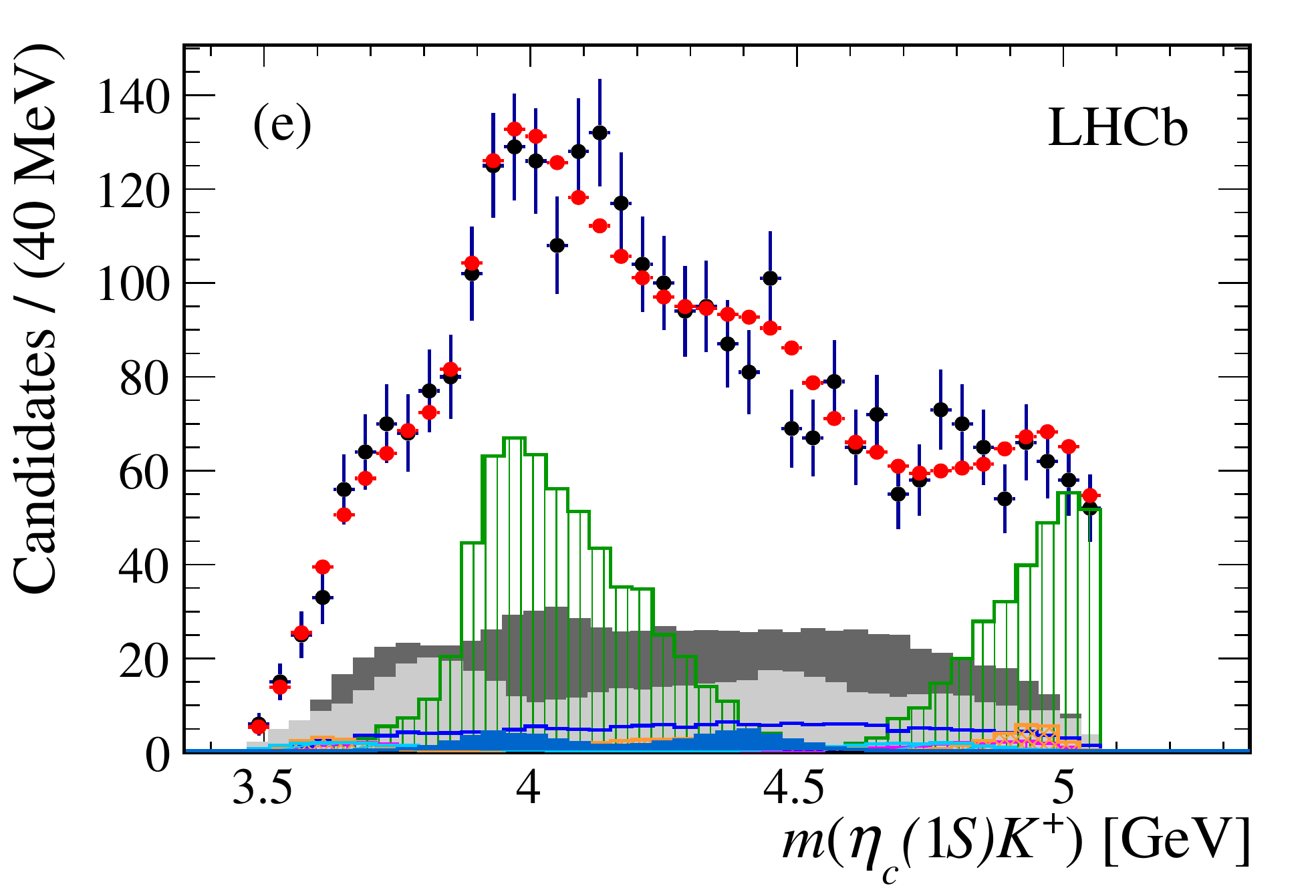}\quad
    \includegraphics[width=0.45\linewidth]{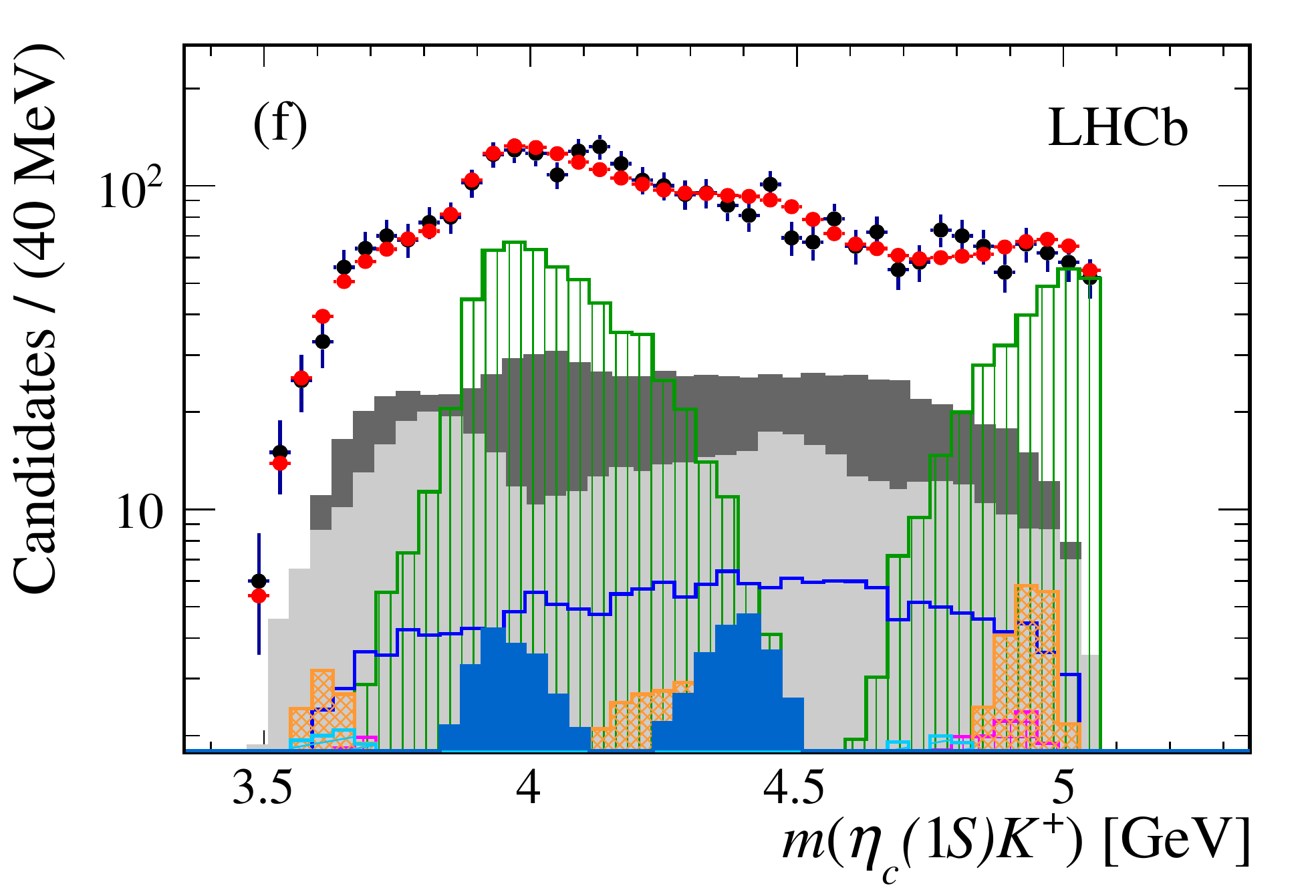}\quad
    \includegraphics[width=0.8\linewidth]{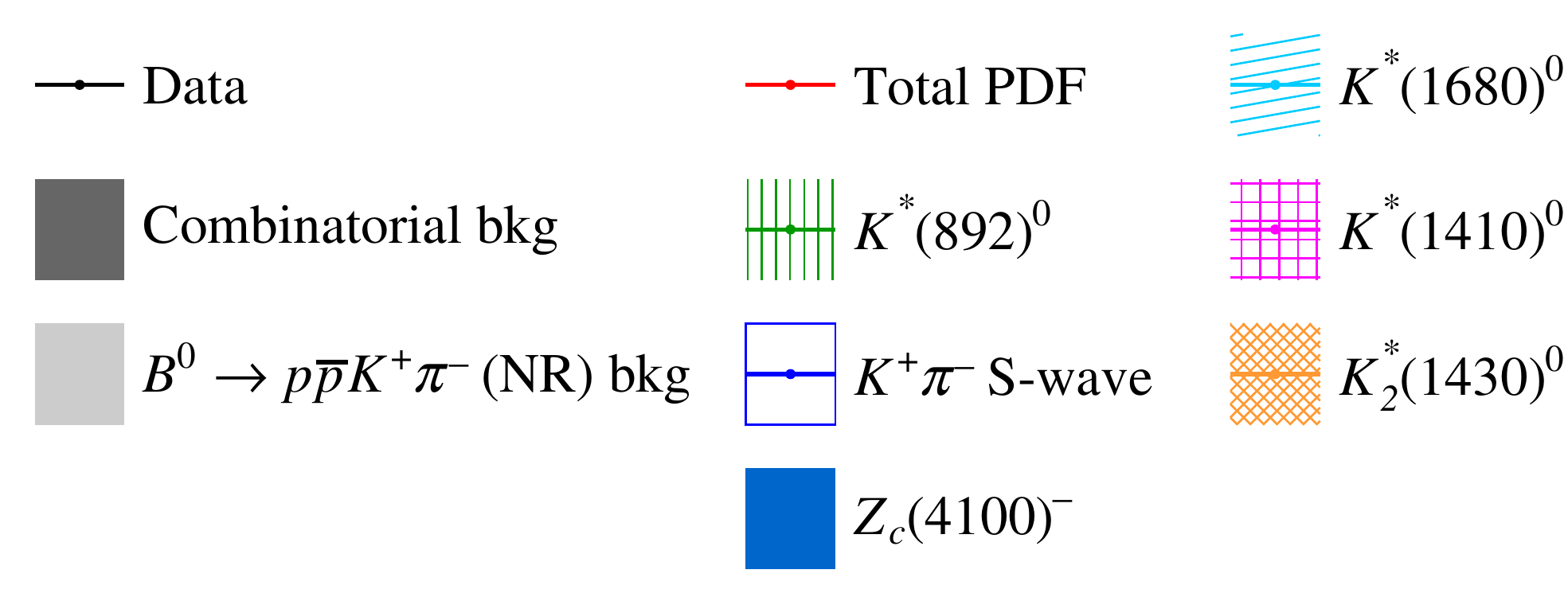}
\end{center}
  \caption{Projections of the data and amplitude fit using the
    nominal model onto~(a) $\mKpi$,~(c) $\metacpi$ and~(e) $\metacK$,
    with the same projections shown in~(b),~(d) and~(f) with a
    logarithmic vertical-axis scale. The veto of
    \decay{\Bd}{\proton\antiproton\Dzb} decays is visible in
    plot~(b). The $\Kpi$ S-wave component comprises the LASS and
    $K^*_0(1950)^0$ meson contributions. The components are
    described in the legend at the bottom.}
  \label{amplitudeProjections}
\end{figure}

Figure~\ref{amplitudeProjections} shows the projections of the nominal
fit model and the data onto $\mKpi$, $\metacpi$ and $\metacK$
invariant masses. A good agreement between the nominal fit model and
the data is obtained. The value of the $\chisq/\text{ndof}$ is
$164/125=1.3$ for the nominal fit model. The fit quality is further
discussed in Appendix~\ref{nominalLegendreMoments}, where a comparison
is reported of the unnormalised Legendre moments between data, the
baseline and nominal models. The 2D pull distributions for the
baseline and nominal models are reported as well.

The significance of the $Z_c^-$ candidate, referred to as the $Z_c(4100)^{-}$ state in the following, is evaluated from the change in the likelihood of the fits with and without the $Z_c(4100)^{-}$ component, assuming that this quantity, $\Delta ( -2 \ln
\mathcal{L})$, follows a $\chi^2$ distribution with a number of degrees of freedom equal to twice the number of free parameters in its
parametrisation~\cite{LHCb-PAPER-2014-014,LHCb-PAPER-2015-029,LHCb-PAPER-2016-018,LHCb-PAPER-2016-019}. This assumption takes into account the
look-elsewhere effect due to the floating mass and width of the
$Z_c(4100)^-$. The validity of this assumption is verified using
pseudoexperiments to predict the distribution of $\Delta ( -2 \ln
\mathcal{L})$ under the no-$Z_c(4100)^{-}$ hypothesis, which is found to be well
described by a $\chi^2$ probability density function with ndof =
8. 
The statistical significance of the $Z_c(4100)^{-}$ is
$4.8\sigma$ in the nominal fit model. The quoted significance does not include the contribution from systematic uncertainties.

To discriminate between various $J^P$ assignments, fits are performed under alternative $J^P$ hypotheses. A lower limit on the significance of rejection of the $J^P=0^+$ hypothesis is determined from the change in the log-likelihood from the preferred hypothesis, assuming a $\chi^2$ distribution with one degree of freedom. The validity of this assumption is verified using
pseudoexperiments to predict the distribution of $\Delta ( -2 \ln
\mathcal{L})$ under the disfavoured $J^P=0^+$ hypothesis. The
statistical rejection of the $J^P=0^+$ hypothesis with respect to the
$J^P=1^-$ hypothesis is $4.3\sigma$.

Systematic effects must be taken into account to report the significance of the $Z_c(4100)^-$ contribution and the discrimination of
 its quantum numbers. The fit variations producing the largest changes in
 the values of the mass, width or isobar coefficients of the exotic candidate
 are used to probe the sensitivity of the significance of the $Z_c(4100)^{-}$
 state to systematic effects, and to determine its quantum numbers, as
 described in Sec.~\ref{sec:Systematics}.

%% file: systematics.tex
\section{Systematic uncertainties}
\label{sec:Systematics}
Systematic uncertainties can be divided into two categories: experimental and model uncertainties. Among the experimental uncertainties, the largest changes in the values of the parameters of the $Z_c(4100)^{-}$ candidate are due to the signal and background yields used in the amplitude fit, the SDP distributions of the background components, and the phase-space border veto applied on the parametrisation of the efficiencies. Among the model uncertainties, the largest effects are due to the treatment of the natural width of the \etac meson within the DP fit and to the $\Kpi$ S-wave parametrisation. The DP fits using the baseline and nominal varied models are used to recompute the significance. 

The signal and background yields used in the amplitude fit are fixed to the values obtained from the 2D mass fit. The statistical uncertainties on the yields are introduced into the amplitude fit by Gaussian constraining the yields within their statistical uncertainties and by repeating the fit.

The systematic uncertainties associated to the parametrisation of the background distributions are evaluated by varying the value in each bin within the statistical uncertainty prior to the spline interpolation. About 300 new background histograms are produced for both the combinatorial and NR background components. The resulting $\Delta ( -2 \ln \mathcal{L})$ distribution follows a Gaussian distribution. The most pessimistic background parametrisation, corresponding to a $\Delta ( -2 \ln \mathcal{L})$ value that is below $3\sigma$ of the Gaussian distribution, is considered when quoting the effect of this source on the significance of the $Z_c(4100)^-$ state.

The phase-space border veto applied on the parametrisation of the efficiencies is removed to check the veto does not significantly affect the result. 

The natural width of the \etac meson is set to zero when computing the DP normalisations, calculated using the \etac meson mass values resulting from the 2D UML fits described in Sec.~\ref{sec:Yields}. In order to associate a systematic uncertainty to the sizeable \etac natural width, the amplitude fits are repeated computing the DP normalisations by using the $m_{\etac} + \Gamma_{\etac}$ and $m_{\etac} - \Gamma_{\etac}$ values, where $m_{\etac}$ and $\Gamma_{\etac}$ are the  mass and natural width of the \etac meson, respectively, obtained from the 2D UML fits.

The LASS model used to parametrise the low mass $\Kpi$ S-wave in the nominal fit is replaced with $K^*_0(1430)^0$ and $K^*_0(700)^0$ resonances parametrised with RBW functions, and a NR S-wave $\Kpi$ component parametrised with a uniform amplitude within the DP.

The effect of the separate systematic sources to the significance of the $Z_c(4100)^-$ are reported in Table~\ref{sigSystSummary}. When including the most important systematic effect, corresponding to the pessimistic background parametrisation, the lowest significance for the $Z_c(4100)^-$ candidate is given by $3.4\sigma$. In order to evaluate the effect of possible correlated or anti-correlated sources of systematic uncertainty, the fits are repeated using the pessimistic background parametrisation together with the alternative $\Kpi$ S-wave model, and with mass values of the \etac meson varied within the corresponding statistical uncertainty resulting from the 2D UML fit. The lower limit on the significance of the $Z_c(4100)^-$ state is found to be $3.2\sigma$.

\begin{table}[tb]
\caption{Significance of the $Z_c(4100)^-$ contribution for the systematic effects producing the largest variations in the parameters of the $Z_c(4100)^-$ candidate. The values obtained in the nominal amplitude fit are shown in the first row.}
\begin{center}
\begin{tabular}{l c c}
\toprule
Source & $\Delta ( -2 \ln
\mathcal{L})$ & Significance\\
\midrule
Nominal fit & $41.4$ & $4.8\sigma$ \\
\midrule
Fixed yields & $45.8$ & $5.2\sigma$ \\
Phase-space border veto & $44.6$ & $5.1\sigma$\\
\etac width & $36.6$ & $4.3\sigma$\\
$\Kpi$ S-wave & $31.8$ & $3.9\sigma$\\
Background & $27.4$ & $3.4\sigma$ \\
\bottomrule
\end{tabular}
\end{center}
\label{sigSystSummary}
\end{table}

The discrimination between the $J^P=0^+$ and $J^P=1^-$ assignments is not significant when systematic uncertainties are taken into account, as reported in Table~\ref{rejSystSummary}. When the S-wave model is varied, the two spin-parity hypotheses only differ by $1.2\sigma$.

\begin{table}[tb]
\caption{Rejection level of the $J^P=0^+$ hypothesis with respect to the $J^P=1^-$ hypothesis for the systematic variations producing the largest variations in the parameters of the $Z_c(4100)^-$ candidate. The values obtained in the nominal amplitude fit are shown in the first row.}
\begin{center}
\begin{tabular}{l D{.}{.}{-1} c}
\toprule
Source & \multicolumn{1}{c}{$\Delta ( -2 \ln
\mathcal{L})$} & Significance\\
\midrule
Default & 18.6 & $4.3\sigma$\\
\midrule
Fixed yields & 23.8 & $4.9\sigma$\\
Phase-space border veto & 24.4 & $4.9\sigma$\\
\etac width & 4.2 & $2.0\sigma$\\
Background & 3.4 & $1.8\sigma$\\
$\Kpi$ S-wave & 1.4 & $1.2\sigma$\\
\bottomrule
\end{tabular}
\end{center}
\label{rejSystSummary}
\end{table} 

Additional sources of systematic uncertainties are considered when evaluating the uncertainty on the mass and width of the $Z_c(4100)^{-}$ resonance, and on the fit fractions obtained with the nominal model. These additional sources are: the efficiency variation across the SDP and a possible bias due to the fitting procedure, contributing to the experimental systematic uncertainties category; and the fixed parameters of the resonances in the amplitude model and the addition or removal of marginal amplitudes, contributing to the model systematic uncertainties category. For each source, the systematic uncertainty assigned to each quantity is taken as the difference between the value returned by the modified amplitude fit and nominal model fit result. The uncertainties due to all these sources are obtained by combining positive and negative deviations in quadrature separately.

The bin contents of the histograms describing the efficiency variation across the SDP are varied within their uncertainties prior to the spline interpolation, as is done for the systematic uncertainty associated to the background parametrisations. A possible source of systematic effects in the efficiency histograms is due to neighbouring bins varying in a correlated way. In order to evaluate this systematic uncertainty, 10 bins of the efficiency histograms are varied within their statistical uncertainty, and the neighbouring bins are varied by linear interpolation. The binning scheme of the control sample used to evaluate the PID performance is varied. 

Pseudoexperiments are generated from the fit results using the nominal model in order to assign a systematic uncertainty due to possible amplitude fit bias.

Systematic uncertainties due to fixed parameters in the fit model are determined by repeating the fit and varying these parameters. The fixed masses and widths of the $\Kpi$ contributions are varied 100 times assigning a random number within the range defined by the corresponding uncertainties reported in Table~\ref{KstarStates}. The Blatt--Weisskopf barrier radii, $r_{\text{BW}}$, are varied independently for $\Kpi$ and $\etacpi$ resonances between $3$ and $\unit[5]{GeV^{-1}}$.

Systematic uncertainties are assigned from the changes in the results when the amplitudes due to the established $K^*_3(1780)^0$ and $K^*_4(2045)^0$ resonances, not contributing significantly in the baseline and nominal models, are included.

The total systematic uncertainties for the fit fractions are given together with the results in Sec.~\ref{sec:Results}. The dominant experimental systematic uncertainty is due to either the phase-space border veto, related to the efficiency parametrisation, or the background distributions across the SDP, while the model uncertainties are dominated by the description of the $\Kpi$ S-wave.

The stability of the fit results is confirmed by several cross-checks. The addition of further high-mass $K^{*0}$ states to the nominal model does not improve the quality of the fit. An additional amplitude decaying to $\etacpi$ is not significant, nor is an additional exotic amplitude decaying to $\etacK$. The \etac meson resonant phase motion due to the sizeable natural width could affect the overall amplitude of Eq.~\eqref{isobarAmplitude}, introducing interference effects with the NR $p\bar{p}$ contribution. In
order to investigate this effect, the data sample is divided in two parts, containing candidates with masses below and above the \etac meson peak, respectively. The results are compatible with those reported in
Sec.~\ref{sec:DalitzFit} using the full data sample, supporting the argument that the effects due to the variation of the \etac phase are negligible.

%% file: results.tex
\section{Results and summary}
\label{sec:Results}
In summary, the first measurement of the $\decaywo$ branching fraction
is reported and gives

\begin{equation*} \label{resultBF}
\mathcal{B}(\decaywo) = (5.73 \pm 0.24 \pm 0.13 \pm 0.66) \times 10^{-4},
\end{equation*}
where the first uncertainty is statistical, the second systematic, and
the third is due to limited knowledge of external branching
fractions. The first Dalitz plot analysis of the $\decaywo$ decay is
performed. A good description of data is obtained when including
a charged charmonium-like resonance decaying to the $\etacpi$ final
state with \mbox{$m_{Z_c^-} = \unit[4096 \pm 20~ ^{+18}_{-22}]{MeV}$} and
\mbox{$\Gamma_{Z_c^-}=\unit[152 \pm 58~^{+60}_{-35}]{MeV}$}. The fit
fractions are reported in Table~\ref{finalFitFrac}. The fit fractions
for resonant and nonresonant contributions are converted into
quasi-two-body branching fractions by multiplying by the $\decaywo$ branching
fraction. The corresponding results are shown in
Table~\ref{finalResults}. The $B^0 \to \eta_c K^*(892)^0$ branching
fraction is compatible with the world-average value~\cite{PDG2018},
taking into account the $K^*(892)^0 \to K^+\pi^-$ branching fraction. The
values of the interference fit fractions are given in
Table~\ref{interferenceFitFractions}.

The significance of the $Z_c(4100)^-$ candidate is more than three
standard deviations when including systematic uncertainties. This is the first evidence
for an exotic state decaying into two pseudoscalars. The favoured
spin-parity assignments, $J^P=0^+$ and $J^P=1^-$, cannot be
discriminated once systematic uncertainties are taken into account,
which prohibits unambiguously assigning the $Z_c(4100)^-$ as one of the states
foreseen by the models described in
Sec.~\ref{sec:Introduction}. Furthermore, the mass value of the
$Z_c(4100)^-$ charmonium-like state is above the open-charm threshold,
in contrast with the predictions of such models. More data will be
required to conclusively determine the nature of the $Z_c(4100)^-$
candidate.

\begin{table}[!h]
\caption{Fit fractions and their uncertainties. The quoted uncertainties are statistical and systematic, respectively.}
\begin{center}
\begin{tabular}{l D{,}{\pm}{-1}}
\toprule
Amplitude & \multicolumn{1}{c}{\text{\hspace{10mm}Fit fraction (\%)}}\\
\midrule
\vspace{1mm}
$B^0 \to \eta_c K^*(892)^0$ & 51.4\hspace{1mm},\hspace{1mm}1.9~^{+1.7}_{-4.8} \\
\vspace{1mm}
$B^0 \to \eta_c K^*(1410)^0$ & 2.1\hspace{1mm},\hspace{1mm}1.1~^{+1.1}_{-1.1}\\
\vspace{1mm}
$B^0 \to \eta_c K^+ \pi^-$ (NR) & 10.3\hspace{1mm},\hspace{1mm}1.4~^{+1.0}_{-1.2}\\
\vspace{1mm}
$B^0 \to \eta_c K^*_0(1430)^0$ & 25.3\hspace{1mm},\hspace{1mm}3.5~^{+3.5}_{-2.8}\\
\vspace{1mm}
$B^0 \to \eta_c K^*_2(1430)^0$ & 4.1\hspace{1mm},\hspace{1mm}1.5~^{+1.0}_{-1.6}\\ 
\vspace{1mm}
$B^0 \to \eta_c K^*(1680)^0$ & 2.2\hspace{1mm},\hspace{1mm}2.0~^{+1.5}_{-1.7}\\
\vspace{1mm}
$B^0 \to \eta_c K^*_0(1950)^0$ & 3.8\hspace{1mm},\hspace{1mm}1.8~^{+1.4}_{-2.5}\\
\vspace{1mm}
$B^0 \to Z_c(4100)^- K^+$ & 3.3\hspace{1mm},\hspace{1mm}1.1~^{+1.2}_{-1.1}\\
\bottomrule
\end{tabular}
\end{center}
\label{finalFitFrac}
\end{table}

\begin{table}[!h]
\caption{Branching fraction results. The four quoted
  uncertainties are statistical, \mbox{$\decaywo$} branching fraction
  systematic (not including the contribution from the uncertainty
  associated to the efficiency ratio, to avoid double counting the
  systematic uncertainty associated to the evaluation of the
  efficiencies), fit fraction systematic and external branching
  fractions uncertainties, respectively.}
\begin{center}
\begin{tabular}{l D{,}{\pm}{-1}}
\toprule
Decay mode & \multicolumn{1}{c}{\text{\hspace{45mm} Branching fraction} ($10^{-5}$)}\\
\midrule
\vspace{1mm}
$B^0 \to \eta_c K^*(892)^0 ( \to K^+\pi^-)$ & 29.5\hspace{1mm},\hspace{2mm}1.6\hspace{1mm}\pm
               \hspace{1mm} 0.6\hspace{3.5mm}^{+1.0}_{-2.8}
               \hspace{3.5mm} \pm 3.4\\
\vspace{1mm}
$B^0 \to \eta_c K^*(1410)^0 ( \to K^+\pi^-)$ & 1.20\hspace{1mm},\hspace{1mm}0.63 \pm
                               0.02 \pm 0.63 \pm 0.14\\
\vspace{1mm}
$B^0 \to \eta_c K^+ \pi^-$ (NR) & 5.90\hspace{1mm},\hspace{1mm}0.84
                   \pm 0.11\hspace{2.5mm}^{+0.57}_{-0.69} \hspace{2mm}\pm 0.68\\
\vspace{1mm}
$B^0 \to \eta_c K^*_0(1430)^0 ( \to K^+\pi^-)$ & 14.50\hspace{1mm},\hspace{1mm}2.10 \pm 0.28\hspace{2.5mm}^{+2.01}_{-1.60} \hspace{2mm}\pm 1.67\\
\vspace{1mm}
$B^0 \to \eta_c K^*_2(1430)^0 ( \to K^+\pi^-)$ & 2.35\hspace{1mm},\hspace{1mm}0.87 \pm 0.05\hspace{2.5mm}^{+0.57}_{-0.92} \hspace{2mm}\pm 0.27\\ 
\vspace{1mm}
$B^0 \to \eta_c K^*(1680)^0 ( \to K^+\pi^-)$ & 1.26\hspace{1mm},\hspace{1mm}1.15 \pm 0.02 \hspace{1.5mm}~^{+0.86}_{-0.97} \hspace{1.5mm}\pm 0.15\\
\vspace{1mm}
$B^0 \to \eta_c K^*_0(1950)^0 ( \to K^+\pi^-)$ & 2.18\hspace{1mm},\hspace{1mm}1.04 \pm 0.04\hspace{2.5mm}^{+0.80}_{-1.43} \hspace{2mm}\pm 0.25\\
\vspace{1mm}
$B^0 \to Z_c(4100)^- K^+$ & 1.89\hspace{1mm},\hspace{1mm}0.64 \pm 0.04\hspace{2.5mm}^{+0.69}_{-0.63} \hspace{2mm}\pm 0.22\\
\bottomrule
\end{tabular}
\end{center}
\label{finalResults}
\end{table}

\begin{sidewaystable}[!h]
\footnotesize
\begin{center}
\resizebox{0.9\textwidth}{!}
{\begin{minipage}{\textwidth}
\caption{Symmetric matrix of the fit fractions (\%) from the amplitude fit using the nominal model. The quoted uncertainties are statistical and
  systematic, respectively. The diagonal elements correspond to the values
  reported in Table~\ref{finalFitFrac}.}
\begin{tabular}{l c c c c c c c c}
\toprule
\vspace{3mm}
& $K^*(892)^0$ & $K^*(1410)^0$ & LASS NR & $K^*_0(1430)^0$ & $K^*_2(1430)^0$ & $K^*(1680)^0$ & $K^*_0(1950)^0$ & $Z_c(4100)^-$\\
\vspace{3mm}
$K^*(892)^0$ & $51.4 \pm 1.9~^{+1.7}_{-4.8}$ & $1.7 \pm 1.9~^{+2.4}_{-1.4}$ & $0$ & $0$ & $0$ & $-2.1 \pm 1.1~^{+1.4}_{-1.5}$& $0$ & $1.4 \pm 1.0~^{+1.2}_{-1.1}$\\
\vspace{3mm}
$K^*(1410)^0$ & & $2.1 \pm 1.1~^{+1.1}_{-1.1}$ & $0$ & $0$ & $0$ & $-2.5 \pm 1.6~^{+1.9}_{-1.7}$ &
                                                                   $0$
                                                                                                        & $-0.4 \pm 0.4~^{+0.7}_{-0.5}$\\
\vspace{3mm}
LASS NR & & & $10.3 \pm 1.4~^{+1.0}_{-1.2}$ & $-5.8 \pm 1.3~^{+2.2}_{-2.0}$ & $0$ & $0$ & $-3.2
                                                               \pm 2.8~^{+4.9}_{-1.4}$ &
                                                                       $1.11
                                                                       \pm
                                                                       0.23~^{+0.54}_{-0.35}$\\
\vspace{3mm}
$K^*_0(1430)^0$ & & & & $25.3 \pm 3.5~^{+3.5}_{-2.8}$ & $0$ & $0$ & $4.7 \pm 0.7~^{+1.3}_{-1.5}$ & $2.8
                                                                 \pm 0.4~^{+0.6}_{-0.4}$ \\
\vspace{3mm}
$K^*_2(1430)^0$ & & & & & $4.1 \pm 1.5~^{+1.0}_{-1.6}$ & $0$ & $0$ & $0.00 \pm 0.31~^{+0.76}_{-0.26}$\\ 
\vspace{3mm}
$K^*(1680)^0$ & & & & & & $2.2 \pm 2.0~^{+1.5}_{-1.7}$ & $0$ & $0.7 \pm 0.5~^{+0.5}_{-0.9}$\\
\vspace{3mm}
$K^*_0(1950)^0$ & & & & & &  & $3.8 \pm 1.8~^{+1.4}_{-2.5}$ & $0.6 \pm 0.5~^{+0.8}_{-1.1}$\\
\vspace{3mm}
$Z_c(4100)^-$ & & & & & & & & $3.3 \pm 1.1~^{+1.2}_{-1.1}$\\
\bottomrule
\end{tabular}
\label{interferenceFitFractions}
\end{minipage}}
\end{center}
\end{sidewaystable}

\clearpage
\newpage

%% file: acknowledgements.tex
\section*{Acknowledgements}
\noindent We express our gratitude to our colleagues in the CERN
accelerator departments for the excellent performance of the LHC. We
thank the technical and administrative staff at the LHCb
institutes.
We acknowledge support from CERN and from the national agencies:
CAPES, CNPq, FAPERJ and FINEP (Brazil); 
MOST and NSFC (China); 
CNRS/IN2P3 (France); 
BMBF, DFG and MPG (Germany); 
INFN (Italy); 
NWO (Netherlands); 
MNiSW and NCN (Poland); 
MEN/IFA (Romania); 
MSHE (Russia); 
MinECo (Spain); 
SNSF and SER (Switzerland); 
NASU (Ukraine); 
STFC (United Kingdom); 
NSF (USA).
We acknowledge the computing resources that are provided by CERN, IN2P3
(France), KIT and DESY (Germany), INFN (Italy), SURF (Netherlands),
PIC (Spain), GridPP (United Kingdom), RRCKI and Yandex
LLC (Russia), CSCS (Switzerland), IFIN-HH (Romania), CBPF (Brazil),
PL-GRID (Poland) and OSC (USA).
We are indebted to the communities behind the multiple open-source
software packages on which we depend.
Individual groups or members have received support from
AvH Foundation (Germany);
EPLANET, Marie Sk\l{}odowska-Curie Actions and ERC (European Union);
ANR, Labex P2IO and OCEVU, and R\'{e}gion Auvergne-Rh\^{o}ne-Alpes (France);
Key Research Program of Frontier Sciences of CAS, CAS PIFI, and the Thousand Talents Program (China);
RFBR, RSF and Yandex LLC (Russia);
GVA, XuntaGal and GENCAT (Spain);
the Royal Society
and the Leverhulme Trust (United Kingdom);
Laboratory Directed Research and Development program of LANL (USA).

%% file: appendixB2etacKpi.tex
\clearpage

{\noindent\normalfont\bfseries\Large Appendix}
\appendix

\section{Blatt-Weisskopf barrier factors}\label{Blatt-Weisskopf}
The Blatt--Weisskopf barrier factors~\cite{Blatt:1952ije} $X(z)$, where $z=|\vec{q}|r_{\text{BW}}$ or
$|\vec{p}|r_{\text{BW}}$ with $r_{\text{BW}}$ being the
barrier radius, are given by 
\begin{flalign}
&L=0:~X(z)=1,&\\
&L=1:~X(z)=\sqrt{\frac{1+z_0^2}{1+z^2}},&\\
&L=2:~X(z)=\sqrt{\frac{z_0^4+3z_0^2+9}{z^4+3z^2+9}},&\\
&L=3:~X(z)=\sqrt{\frac{z_0^6+6z_0^4+45z_0^2+225}{z^6+6z^4+45z^2+225}},&\\
&L=4:~X(z)=\sqrt{\frac{z_0^8+10z_0^6+135z_0^4+1575z_0^2+11025}{z^8+10z^6+135z^4+1575z^2+11025}},
\end{flalign}
where $z_0$ is the value of $z$ when the invariant mass is equal to
the pole mass of the resonance and $L$ is the orbital angular momentum
between the resonance children. Since the latter are scalars, $L$ is
equal to the spin of the resonance. Since the \Bd meson and the
accompanying particle in the decay are scalars as well, $L$ is also
equal to the orbital angular momentum between the resonance and the
accompanying particle in the decay.

\section{Angular probability distributions}\label{Zemach}
Using the Zemach tensor
formalism~\cite{PhysRev.133.B1201,PhysRev.140.B97}, the angular probability distributions $Z(\vec{p},\vec{q})$ are given by
\begin{flalign}
&L=0:~Z(\vec{p},\vec{q})=1,&\\
&L=1:~Z(\vec{p},\vec{q})=-2\vec{p}\cdot\vec{q},&\\
&L=2:~Z(\vec{p},\vec{q})=\frac{4}{3}\left [3(\vec{p}\cdot\vec{q})^2 - (|\vec{p}||\vec{q}|)^2 \right ],&\\
&L=3:~Z(\vec{p},\vec{q})=-\frac{8}{5}\left [5(\vec{p}\cdot\vec{q})^3 - 3(\vec{p}\cdot\vec{q})(|\vec{p}||\vec{q}|)^2 \right ],&\\
&L=4:~Z(\vec{p},\vec{q})=\frac{16}{35}\left [ 35(\vec{p}\cdot\vec{q})^4 - 30(\vec{p}\cdot\vec{q})^2(|\vec{p}||\vec{q}|)^2 + 3\right(|\vec{p}||\vec{q}|)^4 ],
\end{flalign}

\section{Investigation of the fit quality} \label{nominalLegendreMoments}

\begin{figure}[h]
  \centering
    \includegraphics[width=0.45\linewidth]{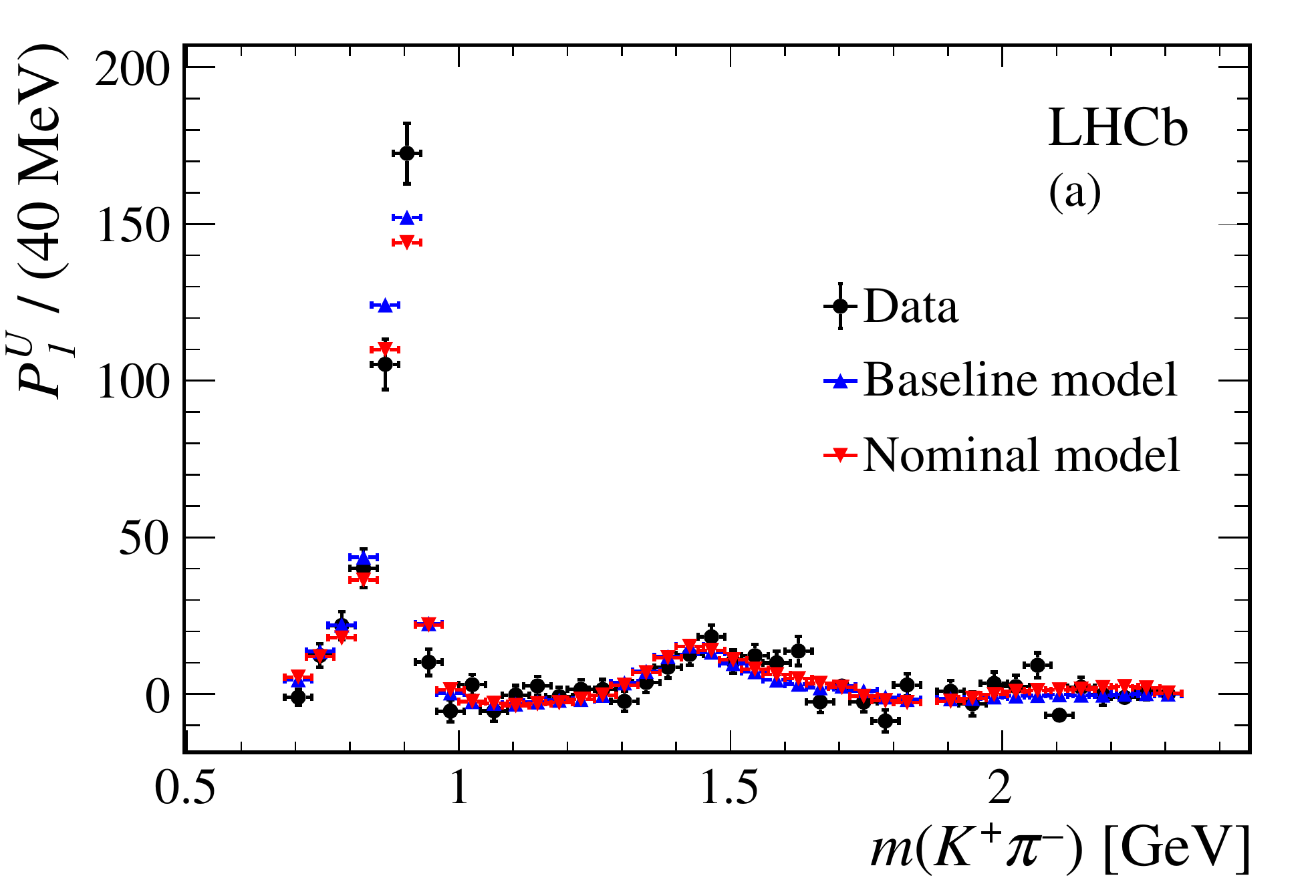}
    \includegraphics[width=0.45\linewidth]{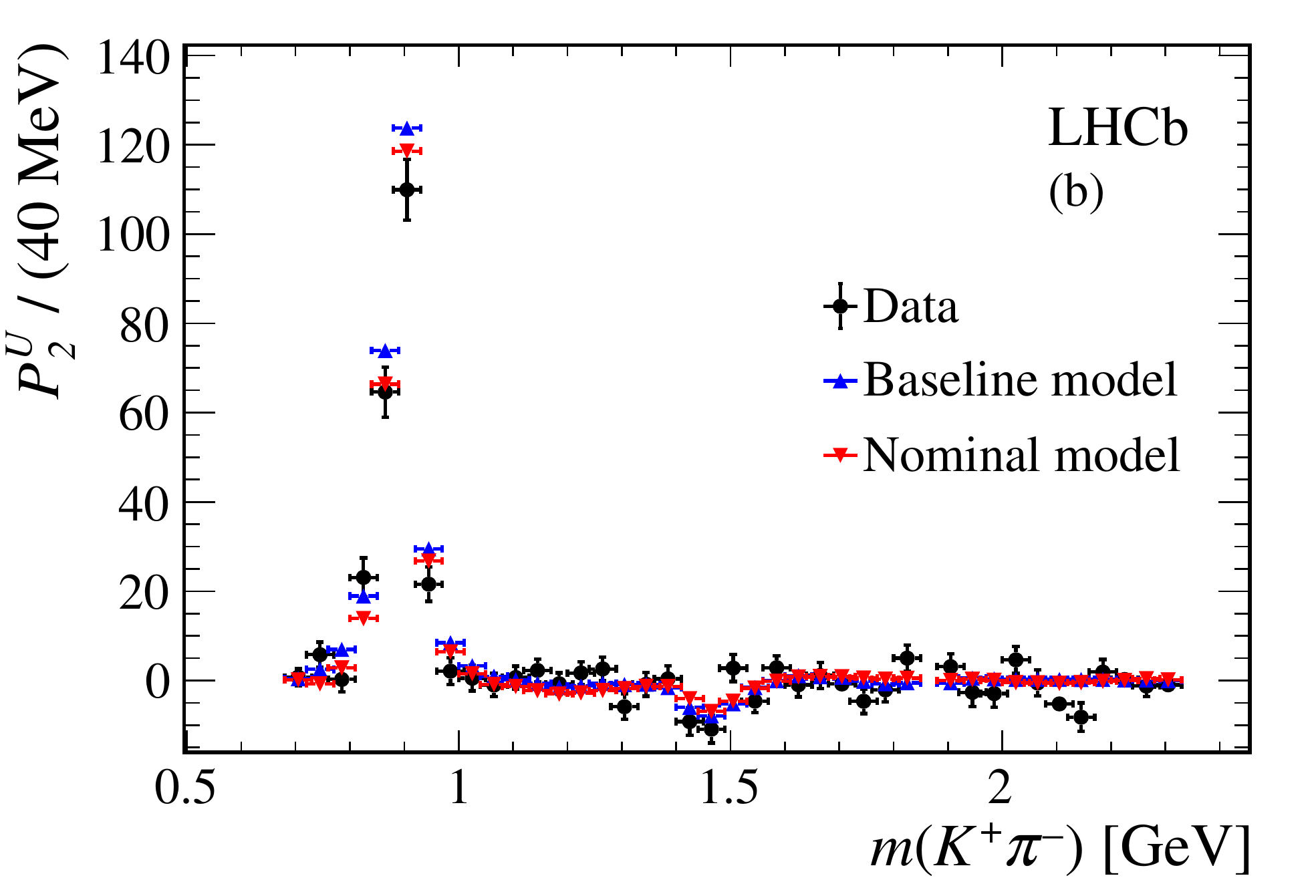}
    \includegraphics[width=0.45\linewidth]{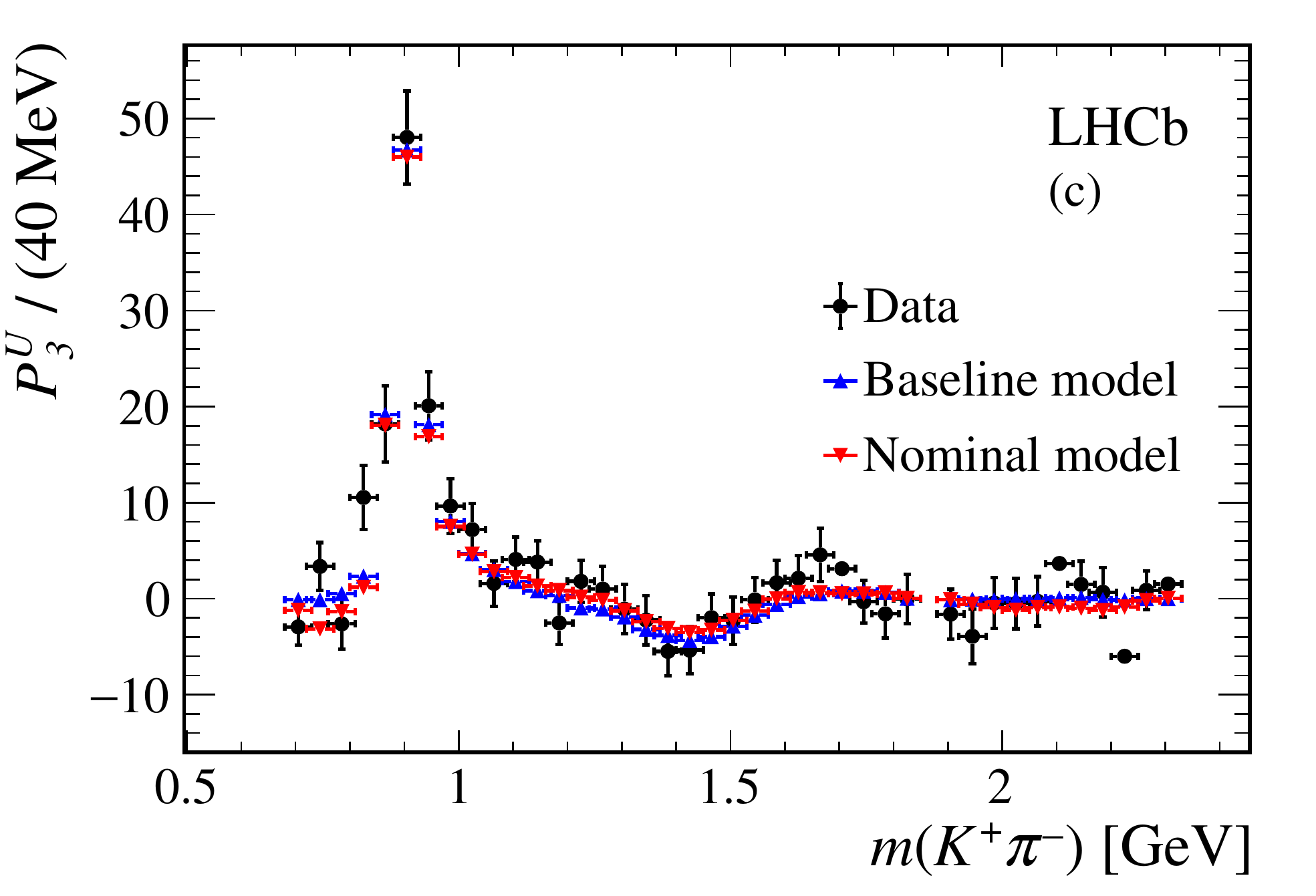}
    \includegraphics[width=0.45\linewidth]{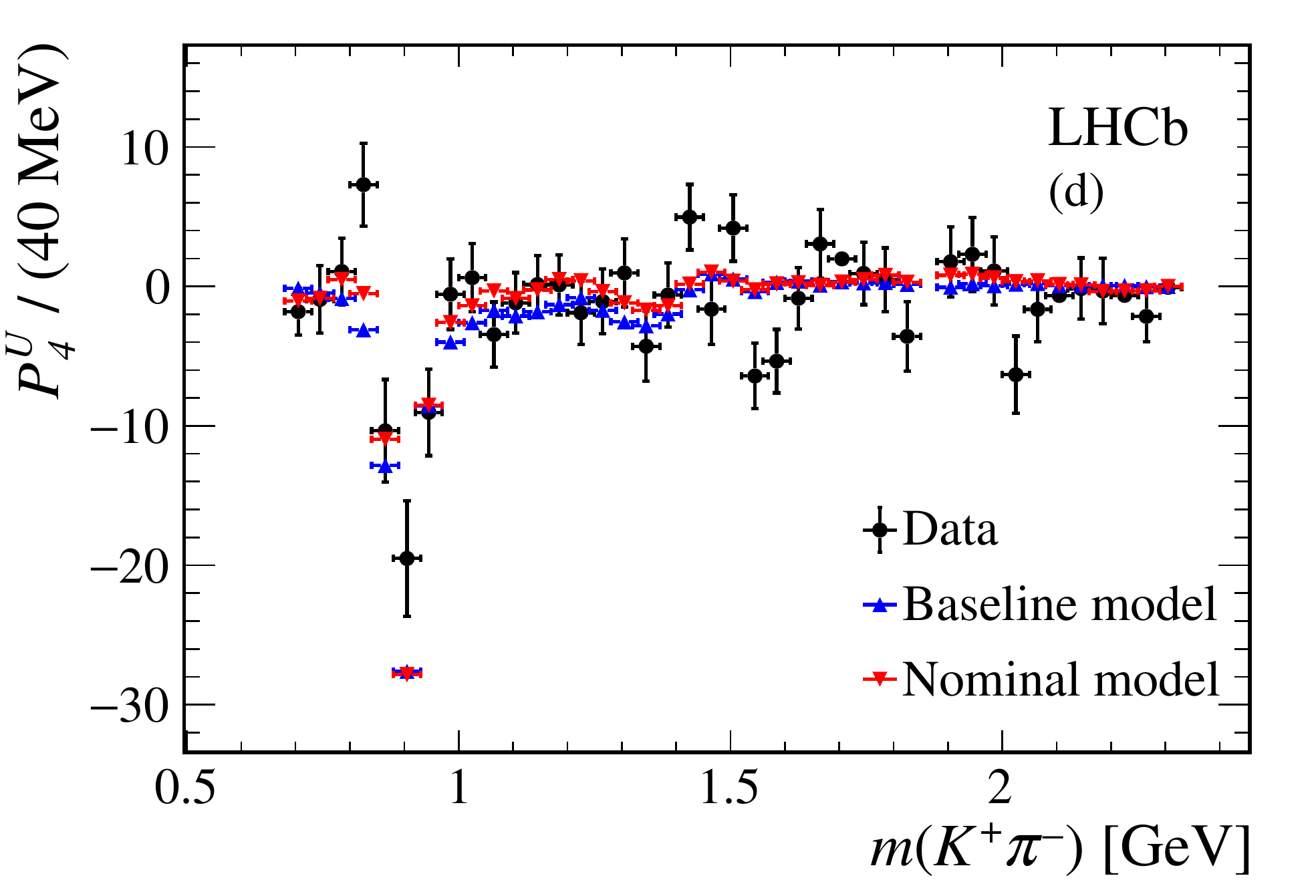}
  \caption{Comparison of the first four $\Kpi$ Legendre moments
    determined from background-subtracted data
    (black points) and from the results of the amplitude fit using the
    baseline model (red triangles) and nominal model (blue triangles) as a function of $\mKpi$.}
  \label{legendreM12base}
\end{figure}

\begin{figure}[h]
  \centering
    \includegraphics[width=0.45\linewidth]{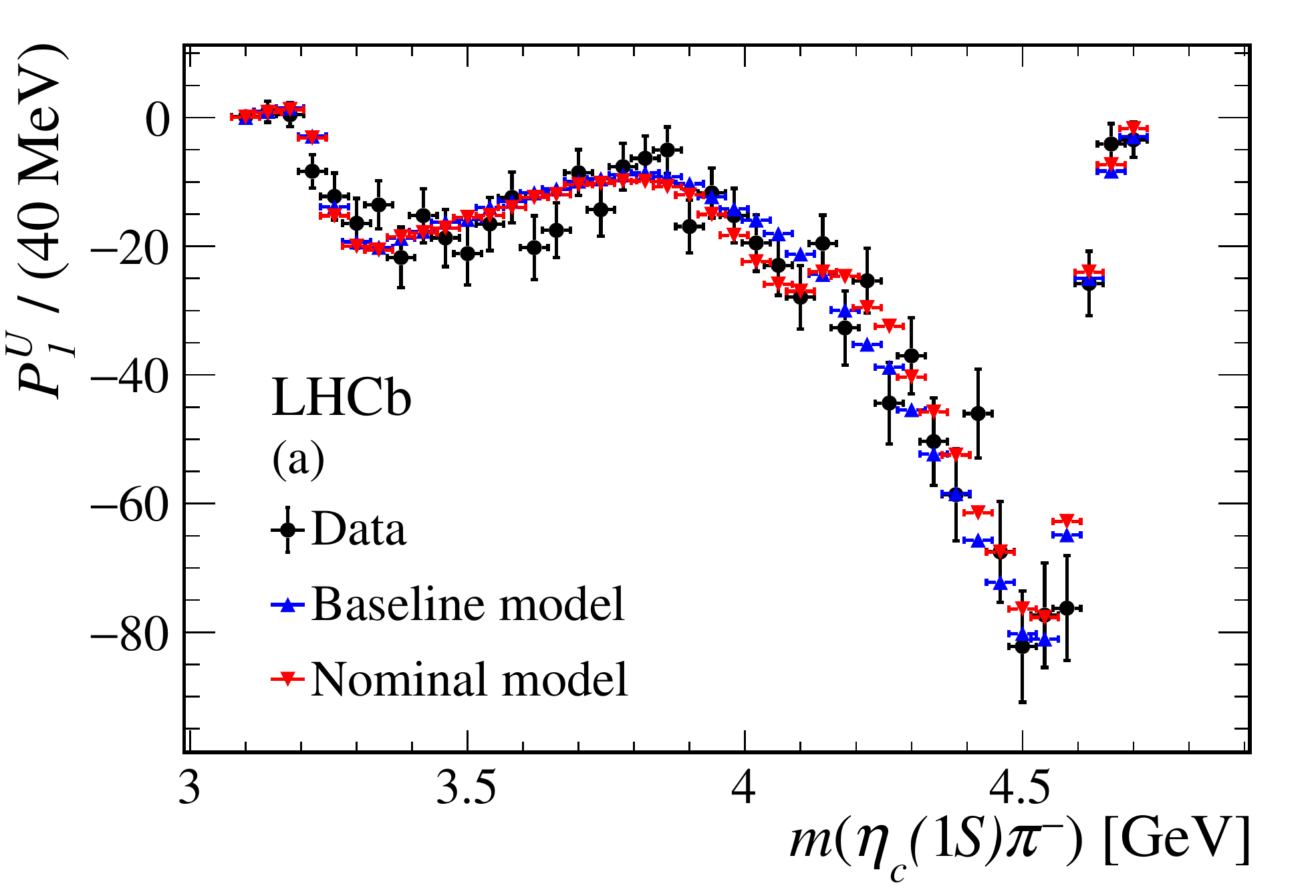}
    \includegraphics[width=0.45\linewidth]{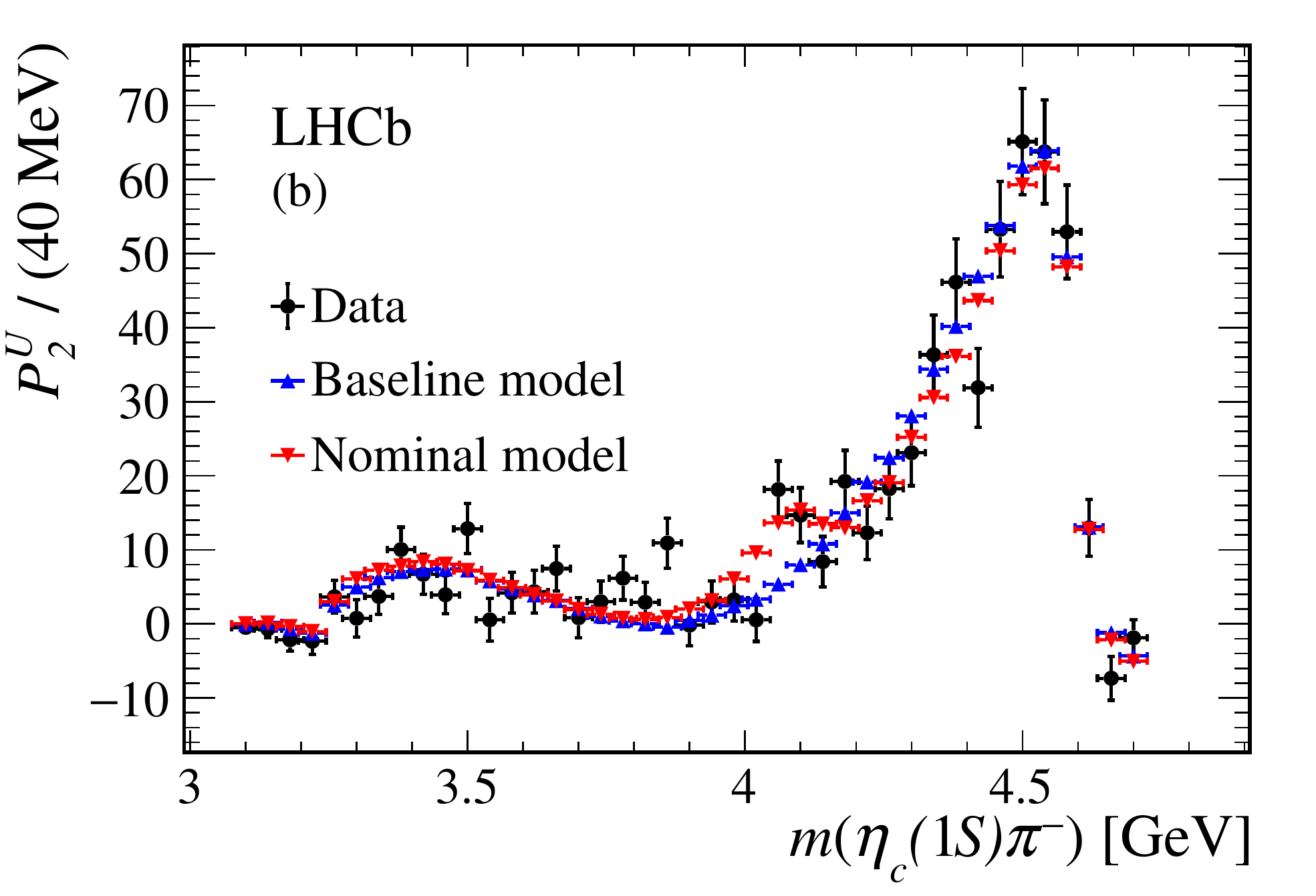}
    \includegraphics[width=0.45\linewidth]{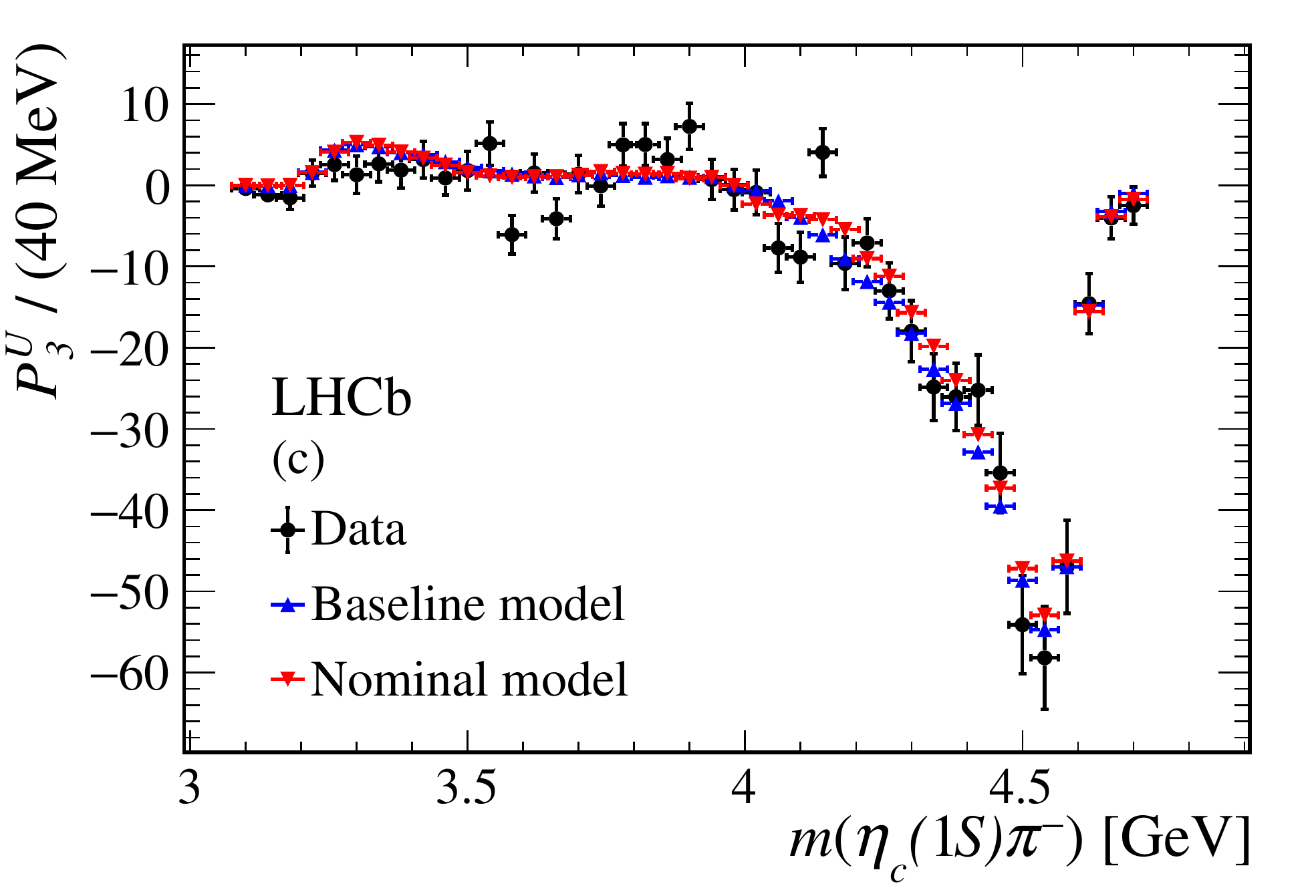}
    \includegraphics[width=0.45\linewidth]{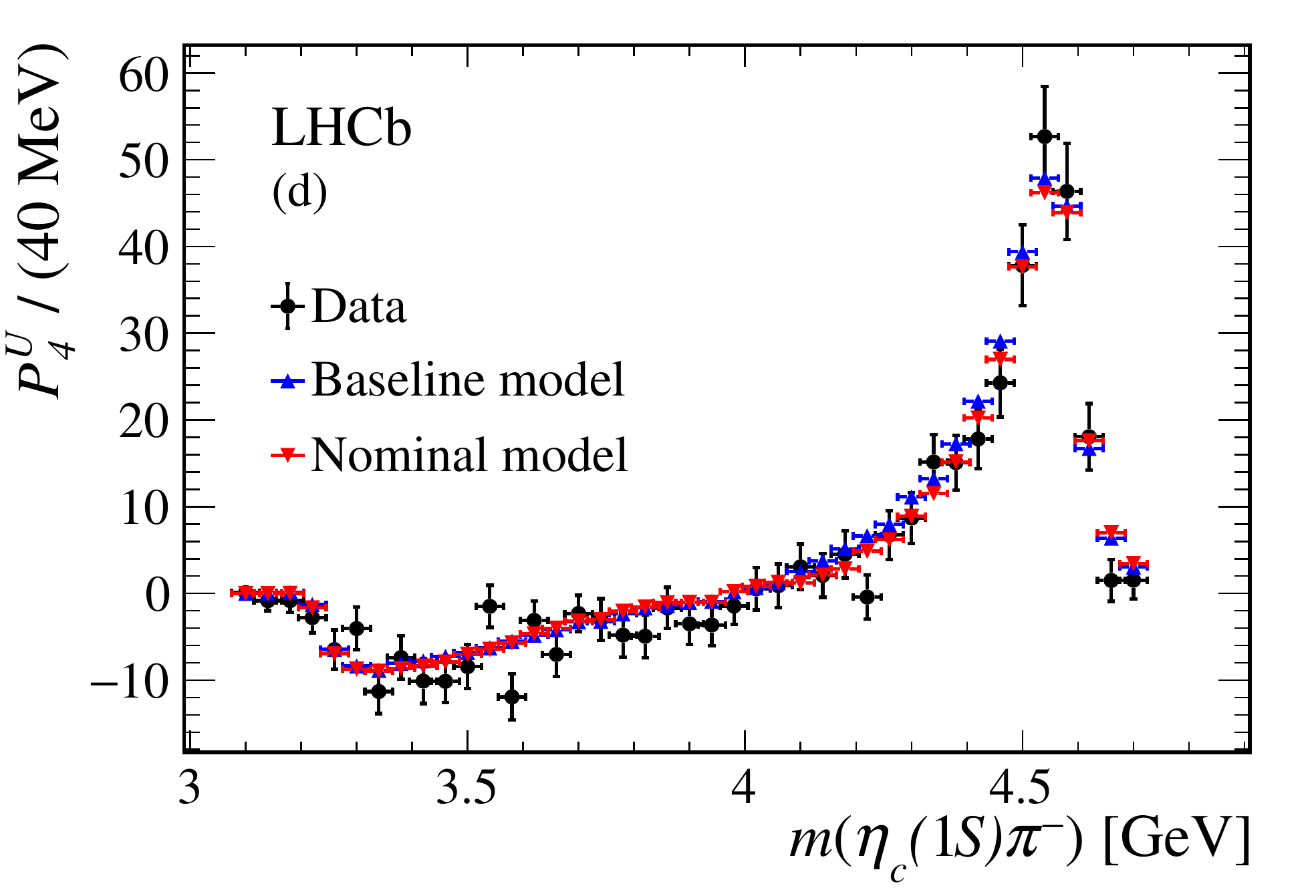}
  \caption{Comparison of the first four $\etacpi$ Legendre moments determined from background-subtracted data
    (black points) and from the results of the amplitude fit using the
    baseline model (red triangles) and nominal model (blue triangles) as a function of $\metacpi$.}
  \label{legendreM23base}
\end{figure}

\begin{figure}[h]
  \centering
    \includegraphics[width=0.45\linewidth]{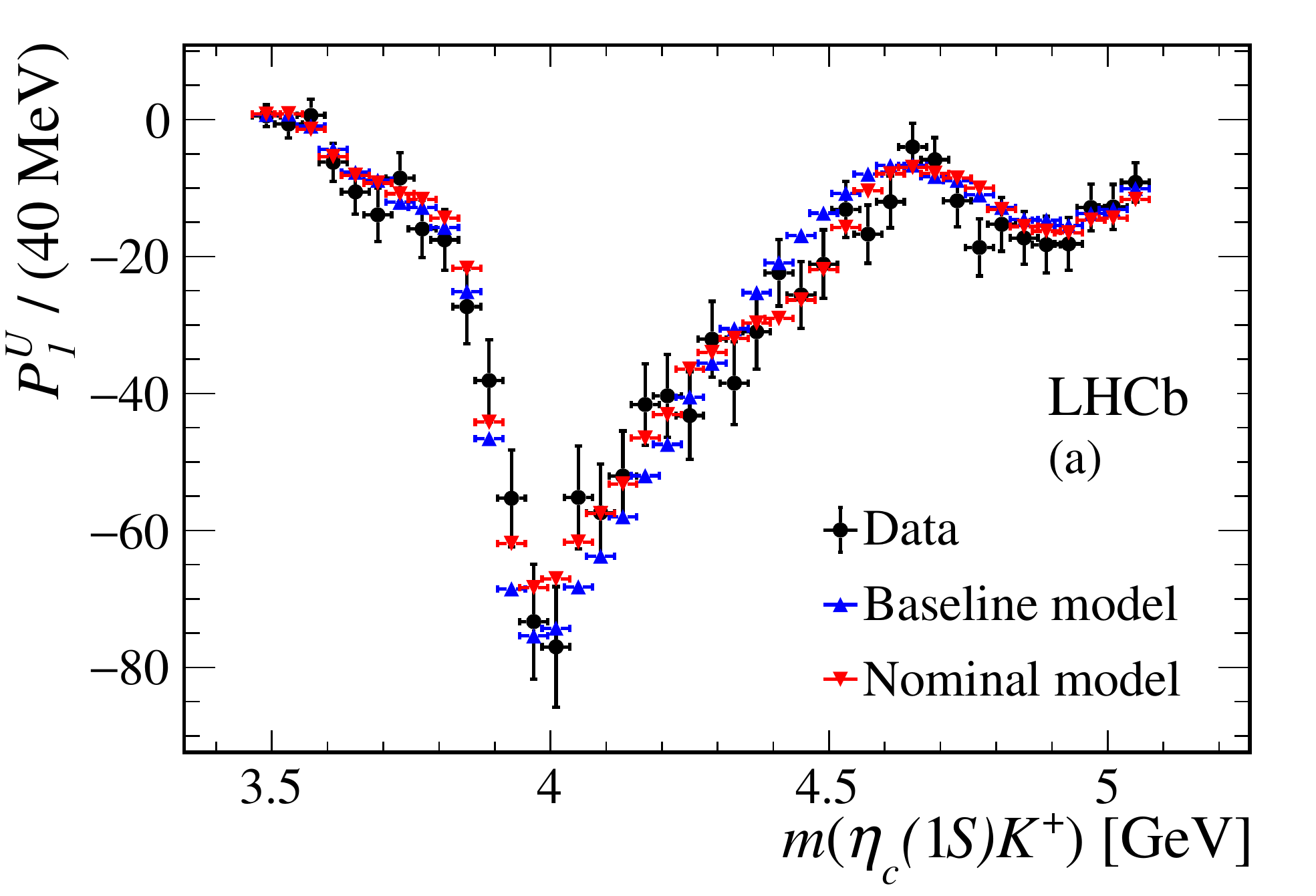}
    \includegraphics[width=0.45\linewidth]{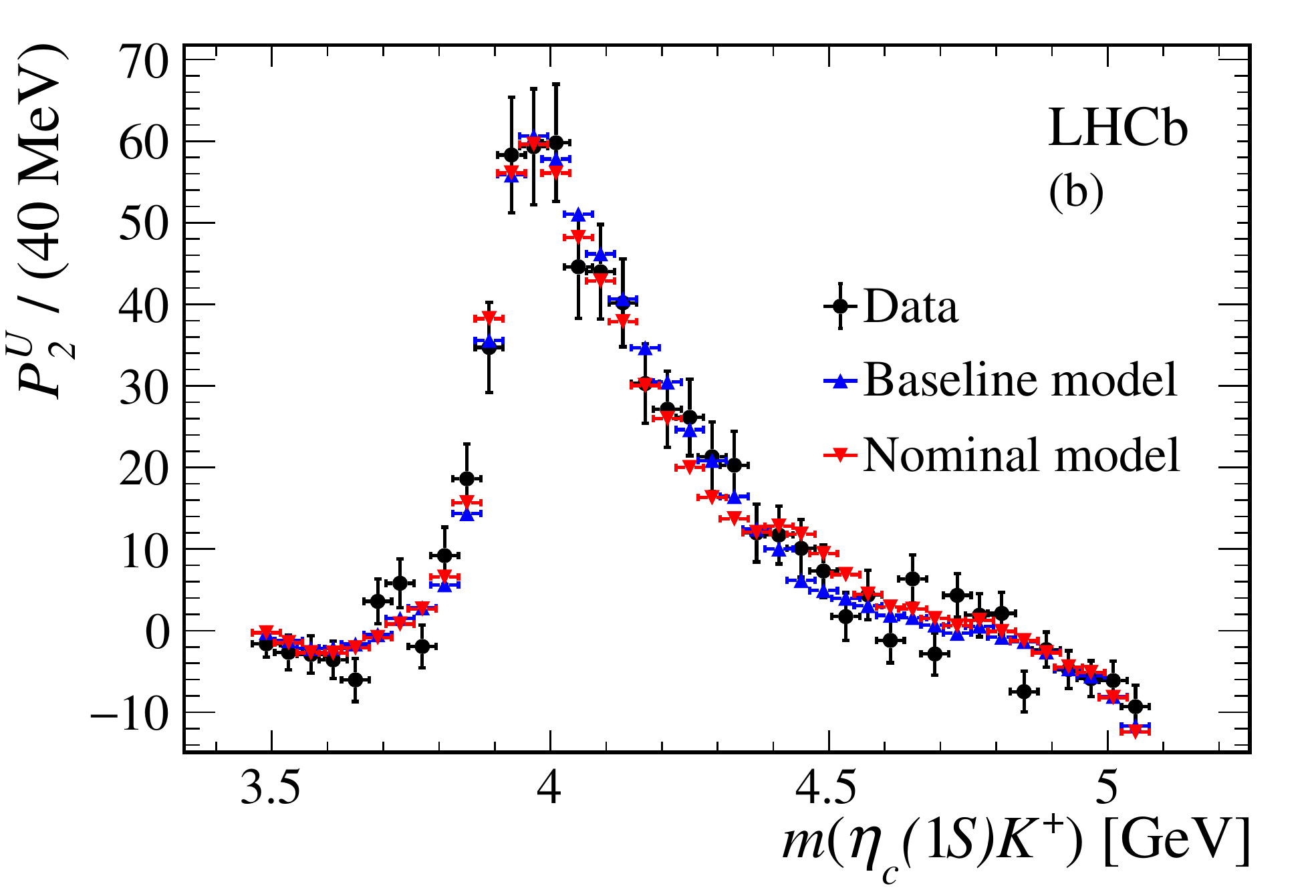}
    \includegraphics[width=0.45\linewidth]{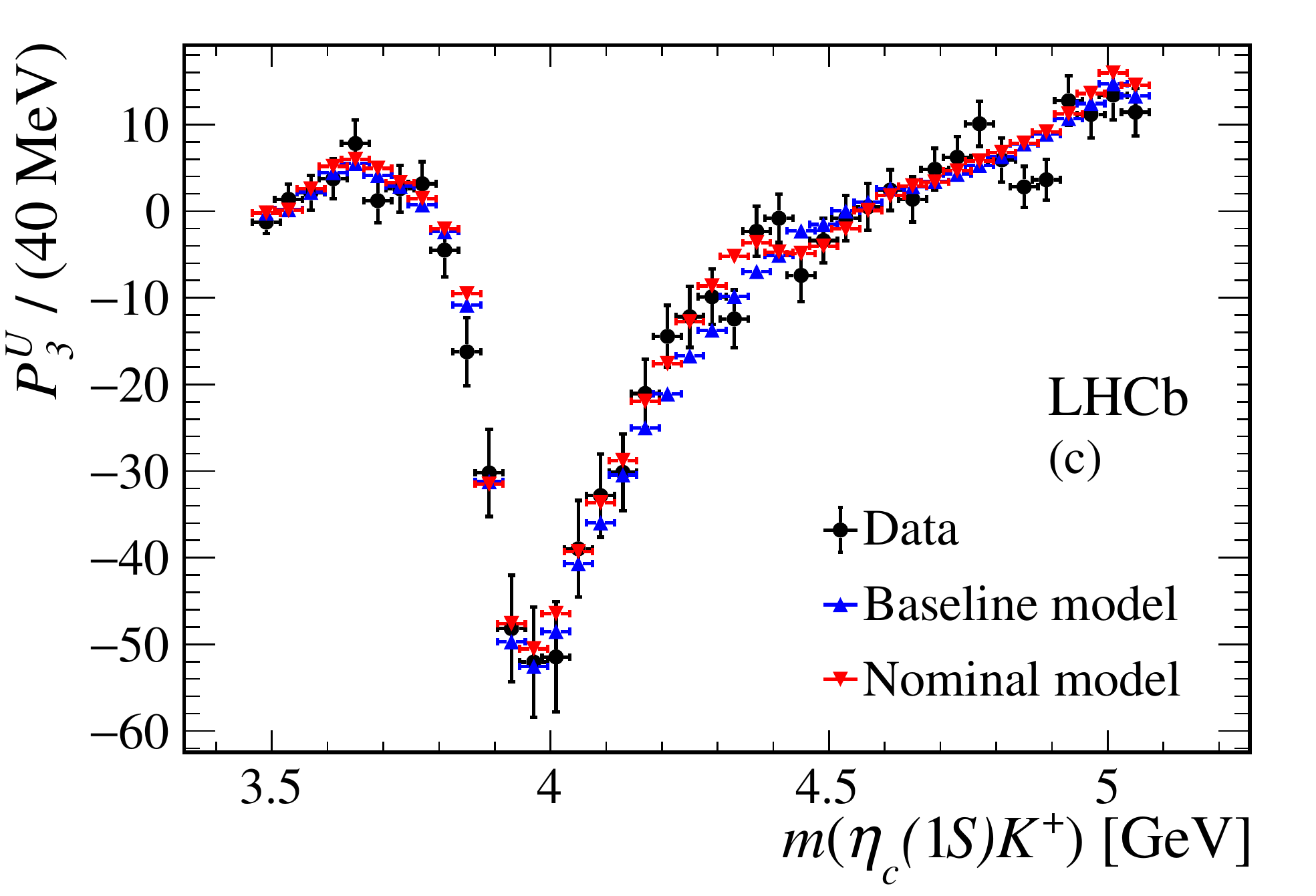}
    \includegraphics[width=0.45\linewidth]{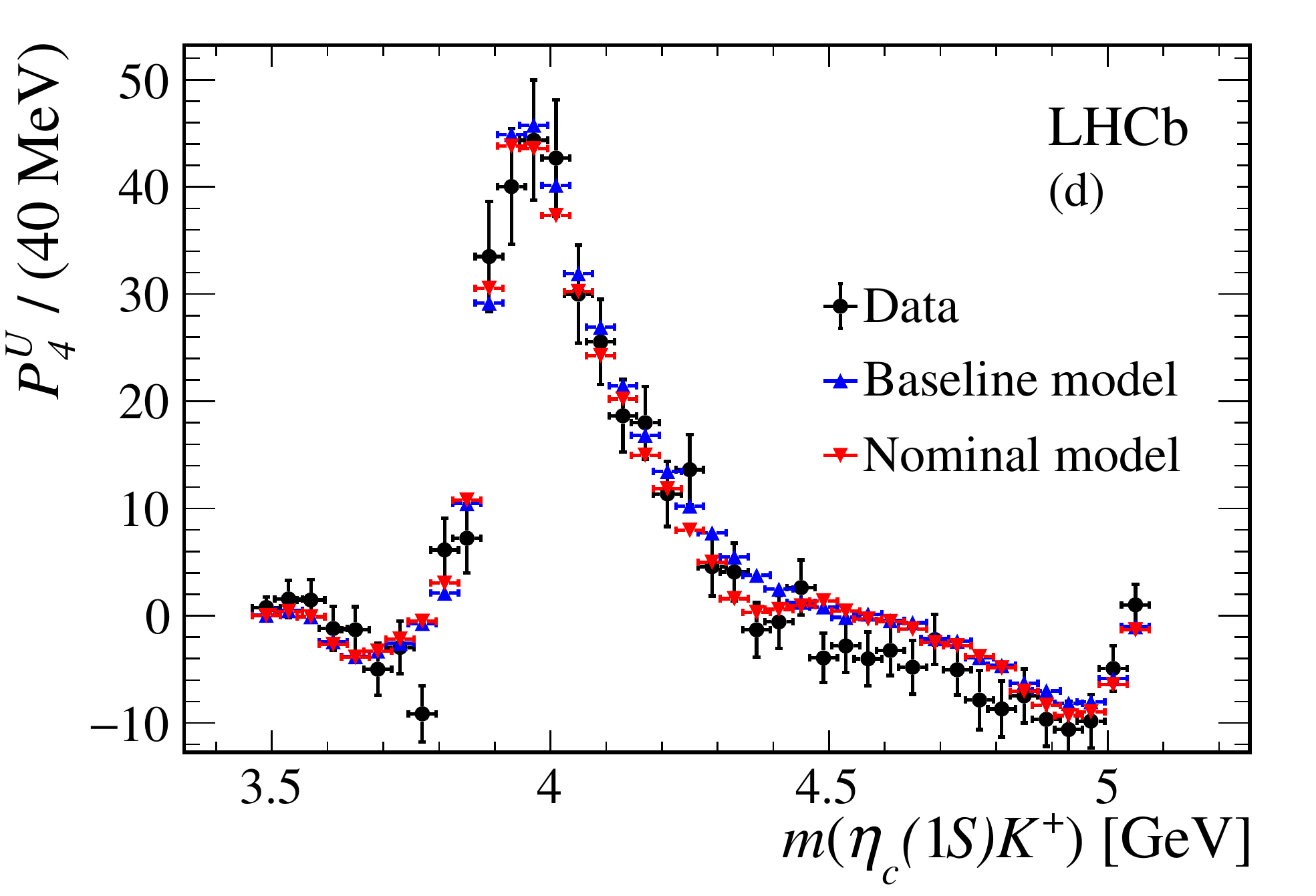}
  \caption{Comparison of the first four $\etacK$ Legendre moments determined from background-subtracted data
    (black points) and from the results of the amplitude fit using the
    baseline model (red triangles) and nominal model (blue triangles) as a function of $\metacK$.}
  \label{legendreM13base}
\end{figure}

\begin{figure}[h]
  \centering
    \includegraphics[width=0.9\linewidth]{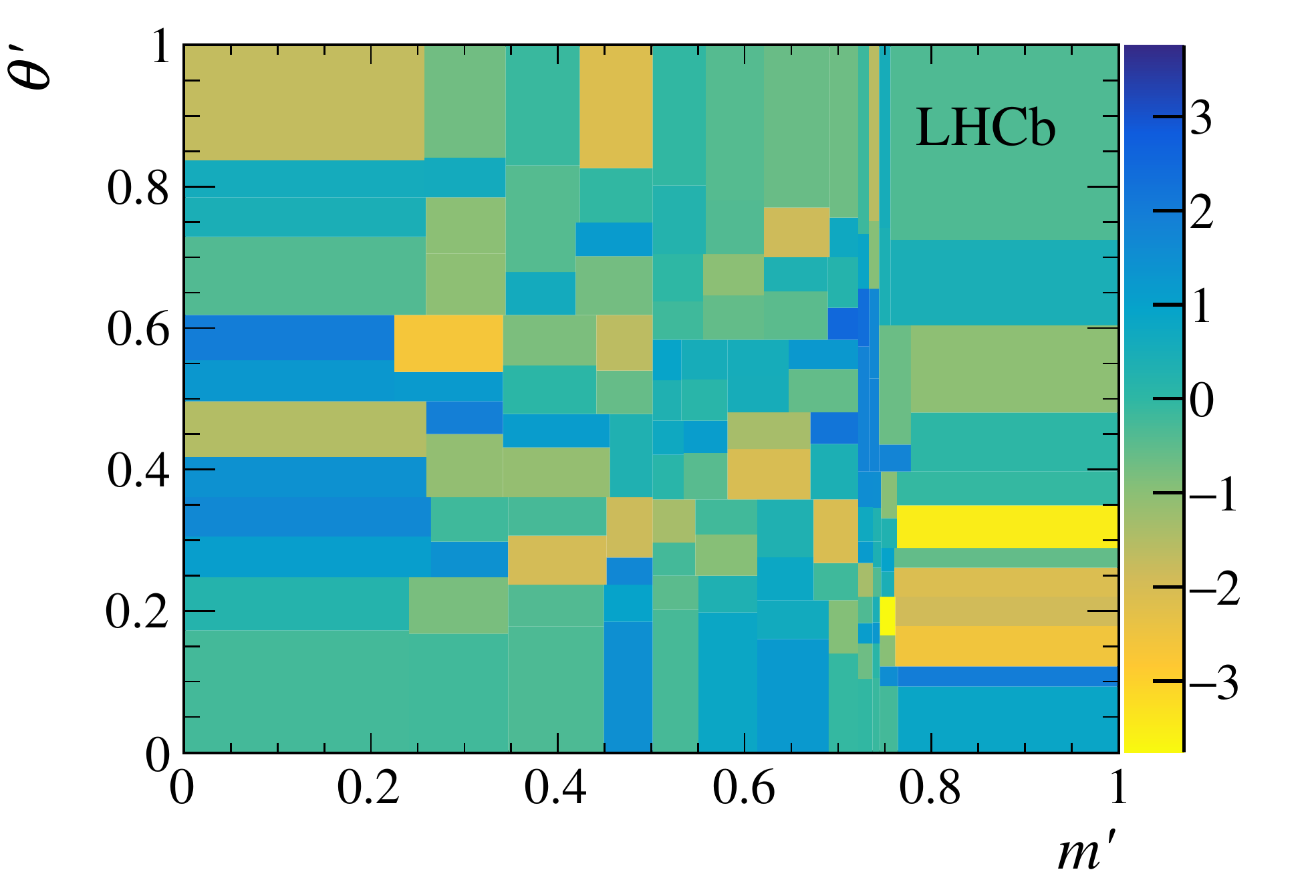}
  \caption{2D pull distribution for to the baseline model.}
  \label{pullBaseline}
\end{figure}

\begin{figure}[h]
  \centering
    \includegraphics[width=0.9\linewidth]{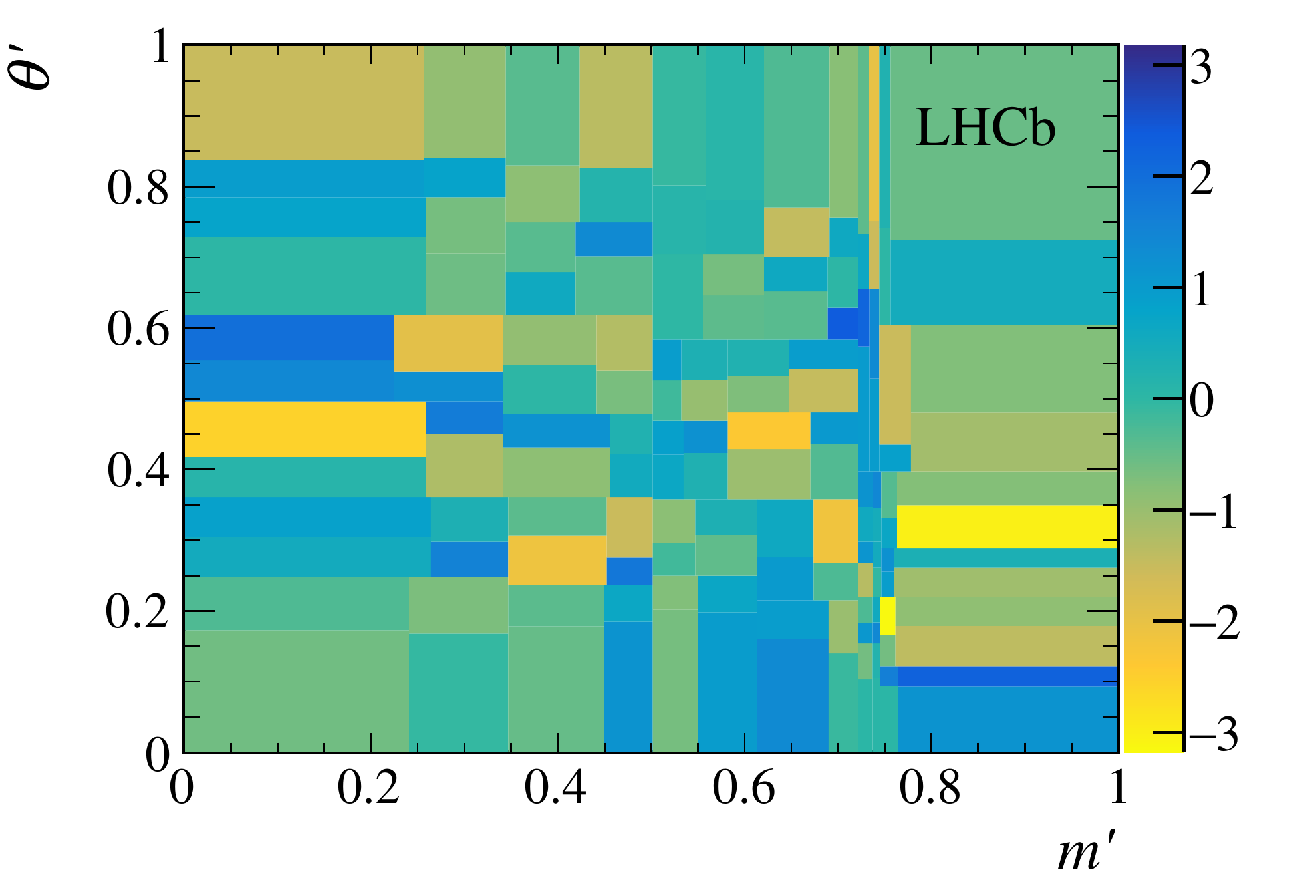}
  \caption{2D pull distribution for to the nominal model.}
  \label{pullNominal}
\end{figure}

Comparisons of the first four Legendre moments determined from
background-subtracted data and from the amplitude fit results using
the baseline and nominal model are reported in
Figs.~\ref{legendreM12base},~\ref{legendreM23base}
and~\ref{legendreM13base} for the $\mKpi$, $\metacpi$ and $\metacK$ projections,
respectively.\\
The 2D pull distributions for the baseline and nominal models are
reported in Figs.~\ref{pullBaseline} and~\ref{pullNominal}, respectively.

\clearpage
\newpage

%% file: LHCb_Authorship_flat_24-Jul-2018.tex
\centerline{\large\bf LHCb collaboration}
\begin{flushleft}
\small
R.~Aaij$^{28}$,
C.~Abell{\'a}n~Beteta$^{45}$,
B.~Adeva$^{42}$,
M.~Adinolfi$^{49}$,
C.A.~Aidala$^{77}$,
Z.~Ajaltouni$^{6}$,
S.~Akar$^{60}$,
P.~Albicocco$^{19}$,
J.~Albrecht$^{11}$,
F.~Alessio$^{43}$
% ...

{\footnotesize \it
% ...
$ ^{6}$Clermont Universit{\'e}, Universit{\'e} Blaise Pascal, CNRS/IN2P3, LPC, Clermont-Ferrand, France\\
% ...
$ ^{11}$Fakult{\"a}t Physik, Technische Universit{\"a}t Dortmund, Dortmund, Germany\\
% ...
$ ^{19}$INFN Laboratori Nazionali di Frascati, Frascati, Italy\\
% ...
$ ^{28}$Nikhef National Institute for Subatomic Physics, Amsterdam, Netherlands\\
% ...
$ ^{42}$Instituto Galego de F{\'\i}sica de Altas Enerx{\'\i}as (IGFAE), Universidade de Santiago de Compostela, Santiago de Compostela, Spain\\
$ ^{43}$European Organization for Nuclear Research (CERN), Geneva, Switzerland\\
% ...
$ ^{45}$Physik-Institut, Universit{\"a}t Z{\"u}rich, Z{\"u}rich, Switzerland\\
% ...
$ ^{49}$H.H. Wills Physics Laboratory, University of Bristol, Bristol, United Kingdom\\
% ...
$ ^{60}$University of Cincinnati, Cincinnati, OH, United States\\
% ...
$ ^{62}$Syracuse University, Syracuse, NY, United States\\
% ...
$ ^{77}$University of Michigan, Ann Arbor, United States, associated to $^{62}$\\
% ...
}
\end{flushleft}